\documentclass[12pt, oneside]{bookstyle}
\renewcommand*{\bibname}{References}
\newcommand{\eqref}[1]{(\ref{#1})}
\def\sn{{\rm sn\,}}
\def\cn{{\rm cn\,}}
\def\dn{{\rm dn\,}}
\def\sech{{\rm sech\,}}
\def\tanh{{\rm tanh\,}}
\def\cosh{{\rm cosh\,}}
\def\sinh{{\rm sinh\,}}
\def\cosec{{\rm cosec\,}}
\usepackage{lscape}
\usepackage{graphics}
\include{graphics}
\begin{document}
\baselineskip 24pt
\flushbottom
\frontmatter
\begin{titlepage}
\null
\begin{center}
{\bf \normalsize QUANTUM HAMILTON - JACOBI SOLUTION FOR SPECTRA OF SEVERAL ONE
  DIMENSIONAL POTENTIALS WITH SPECIAL PROPERTIES} \\
\vfill
{\normalsize A thesis submitted in partial fulfilment of the requirements\\
 for the award of the degree of} \\
\vfill
{\normalsize
{\bf DOCTOR OF PHILOSOPHY}\\
{\it in}\\
{\bf PHYSICS}\\
{\it by}\\
{\bf S. SREE RANJANI}}\\
\vfill
\includegraphics*{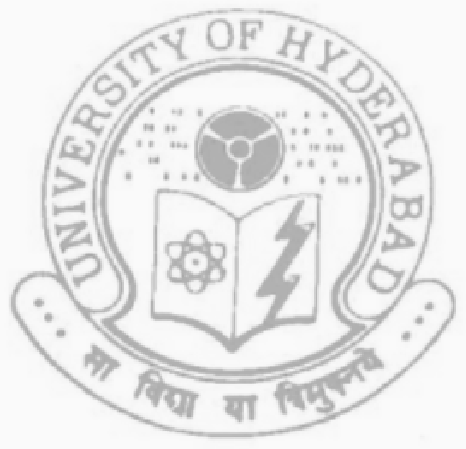}
\vfill
{\bf\small SCHOOL OF PHYSICS\\UNIVERSITY OF HYDERABAD\\ HYDERABAD - 500046,
INDIA\\ JUNE 2004\\}
\end{center}  
\end{titlepage}

\newpage
\begin{center}
\null
\vskip2.3cm
{\bf \Large Declaration}
\end{center}
\vskip1.5cm
     I, \textbf{\textit{S. Sree Ranjani}}, hereby declare that the
     work reported in this dissertation  
titled,  \textbf{\textit{Quantum Hamilton - Jacobi solution for spectra  of 
several one dimensional potentials with special properties}}, is
entirely original and has been carried out by me,  under the supervision of
\textbf{\textit{Prof. A. K. Kapoor}}, Department of Physics, School of
Physics, University of Hyderabad.
   
     To the best of my knowledge, no part of this dissertation was
submitted for any degree of any other institute or university.\\ \\ \\ \\ \\
Place: Hyderabad \\
Date:\hfill S Sree Ranjani

\newpage
\begin{center}
\null
\vskip2.3cm
{\Large \bf Certificate}
\end{center}
\vskip1.5cm
    This is to certify that the report entitled
\textbf{\textit{Quantum Hamilton - Jacobi solution for spectra of
several one dimensional potentials with special properties}}, being
submitted by \textbf{\textit{S Sree Ranjani}}, bearing
\textbf{\textit{Regd. No. 99phph02}}, in partial fulfillment of the
requirements for the award of \textbf{\textit{Doctor of Philosophy
in Physics}} by \textbf{\textit{University of Hyderabad}}, is a
bonafide work carried out at the \textbf{\textit{University of
Hyderabad}} under my supervision. The matter embodied in this report
has not been submitted to any other institute or university for the
award of any degree.\\ \\ \\ \\ \\
\begin{minipage}[b]{4.6cm}
\baselineskip 18pt
{\bf Dean},\\
School of Physics,\\
University of Hyderabad,\\
Hyderabad - 500046.
\end{minipage}
\hfill
\begin{minipage}[b]{4.6cm}
\baselineskip 18pt
\flushright{{\bf Prof. A. K. Kapoor},\\
School of Physics,\\
University of Hyderabad,\\
Hyderabad - 500046.}
\end{minipage}

\newpage
\begin{center}
\null
\vskip11cm
\textbf{\textit{Dedicated to\\
\vskip0.5cm
\hfill My parents and my supervisor}}
\end{center}

\newpage
\begin{center}
\null
\vskip2.3cm
\Large {\bf Acknowledgments}
\end{center}
\vskip1cm
\begin{minipage}[b]{13.7cm} 
\baselineskip18pt
\flushright{\textit{Catching the fleeting moment of a sunset, \\
           feeling the soft caress of the wind,\\
        the world I live for, the love I'd die for,\\
and all because, \textbf{you pass this way but once...}\\ }} 
\end{minipage}
\vskip0.5cm
I am glad, I came this way and I consider it a blessing to have
worked with my supervisor  Prof. A. K. Kapoor. The most beautiful
human being! His humility, his patience and confidence are always my
inspiration. His depth of understanding of the subject and his urge to
teach has always left me amazed. There are not enough words to express
my gratitude for his encouragement, understanding and support
throughout. There were many times, when I kept working only because he
believed in me, even when I lacked faith in myself. ``Dear sir, I
take this opportunity to thank you for everything and it was inspiring
to work under you''.
  
   I convey my sincere gratitude to Dr. P. K. Panigrahi for taking
interest in my work and giving me constant guidance. Discussions
with him were always lively and motivating. His valuable suggestions
allowed me to explore different aspects of my work and in a way
moulded it.

   I thank the Dean, School of Physics, Prof. V. S. S. Sastry and the former
dean, Prof. S. N. Kaul for providing congenial atmosphere for research.

   I thank Prof. V. Srinivasan and Prof. S. Chaturvedi for the
fascinating and enjoyable lectures. Though my interaction with them
is very little, their presence is felt and revered. I thank
Dr. M. Siva kumar, Dr. B. A. Bambah, Prof. C. Mukku and Dr. Anantha
Lakshmi for their encouragement and support.  

   I am thankful to all of the faculty members of the School of Physics for the
interesting courses and lectures in my M. Sc. These have helped me to
enhance my understanding in Physics and nurtured my dream of doing
research. I also thank all the non - teaching staff for their
support, especially I thank Mr. K. Srinivas (Gaddam) for his help with my
latex files and Mr. T. Abraham for his timely and cheerful help.
   
   The long association with the campus has left me with a treasure
trove of memories and experiences. The campus, with its blooming
trees, lovely lakes, beautiful sunsets and skies, has always
enchanted and mesmerized me. My life has become
richer with the friends, I made in the campus.  It's a pleasure to
extend my heartfelt appreciation to my friends, Swati, Chandu, Kuki,
Sarita, Charan, Sunil, Meri, Jo, Latha, Manoj, Madhu, Deepu, Nandu and
many more,
because of whom my days were filled with fun and laughter. The most
memorable were the excited discussions, which flamed and sustained our zeal for
physics. I acknowledge Mr. Seshagiri Rao for his encouragement. I thank my
colleagues and friends, Phani, Ajith, Mani, Raghu, Sathya, Rajneesh,
Rizwan, Chaitanya, Sudarshan, Farzana, Kumar and many others for
providing a friendly atmosphere and extending their help, when
needed. I also acknowledge my seniors, Siddique, Gurappa, Hari Kumar, Sunil,
Solomon, Raishma, Anita, Ramana, Phani Kumar, Naga Srinivas, Prem
Kiran, Geojo and Sandhya for their encouragement and help. I am
grateful to Charan and Rajneesh for the time and energy, they have
spent on going through the manuscripts of my papers.

   A special thanks to valli for always being there, when ever I
needed. I thank her for believing in me and dreaming for me. A note of
thanks to her family for the warmth and affection bestowed on me.
  
   I thank my parents, in - laws, sister, grand parents and all other
family members for their encouragement and support. I am grateful to
them and to my friends for standing by me during the bad phase
of ill health. I thank Mr. Bhaskar Rao, Mr. and Mrs. Dr. Ram Kumar for their
timely help.
   
   Last but not the least, I thank my husband, Ravi, for his unfailing
love and affection for constantly and patiently encouraging me to
push back my limits. Especially, I am deeply indebted to him and my
mother for taking care of me and helping me to build my battered
health and self esteem. Most of all for keeping my spirit
alive.\\

\begin{minipage}[b]{13.2cm}
\flushright{\it Two roads diverged in a wood, and \\
I took the one less travelled by, \\
and that has made all the difference.\\
- Robert Frost\\}
\end{minipage}

\newpage
\tableofcontents
\newpage
\mainmatter
\chapter{Introduction}
     
     It is well known that there are many formulations of classical
mechanics, like the Lagrangian, Hamiltonian, Poisson bracket and
Hamilton - Jacobi formulation [\ref{gold}, \ref{sal}]. Among these, the
first two are the most important and most used formulations. The
Hamilton - Jacobi formalism 
is an alternate formulation, which is not widely used and is more
esoteric.  Although, all these provide the same 
solutions to a given problem in an independent way, one needs to know
about them because they provide different
insights into the physics of the 
problem. Secondly, solving a particular problem might be easy in one 
formalism compared to the other. Hence, depending on the problem, one
needs to choose a method, in which it can be solved easily. Similar to
classical mechanics, there are various formulations of 
quantum mechanics [\ref{sty}]. Of these, some have been developed independently
and some have their origins in classical mechanics. The quantum
transformation theory along with the quantum 
Hamilton - Jacobi (QHJ) formalism, which is the subject of this
thesis, belongs to the later group.
The transformation theory has been developed
analogous to the classical theory, the roots of which, lie in the
works of Dirac 
[\ref{d1}, \ref{d2}], Jordan [\ref{jor}] and Schwinger 
[\ref{sw}]. The QHJ formalism, in its present form, has been proposed by
Leacock and Padgett [ \ref{lea}, \ref{pad}], on the lines of its classical
counterpart, to study stationary states.

     In classical mechanics, the theory of canonical transformation and
the Hamilton - Jacobi (HJ) theory are very well established
[\ref{gold}, \ref{sal}]. These  provided powerful formal techniques,
to deal with 
dynamical systems, which led to a better understanding of the
Hamiltonian formalism and the phase space. The HJ theory provides an
efficient tool to study the systems describing periodic motion. Making use
of the action - angle variables, one can find the frequency of the
periodic motion, without first finding the full solution of the
equation of motion explicitly. Similarly, in the QHJ formalism, one can
find the energy eigenvalues without solving the Schr\"odinger equation
for the stationary states.

   Analogous to the classical theory, the quantum momentum function (QMF)
$p$ is defined as $dS/dx$, where S is analogous to the Hamilton's
characteristic function and will be called quantum action
function. $S$ satisfies the equation 
\begin{equation}
\left(\frac{dS}{dx}\right)^2 - i\hbar\left(\frac{d^2S}{dx^2}\right)
=2m(E-V(x)),     \label{e0}   
\end{equation}
which in the limit $\hbar \rightarrow 0 $ goes to the classical
Hamilton - Jacobi equation [\ref{gold}]. The solution of the
Schr{\"o}dinger equation, $\psi$ is
related to $S$ by 
\begin{equation}
\psi = \exp\left(\frac{iS}{\hbar}\right)   \label {1e1a}
\end{equation}
and one can see that $p$ is the logarithmic derivative of $\psi$,
\begin{equation}
p = -i\hbar\frac{d}{dx}(\ln \psi).    \label{e1b}
\end{equation}
In the QHJ formalism, the quantum momentum function plays a central role.
Here, unlike the conventional wave mechanics, where one has to solve
the Schr{\"o}dinger equation, one makes use of the quantum Hamilton -
Jacobi equation for the QMF 
\begin{equation}
p^2 -i\hbar \frac{dp}{dx} = 2m (E- V(x)) ,    \label{e3}
\end{equation}
which is a Riccati type equation. The method makes use of the
singularity structure of the QMF to compute the energy eigenvalues in a
simple and elegant fashion [\ref{lea}, \ref{pad}]. Thus, for quantum
mechanical systems, which allow periodic motion, one can obtain the
bound state energy eigenvalues by defining 
the quantum action $J$, which is a generalization of the classical
action variable and $J$ satisfies the exact quantization condition,
\begin{equation}
J =J(E) =\frac{1}{2\pi} \oint_C  pdx  = n \hbar ,  \label{e1}  
\end{equation}
where $C$ is the contour in the complex plane enclosing the $n$ poles
of the QMF on the real line. This quantization condition is exact and
is a result of the oscillation theorem [\ref{cou}, \ref{in1}] which states that
the $n^{th}$ excited state wave function has $n$ zeros. From
\eqref{e1b}, one can see that these $n$ zeros correspond to the $n$
poles of the QMF and are called moving poles, as they are dependent on
the initial conditions. It can be shown from \eqref{e3} that these
poles have a residue $-i\hbar$. Thus, applying the Cauchy residue
theorem to the integral in \eqref{e1}, one obtains the above
quantization condition. 
 
Along with these moving poles, the QMF will have other 
singularities originating from the potential which are called fixed
singularities. Here, one makes a simplifying assumption that {\it the
point at infinity is an isolated singular point}. This assumption is
equivalent to saying that the QMF has finite number of moving poles in the
complex domain. Evaluating the integral in \eqref{e1}, in terms of 
these other singular points and using the Cauchy residue theorem, one
can obtain the energy eigenvalue expression. Due to the quadratic
nature of the QHJ equation, all the residues at the fixed poles will have two
values. Hence, one needs to choose the value of the residue, which
will give the 
right physical behaviour. For this, one needs to define the QMF
completely. The QMF is related to the classical momentum function
$p_c$ by the relation  
\begin{equation}
\lim_{\hbar \rightarrow 0} p \rightarrow p_c,   \label{e2}
\end{equation}
which doubles up as a boundary condition for the QMF, as well as
being the correspondence principle. Note that $p_c$ is double valued
and has to be defined precisely. The details of its definition are
given in the next chapter. 

    Bhalla {\it et.al.} [\ref{b1} - \ref{b}] used this formalism to
compute the bound state spectra of one dimensional exactly solvable
(ES) supersymmetric potentials [\ref{susyb}, \ref{supr}]. They have
also demonstrated the applicability of the QHJ 
formalism to the potentials, which exhibit both broken and unbroken phases
of SUSY [\ref{b3}, \ref{b}], depending on the range of the potential
parameters. For a given potential, it was shown that implementing the
boundary conditions correctly, every value of the residue, at a given
fixed pole was  
acceptable in a particular range of parameters and one can obtain the
energy eigenvalues for both 
the phases of SUSY through the QHJ formalism. Extending this 
investigation to complicated potentials like Ginocchio, Natanzon
etc, Jayanthi {\it et. al.} obtained the energy eigenvalues for these
potentials and also obtained the wave functions for a few states [\ref{jth}].

     After successfully obtaining the solutions for bound state
energies, it was natural to look at the other potential models 
like the quasi - exactly solvable (QES) models [\ref{ush}]. These are termed as
QES, because one cannot analytically obtain their entire energy
spectrum but it is possible to obtain a few states analytically, when the
potential parameters satisfy a specific condition known as the quasi
- exact solvability condition. Applying QHJ formalism to these models
it was found that unlike the case of ES models, where one obtains the
energy expressions, one is lead to the quasi - exact solvability
condition by the application of the quantization
condition[\ref{geo}, \ref{geoth}]. In this case, the property of square
integrability was used to choose the right value of residue instead of
the boundary condition \eqref{e2}, due to the difficulty in its
implementation to these models.

  In the above investigations except in [\ref{jth}], the main
emphasis was on obtaining the energy eigenvalues and not much
attention was paid to eigenfunctions. We have shown in [\ref{sree}],
that one can easily obtain the bound state eigenfunctions for one dimensional
exactly solvable models [\ref{susyb}] along with the energy eigenstates
through the QHJ formalism. In addition to these, the
eigenfunctions and eigenvalues for the QES models, discussed in
[\ref{geo}], were obtained in [\ref{geoth}]. 

   Having solved one dimensional ES and QES models, we have taken up the
application of this formalism to more complicated problems like
the periodic potentials [\ref{hill} - \ref{ppsuk}] and the PT
symmetric potentials [\ref{ptkh}, \ref{bag}, \ref{zaf}]. The application of QHJ
formalism to these models is not a straight forward extension of the
method used for the bound state problems in one dimension. In
particular, attention has to be paid to the choice of residues because
square integrability condition cannot be insisted. Also the
quantization condition holds as long as the oscillation theorem
holds. For PT symmetric potentials, it is not clear whether a
generalization of the oscillation theorem can be found. So we had to
find an alternative, which will give the energy values,for which, the
details are given in the later chapters. 

The plan of the thesis is as follows. In the next chapter, we
present an introduction to the QHJ formalism. We will describe in
detail, how one can employ the QHJ formalism to obtain the energy
eigenfunctions and eigenvalues, by explicitly working out the well 
known hydrogen atom potential. In chapter III, the study of Scarf 
[\ref{sc}] and Scarf - I [\ref{susyb}, \ref{b2}] potentials is taken
up. Both the potentials are interesting models because they exhibit
different spectra depending on the range of the potential 
parameters. The Scarf potential exhibits band structure for one
parameter range and ordinary bound state spectrum for another range.
The other potential Scarf - I exhibits the broken and unbroken phases
of SUSY in different regions of its potential parameters. For these
two models, both the regions of the parameters and the corresponding
energy spectra emerge naturally through the QHJ formalism. 

     The next two chapters deal with periodic potentials belonging to
the family of elliptic potentials [\ref{hill} - \ref{ppsuk}], namely
the Lam{\'e} potential, 
\begin{equation}
V(x) = j(j+1) m\,\, \sn^{2}(x,m)    \label{e3a}
\end{equation}
with $j$ being an integer and the associated Lam{\'e} potential (ALP)
[\ref{ppsuk}] 
\begin{equation}
V(x) = a(a+1)m\,\, \sn^{2}(x,m) + b(b+1) m\,\,
\frac{\cn^{2}(x,m)}{\dn^{2}(x,m)}.   \label{e4}
\end{equation}
Here, the functions $\sn(x,m), \cn(x,m)$ and $\dn(x,m)$ are the
elliptic functions with $m$ being the elliptic modulus. The ALP is ES
if $a = b = j$, $j$ being an integer and is QES, when $a \neq b$
with $a+b, a-b$ being integers.  Here, we show, how one can obtain
the band edge solutions of these 
potentials using the QHJ formalism. This way of obtaining the
solutions, turns out to be much simpler compared to the existing
methods [\ref{ars}] and the comparison of the singularity structure
with the ordinary ES and QES models is taken up [\ref{akk},\ref{ppqes}]. 

     Chapter IV deals with ES periodic potentials {\it i.e.} the Lam{\'e}
and the ES case of the ALP. For these potentials, we obtain the
general forms of the band edge 
wave functions in terms of the potential parameters. Explicit
expressions for the energy and wave functions are worked out for specific
examples [\ref{akk}]. In chapter V, we study the QES periodic
potentials {\it i.e.} 
the ALP, when $a \neq b$ and $a, b$ taking non - integer values. Here,
we obtain the quasi exact solvability conditions along with the general forms
of the wave functions [\ref{ppqes}]. 
      
    Chapter VI deals with the study of PT symmetric ES and QES solvable
potentials. In this chapter, the complex Scarf potential and the
Khare - Mandal model are discussed [\ref{pt}]. The comparison of the
singularity structure of the QMF of these 
potentials with the ordinary ES and QES potential models is given. Chapter VII
contains the concluding remarks along with a brief discussion on
further applications of this work.\\

\noindent
{\bf References}

\begin{enumerate}

\item{\label{gold}} H. Goldstein, {\it Classical Mechanics} 
(Addison-Wesley/Narosa, Indian student edition, 1995)

\item{\label{sal}} E. J.  Saletan and  A. H. Cromer  {\it Theoretical
  Mechanics}  (John Wiley, New York, 1971).

\item{\label{sty}} D. F. Styer, {\it et. al. Am. J. Phys.} {\bf 70}
  (3) 288 (2002)

\item{\label{d1}} P. A. M. Dirac, {\it The Principles of Quantum
  Mechanics} (Oxford University Press, London, 1958).

\item{\label{d2}} P. A. M. Dirac,
 {\it Rev. Mod. Phys.} {\bf 17}, 195 (1945) ; P. A. M. Dirac,
 Proc. R. Soc. London, {\bf 113A}, 621 (1927).

\item{\label{jor}} P. Jordan, {\it Z. Phys.} {\bf 38}, 513 (1926). 

\item{\label{sw}} J. Schwinger, {\it Quantum Electrodynamics} (Dover
 Publications, Inc. New York, 1958).

\item {\label{lea}} R. A Leacock and M. J. Padgett   {\it
  Phys. Rev. Lett.} {\bf 50}, 3 (1983). 

\item {\label{pad}} R. A. Leacock and M. J. Padgett   {\it Phys. Rev.}
{\bf D28}, 2491 (1983).

\item{\label{cou}}  R. Courant and D. Hilbert, {\it Methods of
  Mathematical Physics}, Vol 1 (Wiley interscience, New York, 1953).

\item {\label{in1}} E. L. Ince, {\it Ordinary Differential Equations}
  (Dover Publications, Inc, New York, 1956).  

\item {\label{b1}} R. S. Bhalla, A. K. Kapoor and P. K. Panigrahi, {\it
Am. J. Phys.} {\bf 65}, 1187 (1997).

\item {\label{b2}} R. S. Bhalla, A. K. Kapoor and P. K. Panigrahi {\it
Mod. Phys. Lett. A}, {\bf 12}, 295 (1997).

\item {\label{b}} R. S. Bhalla, {\it The quantum Hamilton - Jacobi
 approach to energy spectra of potential problems}, {\it Ph. D. thesis}
 submitted to University of Hyderabad (1996).

\item {\label{susyb}} F. Cooper, A. Khare and U. Sukhatme, {\it
Supersymmetry in Quantum Mechanics} (World Scientific, Singapore, 2001)  
and references therein.

\item {\label{supr}} F. Cooper, A. Khare and U. Sukhatme, {\it
  Phys. Rep.} {\bf 251}, 267 (1995).

\item {\label{b3}} R. S. Bhalla, A. K. Kapoor and P. K. Panigrahi,
  {\it Int. J. Mod. Phys. A}, {\bf 12}, No. {\bf 10}, 1875 (1997). 

\item{\label{jth}} S. Jayanthi, {\it Study of one
  dimensional potential problems using quantum Hamilton - Jacobi
  formalism}, {\it M. Phil thesis} submitted to University of Hyderabad (1998).

\item {\label{ush}} A. G. Ushveridze, {\it Quasi-exactly solvable models in
quantum mechanics}, (Institute of Physics Publishing, Bristol, 1994). 

\item {\label{geo}} K. G. Geojo, S. Sree Ranjani and A. K. Kapoor,
  {\it J. Phys A : Math. Gen.} {\bf 36}, 4591 (2003); preprint quant
  - ph/0207036. 

\item{\label{geoth}} K. G. Geojo, {\it Quantum Hamilton - Jacobi
  study of wave functions and energy spectrum of solvable and quasi -
  exactly solvable models}, {\it Ph. D. thesis} submitted to University of
  Hyderabad (2004).

\item {\label{sree}} S. Sree Ranjani, K. G. Geojo, A. K Kapoor and
  P. K. Panigrahi, to be published in {\it Mod. Phys. Lett. A.} {\bf
  19}, No. {\bf 19}, 1457 ( 2004);  preprint quant - ph/0211168. 

\item{\label{hill}} W. Magnus and S. Winkler, {\it Hills Equation}
  (Interscience Publishers, New York, 1966).

\item{\label{ars}} F. M. Arscot, {\it Periodic Differential equations}
  (Pergamon, Oxford, 1964).

\item{\label{ppkh}} A. Khare and U. Sukhatme, {\it J. Math. Phys.} {\bf 40},
  5473 (1999); preprint quant - ph/9906044.

\item{\label{ppsuk}} A. Khare and U. Sukhatme, {\it J. Math. Phys.} {\bf 42},
  5652 (2001); preprint quant - ph/0105044.

\item{\label{ptkh}} A. Khare and B. P. Mandal; quant-ph/0004019.

\item{\label{bag}}  B. Bagchi, S. Mallik, C. Quesne and R. Roychoudhury;
  quanth-ph/ 0107095.

\item{\label{zaf}} Z. Ahmed, {\it Phys. Lett. A.} {\bf 282}, 343 (2001).

\item{\label{sc}} F. L. Scarf, {\it Phys. Rev.} {\bf D57}, 1271 (1998).

\item {\label{akk}} S. Sree Ranjani, A. K. Kapoor and P. K. Panigrahi,
  to be published in {\it Mod. Phys. Lett. A.} {\bf 19}. No. {\bf 27} 2047 (2004); preprint quant
  - ph/0312041 

\item{\label{ppqes}}  S. Sree Ranjani, A. K. Kapoor and P. K. Panigrahi;
  preprint quant - ph/0403196.

\item{\label{pt}}  S. Sree Ranjani, A. K. Kapoor and P. K. Panigrahi;
  preprint quant - ph/0403054. 

\end{enumerate}

\newpage
\chapter{Quantum Hamilton - Jacobi Formalism}
      
      In this chapter, we will briefly discuss the classical Hamilton
- Jacobi (HJ) formalism, followed by a detailed description of the QHJ
theory. By means of an example, the singularity structure of the QMF is
explained and we will show, how one can obtain the energy eigenvalues
from the exact quantization condition and simultaneously obtain the
eigenfunctions. The example, which we use to elucidate the method, is
the well known hydrogen atom, for which, we obtain both the eigenvalues
and eigenfunctions. 

\section{Classical Hamilton - Jacobi theory}

    The classical HJ theory deals with continuous canonical
transformations, which relates the old canonical coordinates $(q_i,p_i)$ at
time $t$ to the new canonical coordinates $(Q_i,P_i)$, which are constant
and may have 
initial values $(q_0,p_0)$ at time $t_0$ [\ref{2gol}, \ref{2sa}]. The equations
of transformations are of the following form        
\begin{equation}
q_i = q_i(Q_i,P_i,t_0) \,\,\, ;\,\,\, p_i = p_i(Q_i,P_i,t_0).    \label{2c1}
\end{equation}
One can go from the old coordinates
to the new coordinates with the help of a generating function. There
are four types of generating functions and in our discussion, we
consider a generating function $F_2(q_i,P_i,t)$, with $q_i$ and $P_i$
as the independent variables, which will be useful for our present
discussion. The new Hamiltonian $K$ is related to the old Hamiltonian
$H$ by   
\begin{equation}
K = H +\frac{\partial }{\partial t}F_2(q_i,P_i,t).  \label{2c2}
\end{equation}
If $K$ is equal to zero, then
\begin{eqnarray}
\dot{Q_i} & =  & \frac{\partial K}{\partial P_i} =0 , \nonumber \\
\dot{P_i} &=&   -\frac{\partial K}{\partial Q_i} = 0,  \label{2c3}
\end{eqnarray}
such that, the new canonical coordinates are constants and \eqref{2c2} becomes 
\begin{equation}
 H +\frac{\partial }{\partial t}F_2(q_i,P_i,t) =0 .  \label{2c4}
\end{equation}
The transformation equations in terms of the generating function are
\begin{eqnarray}
p_i &=& \frac{\partial }{\partial q_i}F_2(q_i, P_i, t) \nonumber \\
Q_i &=&\frac{\partial }{\partial P_i}F_2(q_i,P_i,t).    \label{2c5}
\end{eqnarray}
Using the above equations in \eqref{2c4}, one obtains 
\begin{equation}
H(q_i, \frac{\partial}{\partial q_i} F_2(q_i,P_i,t),t) + \frac{\partial
  }{\partial t}F_2(q_i,P_i,t) =0,  \label{2c6}
\end{equation}
which is known as the HJ equation, which constitutes a partial
differential equation of $F_2(q_i,P_i,t)$ in $(n+1)$ variables. The
solution of \eqref{2c6} is termed as the Hamilton's principle
function and depends on $n$ constants of integration $\alpha_i$ so
that, we may write 
\begin{equation}
F_2 \equiv S(q_i, \alpha_i,t) \label{2c7}
\end{equation}
and the new momenta $P_i$ can be chosen to be the independent constants
of integration $\alpha_i$, {\it i. e.} $P_i = \alpha _i$. Now the
transformations in \eqref{2c5} can be written as  
\begin{equation}
p_i= \frac{\partial }{\partial q_i}S(q_i, \alpha_i, t),    \label{2c8}
\end{equation}
which is the classical momentum function and
\begin{equation}
Q_i = \frac{\partial }{\partial \alpha_i}S(q_i,\alpha_i,t) = \beta_i
. \label{2c9} 
\end{equation}
where $\beta_i$ is a constant. At time $t_0$, these constitute $2n$
equations relating $n \, \alpha's $ and $n \, \beta's$ to the initial $q_i,p_i$
values, using which, one can evaluate the constants of integration in
terms of the initial coordinates. Thus, the above equation can be used
to obtain $q_i,p_i $ in terms of $\alpha, \beta$ and $t$ as
\begin{equation}
q_i = q_i(\alpha_i, \beta_i,t)\,\,\, , p_i = p_i(\alpha_i, \beta_i,t).
\label{2c10} 
\end{equation}
Therefore, when solving the HJ equation, one not only obtains the
generating function for the canonical transformation but also obtains
the solution of the mechanical problem.
 
      When $H$ is not an explicit function of $t$, then the HJ
equation for $S$ can be written as
\begin{equation}
H \left(q_i, \frac{\partial S}{\partial q_i} \right) + \frac{\partial
  S}{\partial t} = 0,   \label{2c11}
\end{equation}
where the first term involves only the $q$ dependence and the second
term involves time. Hence, one can assume a solution of the form
\begin{equation}
S(q_i, \alpha_i,t) = W(q_i, \alpha_i) - \alpha_1 t,   \label{2c12}
\end{equation}
where $W$ is the Hamilton's characteristic function. Substituting this
in \eqref{2c11} gives the HJ equation for the characteristic function as
\begin{equation}
H \left(q_i, \frac{\partial W}{\partial q_i} \right) =\alpha_1.  \label{2c13}
\end{equation}
The transformation equations will be 
\begin{eqnarray}
p_i &=& \frac{\partial W}{\partial q_i}, \nonumber \\
Q_i &=& \frac{\partial W}{\partial P_i}  = \frac{\partial W}{\partial
  \alpha_i}.  \label{2c14} 
\end{eqnarray}
Since, $H$ and $W$ are independent of $t$, from \eqref{2c2}, one obtains 
\begin{equation}
K = \alpha_1   \label{2c15}
\end{equation}
and the equations of motion are
\begin{equation}
\dot{P}_i = - \frac{\partial K}{\partial{Q_i}} = 0 \label{2c16}
\end{equation}
with solution $P_i = \alpha_i$ and
\begin{eqnarray}
\dot{Q}_i =  \frac{\partial K}{\partial{\alpha_1}} = 1 \,\,\,i=1  \nonumber \\
                                                   = 0 \,\,\, i\neq 1
                                                   \label{2c17} 
\end{eqnarray}
with solutions 
\begin{eqnarray}
Q_1 & = & t + \beta_1 \equiv \frac{\partial W}{\partial{\alpha_1}}
\,\,\,, i = 1 \nonumber \\
Q_i &=  & \,\,\,\,\, \beta_i\,\,\,\, \equiv \frac{\partial W}{\partial{\alpha_i}} \,\, ,\,\,
i \neq 1   \label{2c18}
\end{eqnarray}
From the above equation, we see that only $Q_1$ is not a constant, but is
a function of time. Thus, with HJ equation of the characteristic
function, one can solve the systems in which $H$ is independent of
$t$.

\subsection{Action - angle variables}
    
      As a variation to the above discussed HJ method, one can deal
with systems with periodic motion. Here, for each periodic motion, one
introduces the action variable defined in terms of the momentum $p$
as
\begin{equation}
   J = \oint p dq ,\label{2c19}
\end{equation}
where the integration is carried over a complete period of the periodic
motion in the phase space. The Hamilton's characteristic function in
terms of this new variable is
\begin{equation}
W = W(q,J).    \label{2c20}
\end{equation}
The coordinate conjugate of $J$, known as the angle variable is
defined as
\begin{equation}
\omega = \frac{\partial W}{\partial J}    \label{2c21}
\end{equation}
and the corresponding equation of motion is
\begin{equation}
\dot{\omega} = \frac{\partial H(J)}{\partial J} \equiv \nu  \label{2c22}
\end{equation}
with the solution
\begin{equation}
\omega = \nu t +\beta \label{2c23}
\end{equation}
where $\nu$ is the frequency.
Equation \eqref{2c21} can be solved for $q$ as a function of $\omega$
and $J$, which, in combination with \eqref{2c23}, will give the desired
solution as a function of time.

      In order to see how these variables are advantageous, consider
the change in $\omega$, as $q$ goes through a complete cycle of
rotation, which is given by
\begin{equation}
\Delta \omega = \oint \frac{\partial \omega}{\partial q}dq.   \label{2c24}
\end{equation}
using \eqref{2c21}, one gets
\begin{equation}
\Delta \omega = \oint \frac{\partial^2 W}{\partial q \partial J}dq.
\label{2c25} 
\end{equation}
Since $J$ is a constant, one can write
\begin{equation}
\Delta \omega = \frac{d}{dJ}\oint \frac{\partial W}{\partial q}dq
\equiv  \frac{d}{dJ}\oint pdq =1,    \label{2c26}
\end{equation}
where \eqref{2c19} has been put to use. From \eqref{2c23}, it follows
that, if $\tau$ is the period for complete cycle of $q$, then
\begin{equation}
\Delta  \omega  = 1 = \nu \tau.  \label{2c27}
\end{equation}
Hence, the constant $\nu$ can be identified as the reciprocal of the
time period
\begin{equation}
\nu = \frac{1}{\tau} \label{2c28}
\end{equation}
and therefore, the frequency is associated with the periodic motion of
$q$. Thus for a
one dimensional periodic system, the frequency of the system can be
found by determining $H$ as a function of $J$ and using \eqref{2c22}.
Hence the use of action - angle variables, provides a powerful
technique for obtaining the frequency of the periodic motion, without
finding a complete solution of the motion of the system. 

\section{Quantum Hamilton - Jacobi Formalism}

     The QHJ formalism draws its inspiration from the classical HJ
theory and was first initiated in this form by Leacock and Padgett, in
1983 [\ref{2le},\ref{2pa}]. As mentioned in the introduction, the
central entity in this formalism is the quantum
momentum function p, which is defined in three dimensions as
\begin{equation}
\vec{p} = \vec{\nabla} S \label{2c29}
\end{equation}
where $S$ is analogous to the classical characteristic function $W$. $S$
is related to the solution of the Schr\"odinger equation,
\begin{equation}
- \frac{\hbar^2}{2m}\nabla^2\psi(x,y,z)+ V(x,y,z) \psi(x,y,z) = E
  \psi(x,y,z)    \label{2c30} 
\end{equation}
by the relation 
\begin{equation}
\psi(x,y,z) = \exp\left(\frac{iS}{\hbar}\right)       \label{2c31}
\end{equation}
which, when substituted in \eqref{2c30}, gives
\begin{equation}
(\vec{\nabla}S)^2 -i \hbar \vec{\nabla}.(\vec{\nabla}S) = 2m (E
  - V(x,y,z)).   \label{2c32} 
\end{equation}
Substituting \eqref{2c29} in \eqref{2c32} gives the QHJ equation
for $\vec{p} $ as 
\begin{equation}
(\vec{p})^2 - i \hbar \vec{\nabla}.\vec{p} = 2m (E - V(x,y,z))
  \label{2c33} 
\end{equation}
and from \eqref{2c29} and \eqref{2c31}, one can see that $\vec{p}$ 
is the the logarithmic derivative of $\psi(x,y,z)$ {\it i. e}. 
\begin{equation}
\vec{p} = -i \hbar \frac{\vec{\nabla}\psi(x,y,z)}{\psi(x,y,z)}   \label{2c34}
\end{equation}
The above discussion of the QHJ formalism is done in three
dimensions but, from here on, we proceed to one
dimension, as we will be discussing only one dimensional potentials in
this thesis. One can see that in one dimension, the QHJ equation
\eqref{2c33} takes the form of the Riccati equation
\begin{equation}
p^2 - i \hbar \frac{dp}{dx} = 2m (E - V(x)).       \label{2c34a}
\end{equation}
In the QHJ formalism, $x$ is treated as a complex variable, there by
extending the definition of p to the complex plane.  
To obtain the full solution of the Riccati equation, one needs to know
$p$ completely in the complex plane. From \eqref{2c33}, one
observes that   
\begin{equation}
\lim_{\hbar \rightarrow 0} \,\,\,\,\, p \rightarrow p_{c} ,        \label{2c35}
\end{equation}
where $p_c = \sqrt{2m(E-V(x))}$ is the classical momentum, which can be
used as a boundary condition on $p$. When $x$ is a complex variable,
$p_c$ needs to be defined carefully, since, it is a double valued function.
In the phase space, $p_c$ vanishes at points where $E =
V(x)$. In the complex $x$ plane, the classical region is on the real
axis, in between the turning points $x_1$  and
$x_2$. Correspondingly,  $p_c$ is given a branch cut in between the
turning points and is defined to be the branch, which is positive
just below the branch cut [\ref{2le}, \ref{2pa}]. This 
completes the definition of $p_c$  and the role of \eqref{2c35}
is not only that of a boundary condition on $p$, but also as a
correspondence principle.

\subsection{Singularities of the QMF} 
     
     It has already been mentioned that the singularity structure of 
the QMF plays an important role in obtaining the eigenfunctions
and eigenvalues through the QHJ formalism. The location of
singularities of $p$ and the residues at these singular points are
the essential ingredients required in this formalism. 

   The QMF has two types of singularities, the moving and the fixed
singularities.  
It is well known from the oscillation theorem, that the $n^{th}$
excited state wave function has $n$ zeros on the real line. These
zeros correspond to the poles of $p$, as seen from \eqref{2c34}. The
location of these poles depends on the energy and the 
initial conditions. Hence, these are called as moving poles. For
Riccati equation, it is known that only poles can appear as moving
singularities [\ref{2in}]. It is easy to see that the residue at
these moving poles is $-i\hbar$. Let $x_0$ be a point, where $V(x)$ is
analytic and $p$ has a  moving
pole.  Assuming a moving pole of order $m$, we proceed to calculate
the residue of $p$ at $x_0$, by expanding $p$ in Laurent expansion
around $x_0$ as 
\begin{equation}
p = \sum_{k=1}^m \lambda_k(x-x_0)^{-k} +
\sum_{k=0}^{\infty}\gamma_k (x-x_0)^k.       \label{2c35a}
\end{equation}
Substituting \eqref{2c35a} in the QHJ equation \eqref{2c34a} and
comparing the coefficients of different powers of $x-x_0$, one gets $m=1$
and 
\begin{equation}
\lambda_1 = - i \hbar.    \label{2c36}
\end{equation}
Thus, the moving poles are simple poles with residue $-i\hbar$. This
information of the moving 
poles is used to obtain the quantization condition satisfied by the
quantum action 
\begin{equation}
J = J(E) \equiv \frac{1}{2\pi} \oint_C{pdx}.  \label{2c37}
\end{equation}
Here, $C$ is the contour, which encloses the moving poles of
$p$ as shown in the fig. 2. 1. 
\begin{center}
\includegraphics*{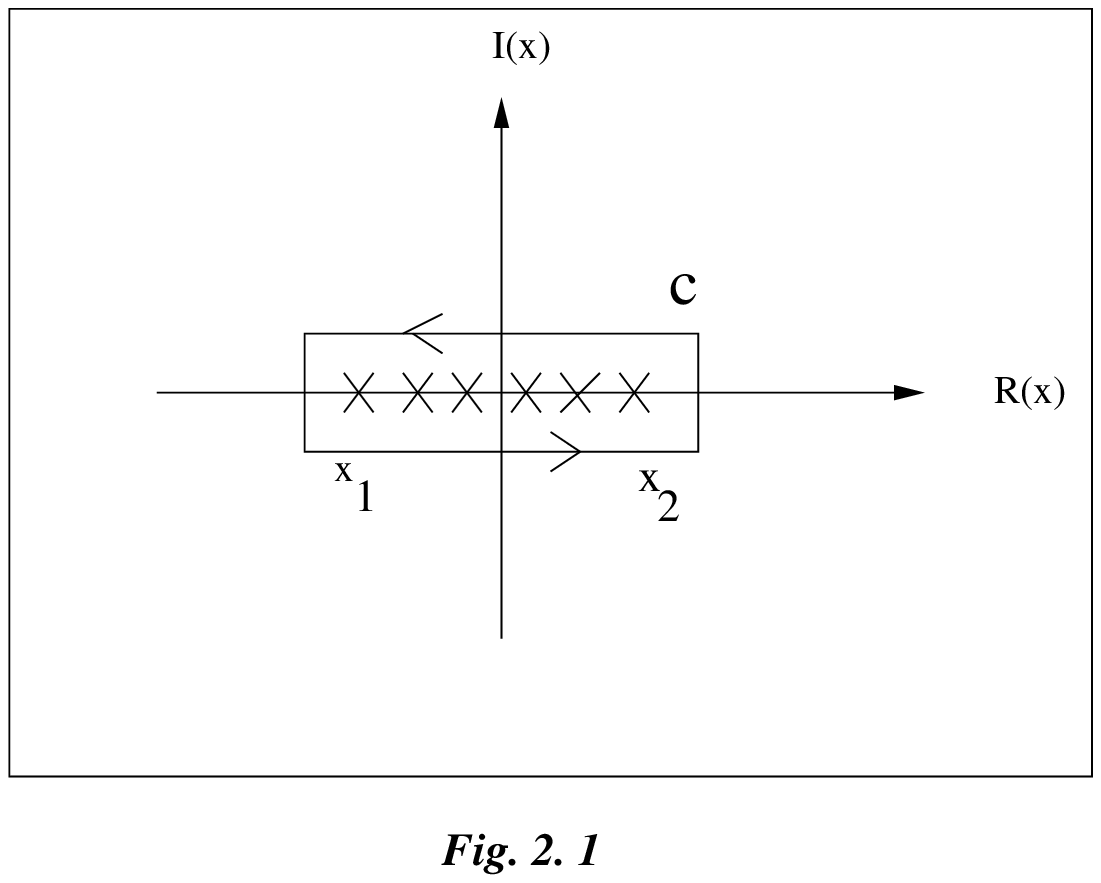}
\end{center}
 Applying the Cauchy residue theorem to
the $n$ poles enclosed by $C$, one obtains the quantization condition 
\begin{equation}
J(E) \equiv  \frac{1}{2\pi} \oint_C{pdx} = n\hbar,       \label{2c38}
\end{equation}
which can be viewed as a generalization of the Wilson - Sommerfield
quantization. Unlike the WKB quantization and the Wilson - Sommerfield
quantization, this is an exact  quantization condition and is a result of the
oscillation theorem and is true only for the bound states [\ref{2in},
\ref{2cou}]. Hence one cannot apply \eqref{2c38} to continuous energy states
with wave functions non vanishing at infinity.

   The power of the classical HJ theory lies in the fact that one obtains
 the frequencies of the system directly from the classical action -
 angle variables. Analogously, in the QHJ formalism, one can obtain the energy
 eigenvalues of the stationary states directly from the quantum action
 variable making use of the exact quantization condition
 \eqref{2c38}. To arrive at the energy eigenvalues, one needs to know
 the location of other singularities of the QMF, which lie outside the
 contour $C$ and their corresponding residues. The other singularities
 of the QMF arise from the potential $V(x)$ and are called as fixed
 singularities. The locations of these singular points are energy
 independent and appear in all the solutions of the QHJ
 equation. Thus, if the potential is meromorphic, the QMF will also be
 meromorphic. The residue at any of these fixed poles can also be
 computed by using the QHJ equation \eqref{2c33}. 
 Due to the quadratic nature of the QHJ equation, one obtains two
 values for each of these residues at the fixed poles. Of these two
 values, one needs to check which of the values satisfy the boundary
condition \eqref{2c35}. Since, $p_{c}$ is defined to be positive
 just below the branch cut, we choose those values of residues, which
 in the limit $\hbar \rightarrow 0$, correspond
 to positive $p_{c}$ below the branch cut. This method of choosing
 the residue is explained in the study of hydrogen atom, later
 in the chapter.
      
    In all our investigations, we have assumed that the QMF has only
finite number of singularities in the complex plane, which turned out
to be true for all the ES and QES models studied. This assumption is 
equivalent to saying that the point at infinity is an isolated
singularity. This property is shared by both the ES and QES models and the
only difference is the location of the moving poles. For ES
models, the QMF has poles only on the real line, whereas for QES models,
the moving poles are in both 
real and complex locations. In case of periodic potentials, it will be
shown that both ES and QES cases share the same kind of singularity
structure. The knowledge of the singularity 
structure allows one to  write the QMF as a function of $x$ and hence compute 
the eigenvalues and the eigenfunctions. The integral $J(E)$ in
\eqref{2c38} can be computed in terms of the residues at the
singularities, outside the contour $C$. This gives the energy
eigenvalues for one dimensional ES models [\ref{2sre}] and for QES
models, one is lead to the QES [\ref{2ge}, \ref{2geot}] condition by
the same condition.

In the next section, we clearly explain all the above discussed
points using the example of the hydrogen atom. We also show directly
from the QHJ formalism that without using the quantization condition,
we can arrive at the form of the QMF and hence, the values of the
eigenvalues and eigenfunctions by using well known results in complex
variables. 
\section{Hydrogen atom}

In this section, we obtain the bound state wave functions for the
radial part of the Coulomb problem [\ref{2sc}]. We consider the
supersymmetric potential
\begin{equation}
V_{-}(r) = -\frac{e^2}{r} +\frac{l(l+1)}{r^2} + \frac{e^4}{4(l+1)^2}   \label{h1}
\end{equation}
which is equal to the original coulomb potential plus the constant
$\frac{e^4}{4(l+1)^2}$, which has been added to make the ground state
energy equal to zero. The radial Schr\"odinger equation with $\hbar
= 2m = 1$ is given by
\begin{equation}
\frac{d^2 R}{dr^2} + \frac{2}{r}\frac{dR}{dr} + \left( E + \frac{e^2}{r}
- \frac{l(l+1)}{r^2} - \frac{e^4}{4(l+1)^2} \right)R =0.   \label{eh1a}
\end{equation}
Using the transformation $R(r) = u(r)/r$, $ p =
-i(u^{\prime}(r)/u(r))$, one obtains the QHJ equation as   
\begin{equation}
p^2 - i\frac{dp}{dr} - \left(E +\frac{e^2}{r} - \frac{l(l+1)}{r^2} -
\frac{e^4}{4(l+1)^2}\right) = 0.       \label{h2} 
\end{equation}
The wave function $u(r)$ should vanish at  $r=0$ and hence the QMF has fixed
pole at the origin along with the $n$ moving poles with residue $-i$ on the
positive real line. We assume that there are no other singular points
of $p$ in the complex plane, which makes {\it the point at infinity,  an
isolated singularity of $p$}.  With this singularity structure of $p$, we, now
proceed to find the energy eigenvalues. 

     From this information about the singularity structure
of the QMF, we know that $p$ is meromorphic and has the above
mentioned singular points. Hence, one can write $p$ as a sum of
singular and analytic parts as written below  
\begin{equation}
p = \sum_{k=1}^{n} \frac{-i}{r - r_k} +\frac{b_1}{r} +Q(r).     \label{h3}
\end{equation}
In the above equation, the summation term describes the sum of all the
principal parts coming from  all the individual Laurent expansions of
$p$ taken around each moving pole with $-i$ as the residue. 
The second term in \eqref{h3} represents the singular part in the
Laurent expansion of $p$ around $r=0$, with $b_1$ as the residue and
$Q(r)$ represents the analytic part of $p$. From the QHJ equation
\eqref{h2}, one can  see that for large $r$, $p$ goes to zero. Thus
$Q(r)$ is an entire function and bounded at infinity. Thus, from Louville's
theorem, $Q(r)$ is a constant, say $C$. Thus, \eqref{h3} becomes
\begin{equation}
p = \sum_{k=1}^{n} \frac{-i}{r - r_k} + \frac{b_1}{r} + C,    \label{h5}    
\end{equation}
which gives the form of $p$ in the entire complex plane.

To obtain the energy eigenvalues one only needs to calculate the value
of the residue $b_1$ and the value of the constant $C$. On the other hand, the
location $r_k$ of the moving poles must be found in order to obtain
the wave function. 

\subsection{Definition of $p_c$ }
As mentioned in the previous section, due to the quadratic nature of the
QHJ equation, $b_1$ will have two values. To select the right value of
residue, we make use of the boundary  condition \eqref{2c35}, which uses the
classical momentum $p_c$. Hence, we first proceed to define 
$p_c$ which is given by 
\begin{equation}
p_c = \sqrt{E +\frac{e^2}{r} - \frac{l(l+1)}{r^2} -
\frac{e^4}{4(l+1)^2}}     \label{h9a}
\end{equation}
in the complex plane and which is equivalent to
\begin{equation}
p_c  \equiv \frac{\pm i \alpha}{2r}\left( r^2 +\frac{e^2
      r}{\alpha^2} -\frac{l(l+1)}{\alpha^2}\right)^{1/2}.     \label{h10} 
\end{equation}
where $\alpha = \sqrt{\frac{e^4}{4(l+1)^2}-E}$, is a positive
constant. Let $r_1$ and $r_2$ be the two turning points and in the   
complex $r$ plane, $p_c$ is given a branch cut between the two turning
points. $p_c$ is defined to be positive just below the branch cut.
Hence, one needs to choose that branch of $p_c$ from \eqref{h10} which
satisfies this condition. In order to be able to do that we introduce
the variables, $\rho_1, \rho_2, \theta_1$ and $ \theta_2$ such that
\begin{equation}
r-r_1 = \rho_1 \exp({i\theta_1}) ,\,\,\,\,r-r_2 = \rho_2 \exp({i\theta_2})
\label{h11} 
\end{equation}
where $\theta_1$ and $\theta_2$ are the angles as shown in the
fig 2.2 below
\vskip1cm
\begin{center}
\includegraphics*{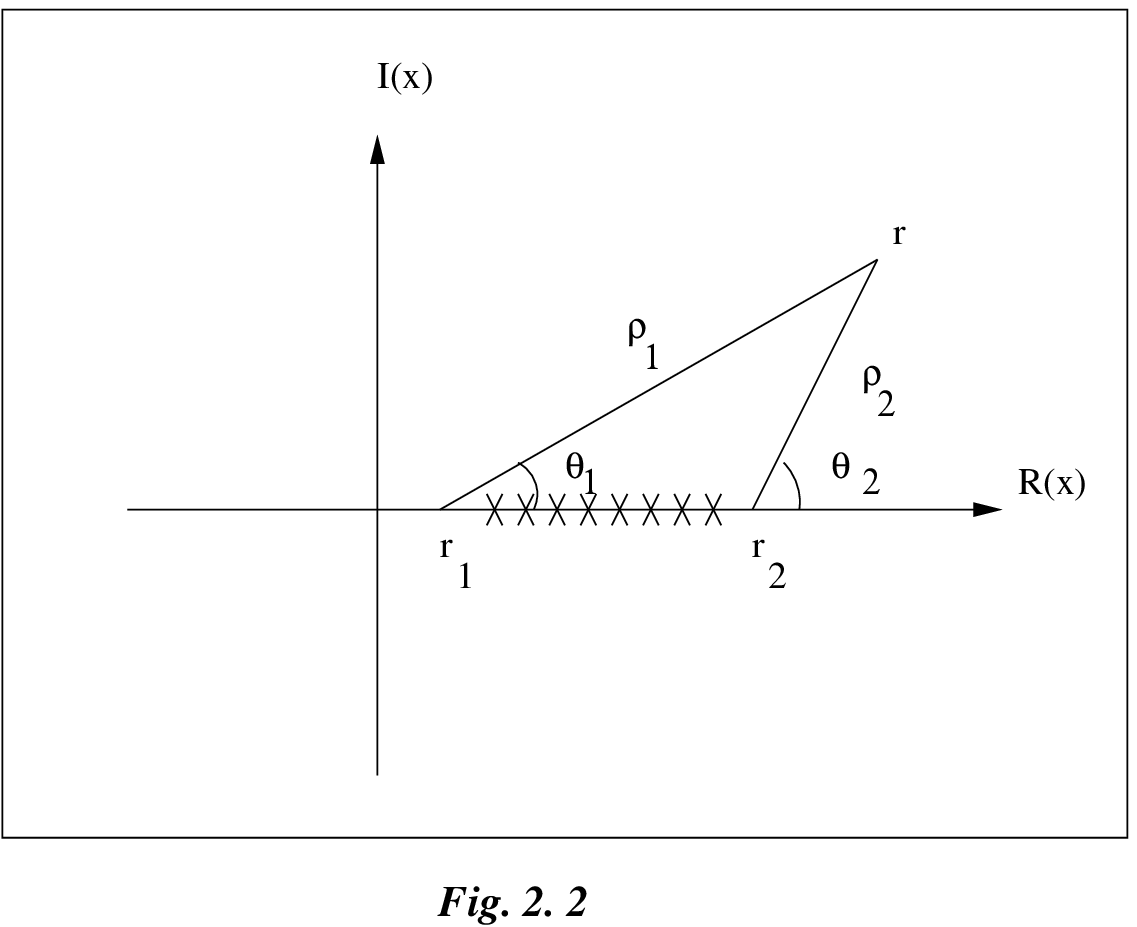}
\end{center}
and are chosen to lie in the open interval $(0,2\pi)$. This gives a
branch cut to $p_c$, for $r_1 <r <r_2$. Thus, one can
write $p_c$ as 
\begin{equation}
p_c  \sim  \frac{i\lambda\alpha}{2r}\sqrt{(r-r_1)(r-r_2)},
\label{h12}
\end{equation}
where $\lambda = \pm 1$ and using \eqref{h11}, $p_c$ becomes
\begin{equation}
p_c \equiv \frac{i\lambda\alpha}{2r}(\rho_1
\rho_2)^{1/2} \exp\left(i\frac{(\theta_1 +\theta_2)}{2}\right).  \label{h13}
\end{equation}
 We need to fix the sign of $\lambda$, which will fix the branch of
 $p_c$ to be selected.  For this, we consider a point just below the
 branch cut which  corresponds to
$\theta_1 \sim 2\pi, \,\,\,   \theta_2 \sim \pi $, 
which gives
\begin{equation}
p_c = \frac{\lambda\alpha}{2r}(\rho_1\rho_2)^{1/2}. \label{h15}
\end{equation}
Therefore, for $p_c$ to be positive just below the branch cut, we choose
$\lambda = +1$. Thus, we obtain the definition of $p_c$ in the entire complex
plane as
\begin{equation}
p_c = \frac{i\alpha}{2r}(\rho_1\rho_2)^{1/2}\exp\left(i\frac{(\theta_1
  +\theta_2)}{2}\right),  \label{h16} 
\end{equation}
with $0 < \theta_1 < 2\pi$ and $ 0 < \theta_2 < 2\pi$.

\subsection{Residue at the fixed pole $r=0$ }

     With the complete definition of $p_c$ obtained, we proceed to calculate
the residue of both $p_c$ and $p$ at $r=0$. Comparing  the residues,
one can fix the correct value of $b_1$. First, we calculate the residue
of $p_c$. At the origin $\theta_1 \sim \pi$ and $\theta_2 \sim \pi$,
which, when substituted in \eqref{h16}, give the residue at $r=0$ as 
\eqref{h16} $p_c$ at $r=0$
\begin{equation}
  - \frac{i\alpha}{2}(r_1
r_2)^{1/2}.    \label{h17}
\end{equation}
Now to calculate the residue of $p$, we perform a Laurent expansion of
$p$ around $r=0$ as
\begin{equation}
p = \frac{b_1}{r} + a_0 + a_1 r +.........\label{h6}
\end{equation}
Substituting  \eqref{h6} in \eqref{h2} and comparing the
coefficients of $1/r^2$, one obtains the quadratic equation
\begin{equation}
b_1^2 + ib_1 + l(l+1) = 0,    \label{h7}
\end{equation}
 which gives the two values of $b_1$ as
\begin{equation}
b_1 = il \,\, , \,\, -i(l+1).     \label{h8}
\end{equation}
Comparing the values of $b_1$ with the residue of $p_c$ in
\eqref{h17}, we see that the sign of the imaginary part must be
negative. Hence, we must select   
\begin{equation}
b_1 = -i(l+1),    \label{h19}
\end{equation}
which satisfies the boundary condition \eqref{2c35}.
 
\subsection{Calculation of energy eigenvalues and eigenfunctions }
With the selection of the right value of the residue at the fixed pole,
one can write $p$ as 
\begin{equation}
p = -i\frac{P^{\prime}_n(r)}{P_n(r)} - \frac{i(l+1)}{r} + C  \label{h19a}
\end{equation} 
where $\sum_{k=1}^{n}
\frac{-i}{r - r_k} = -i \frac{P^{\prime}_n(r)}{P_n(r)}$  and $P_n(r)$ is
an $n^{th}$ degree polynomial.
Substituting $p$ from \eqref{h19a}  in \eqref{h2} one obtains 
\begin{eqnarray}
\frac{P^{\prime \prime}_n(r)} {P_n(r)} -
2iC\left(\frac{P^{\prime}_n(r)}{P_n(r)}+\frac{(l+1)}{r} \right)
-\frac{2(l+1)}{r}\frac{P^{\prime}_n(r)}{P_n(r)} \nonumber \\
+C^2 - E -\frac{e^2}{r}+\frac{e^4}{4(l+1)^2} = 0.
\label{h19b} 
\end{eqnarray}
In order to fix $C$ and obtain the energy eigenvalues, we look at the
behaviour of each term of \eqref{h19b}, for large $r$. Noting that for
large $r$,  
\begin{equation}
\frac{P^{\prime\prime}_n(r)}{P_n(r)} \sim \frac{n(n-1)}{r^2} \,\,\, ;
  \,\,\, \frac{P^{\prime}_n(r)}{P_n(r)} \sim  \frac{n}{r},  \label{h19c}
\end{equation}
we equate the coefficients of the constant terms and
coefficients of $1/r$ in \eqref{h19b} to zero individually and get  
\begin{equation}
C = \frac{ie^2}{2(n+l+1)} ; \,\,\,  E = C^2 + \frac{e^4}{4(l+1)^2}
\label{h20} 
\end{equation}
which in turn give
\begin{equation}
 E = -\frac{e^4}{4(n+l+1)^2} + \frac{e^4}{4(l+1)^2},     \label{h20a}
\end{equation}
which is the required energy eigenvalue expression and one is left
with the second order differential equation
\begin{equation}
P^{\prime\prime}_n(r) + P^{\prime}_n(r)\left(\frac{2(l+1)}{r} -
\frac{e^2}{n+l+1}\right) + e^2\left(\frac{1}{r^2} -
\frac{l+1}{(n+l+1)r}\right)P_n(r)   =0. \label{h21}
\end{equation}
Defining $ \frac{e^2}{n+l+1}r = y$ in \eqref{h21}, we get
\begin{equation}
yP^{\prime\prime}_n(y) + ((2l+1)+1 -y)P^{\prime}_n(y) +n P_n(y) = 0
\label{h22} 
\end{equation}
which is the associated Laguerre equation, and $P_n(y)$ is proportional
to the
Laguerre polynomial, denoted by $L^{2l+1}_n(y)$. The bound state wave
function  can be obtained by using \eqref{2c34}, which gives
\begin{equation}
u(r) = \exp\left( i\int p dx\right)    \label{h23}
\end{equation}
Substituting the value of $p$ from \eqref{h19a} in the above equation
and changing the variable from $r$ to $y$, one obtains the wave function
in terms of $y$ as
\begin{equation}
u_n(y) = y^ {l+1}\exp \left(-\frac{y}{2}\right) L^{2l+1}_n(y), \label{h24}
\end{equation}
which matches with the known result [\ref{2susb}].

   We would like to mention here that one can obtain the energy
eigenvalues by using the exact quantization condition \eqref{2c38} discussed in
\S2.2 without solving the QHJ equation. For this, one needs to
calculate the residue at infinity ,
which will again have two values. The right value is chosen by applying
the boundary condition \eqref{2c35} and by using the Cauchy residue
theorem, one is led to the expression for the energy eigenvalues, the
details of which are given in [\ref{2bt}]. 
Instead of using the above mentioned
boundary condition, one may note that $p$ \eqref{h5}is a rational
function of $x$ 
and hence, the sum of all the residues (including the
residue at infinity) is zero . This is equivalent to what one gets
from the  exact quantization condition and is widely used in our
analysis of periodic and PT symmetric potentials.

\section{Change of variables in QHJ Formalism}
    We would like to emphasize here that, locating the singular points
of the QMF and imposing the boundary condition \eqref{2c35} are
two crucial steps in arriving at the correct physical
solutions. For many problems, it is necessary to perform a change of
variable  
\begin{equation} 
y = f(x).    \label{2c40}
\end{equation}
Thus, for a general
potential $V(x)$, the QHJ equation \eqref{2c34a} with $\hbar = 2m = 1$ is
\begin{equation}
p^2 - i \frac{dp}{dx}  - (E - V(x)) = 0,     \label{2c41}
\end{equation}
which, after a change of variable \eqref{2c40} and introduction of $q$ by
$q(y) \equiv ip(x(y))$, becomes 
\begin{equation}
q^2(y) + F(y) \frac{dq(y)}{dy} + E - \tilde{V}(y) = 0    \label{2c42}
\end{equation}
Here $F(y)$ equals the derivative $\frac{df(x)}{dx}$ expressed as a
function of $y$ and $\tilde{V}(y) = V(x(y))$. In order to calculate the
residue at the moving poles, removing the complexity caused due to
the presence of $F(y)$, we bring \eqref{2c42} to a
simple form, by introducing $\chi$ through the following
transformation equations.
\begin{equation}
q = F(y) \phi,\,\,\,\,\,\,\,\,\,\,\,  \chi = \phi +
\frac{1}{2}\frac{F^\prime(y)}{F(y)},  \label{2c43} 
\end{equation}
when substituted in \eqref{2c42}, give
\begin{equation}
\chi^2 + \frac{d\chi}{dy} + \frac{E - \tilde{V}(y)}{(F(y))^2} -
\frac{1}{2}\left(\frac{F^{\prime \prime}(y)}{F(y)}\right) +
\frac{1}{4}\left(\frac{F^\prime(y)}{F(y)}\right)^2 = 0,     \label{2c44}
\end{equation}
where, the residue at moving poles is one. The function $\chi$ will
also be called as the QMF and equation \eqref{2c44} as 
the QHJ equation. From \eqref{2c44}, one can
again see that the residue at the  moving poles of $\chi$ is
unity. 
   The expression for the wave function, in the
new variable $y$ can be obtained as follows. From \eqref{2c34}, one obtains 
$\psi$  with $\hbar =1$ as
\begin{equation}
\psi(x) = \exp\left( i \int p dx\right)    \label{2c43a}
\end{equation}
in which, substituting $p=-iq$, followed by the change of variable
\eqref{2c40} gives
\begin{equation}
\psi(y) = \exp \left( \int \frac{q}{F(y)}dy \right).     \label{2c44a}
\end{equation}
Using the transformation equations given in \eqref{2c43}, one is led
to the expression for $\psi(y)$ in terms of $\chi$ and $F(y)$ as 
\begin{equation}
\psi(y) = \exp  \int\left( \chi - \frac{1}{2}\frac{d}{dy}(\ln F(y)) \right)dy.
\label{2c45} 
\end{equation}

It is not always easy to impose the boundary condition
\eqref{2c35}. We, therefore, describe a set of constraints, which will
be found useful in restricting the choice of residues at the fixed
poles. Firstly, the wave functions should be
finite every where for the physical values of $x$. If one is looking
for bound state solutions, one must have the condition that the
wave function vanishes for large $x$. Since, in this
thesis, we are dealing only with the one dimensional potentials, we
can also use 
the fact that vanishing of wave function for large $x$,
implies that the  bound state levels are non -  
degenerate. So, for example, parity invariance will imply  $\psi(-x) =
\pm \psi(x)$ and QMF will satisfy $p(-x) = -p(x)$. 
 Also the bound state wave functions will be real and hence $p^{\ast}(x)
 =p(x)$.
All possible values of the residues, consistent with these and other
similar conditions, must be accepted.
In all the cases studied here, the QMF turns out to be a rational
function and hence, the quantization condition becomes equivalent to
{\it the sum of all the residues of QMF, including the residue at
infinity, being equal to zero}. This leads to a relation between the
residues and the number of moving poles $n$. As we  
will see in later chapters, some combinations of residues get
eliminated because the number $n$ has to be positive.
\newpage

{\bf References}
 
\begin{enumerate}

\item {\label{2gol}} H. Goldstein, {\it Classical Mechanics} 
(Addison-Wesley/Narosa, Indian student edition, 1995)

\item {\label{2sa}} E. J.  Saletan and
  A. H. Cromer, {\it Theoretical Mechanics} (John Wiley, New York, 1971).

\item {\label{2le}} R. A Leacock and M. J. Padgett, {\it
  Phys. Rev. Lett.} {\bf 50}, 3 (1983).

\item {\label{2pa}} R. A. Leacock  and M. J. Padgett, {\it Phys. Rev.}
 {\bf D28}, 2491 (1983).

\item {\label{2in}} E. L. Ince, {\it Ordinary Differential Equations}
  (Dover Publications, Inc, New York, 1956).  

\item{\label{2cou}}  R. Courant and D. Hilbert, {\it Methods of
  Mathematical Physics}, Vol 1 (Wiley interscience, New York, 1953).

\item {\label{2sre}} S. Sree Ranjani, K. G. Geojo, A. K Kapoor and
  P. K. Panigrahi, to be published in {\it Mod. Phys. Lett. A.} {\bf
  19}, No. {\bf 19}, 1457 (2004);  preprint quant - ph/0211168. 

\item {\label{2ge}} K. G. Geojo, S. Sree Ranjani and A. K. Kapoor,
  {\it J. Phys. A : Math. Gen.} {\bf 36}, 4591 (2003); quant - ph/0207036.

\item{\label{2geot}} K. G. Geojo, {\it Quantum Hamilton - Jacobi
  study of wave functions and energy spectrum of solvable and quasi -
  exactly solvable models}, {\it Ph. D. thesis} submitted to University of
  Hyderabad (2004).

\item{\label{2sc}} L. I. Schiff, {\it Quantum Mechanics} (McGraw -
  Hill, Singapore, 1986). 

\item{\label{2susb}} F. Cooper, A. Khare  and U. Sukhatme, {\it
  Supersymmetry in Quantum Mechanics} (World Scientific, Singapore,
  2001) and references therein. 

\item {\label{2bt}} R. S. Bhalla, {\it The quantum Hamilton - Jacobi
    approach to energy spectra of potential problems}, {\it Ph. D. thesis}
    submitted to University of Hyderabad (1996).
\end{enumerate}

\newpage
\chapter{Study of potentials exhibiting different spectra} 

\section{ Introduction}

    This chapter is devoted to the study of potentials, which exhibit
different spectra for different ranges of the potential
parameters. The first potential investigated, here, is the Scarf - I potential,
which has been well studied [\ref{3susb}, \ref{3supr}] and is known to
exhibit both phases of SUSY, namely the 
broken and the unbroken phases of SUSY, for different parameter
ranges. In [\ref{3bph}], Bhalla {\it et. al.} 
have studied this potential using the QHJ formalism and obtained the
the energy eigenvalue expressions in both the phases. In the present
study, we obtain both the eigenvalues and eigenfunctions along with the
different ranges. 

    The second potential  to be discussed is the Scarf
potential[\ref{3scf}, \ref{3li}], which  exhibits band structure in
one range and ordinary 
bound state spectrum in the other range. In [\ref{3li}], a group
theoretical treatment of Scarf potential has been 
discussed with the bound state and scattering state problem, being
extended to include the band structure. Using the QHJ formalism, we
show that one can arrive at the bound state solutions  and the band
edge solutions of this potential in a simple and straight forward
fashion. The next section  of the chapter discusses the Scarf - I
potential, followed by a section on Scarf potential.

\section{Phases of Supersymmetry}
    The expression for the supersymmetric Scarf - I potential is 
\begin{equation}
V_{-}(x) = - A^2 + (A^{2}+B^{2}-A\alpha \hbar)\sec^{2}\alpha x -
B(2A-\alpha \hbar)\tan\alpha x\sec\alpha x   \label{3e53}
\end{equation}
and it exhibits broken and unbroken phases of SUSY for the following
ranges of potential parameters [\ref{3susb}].  In the parameter range,
\begin{equation}
(A-B)>0,\,\, (A+B)>0,    \label{3e54}
\end{equation}
SUSY is exact and SUSY is broken for the range
\begin{equation}
(A-B)>0 ,\,\, (A+B)<0.   \label{3e55}
\end{equation} 
The supersymmetric phases of this potential are well studied and their
corresponding bound state solutions are available  in the
literature [\ref{3susb} \ref{3bph}]. In this chapter, we will show
that the ranges of the two different phases and their corresponding
solutions emerge naturally from the QHJ analysis [\ref{3sree}]. 
    
    Putting $2m =1$ and writing the QHJ equation in terms of $q= ip$ we get 
\begin{equation}
q^2 + \hbar q^{\prime}+E+ A^2 - (A^{2}+B^{2}-A\alpha
 \hbar)\sec^{2}\alpha x 
+  B(2A-\alpha \hbar)\tan\alpha x\sec\alpha x = 0  \label{3e57}
\end{equation} 
As discussed in the end of \S 2.4, we perform a change of variable
\begin{equation}
y = \sin\alpha x,   \label{3e58}
\end{equation} 
which  gives $F(y) = \alpha\sqrt{1-y^2}$. The transformation
equations \eqref{2c43} of \S 2.4 take the form, 
\begin{equation}
q = \alpha \sqrt{1-y^2}\phi\,\,\, ;\,\,\,\, \phi = \chi +\frac{y
  \hbar}{2(1-y^2)}.    \label{3e58b} 
\end{equation}
and the QHJ equation in terms of $\chi$  is obtained as,
\begin{eqnarray}
\chi^{2}+\hbar \chi^{\prime}+\frac{y^{2}\hbar^{2}}{4(1-y^2)^2}
+\frac{E+A^{2}}{\alpha^{2}(1-y^{2})}
&+&\frac{\alpha^{2}\hbar-2(A^{2}+B^{2}-A\alpha\hbar)}{2\alpha^{2}(1-y^2)^2}
\nonumber \\
&+&\frac{B(2A-\alpha \hbar)y}{\alpha^{2}(1-y^2)^2} = 0.  \label{3e59}
\end{eqnarray}

\subsection{Form of QMF ($\chi$)  }
       From \eqref{3e59}, one can see that $\chi$ has fixed poles at
$y = \pm 1$. In addition, there will  $n$ moving
poles, with residue one, on the real line  
corresponding to the $n$ nodes of the $n^{th}$ excited
state. Proceeding in the same way as in 
the case of hydrogen atom, we assume that $\chi$ has only the above mentioned
singularities  in the complex plane. Hence, we write $\chi$  as   
\begin{equation}
\chi= \frac{b_{1}}{y-1}+\frac{b_{1}^{\prime}}{y+1}
+\hbar \frac{P^{\prime}_{n}(y)}{P_n(y)} + C,   \label{3e60}
\end{equation} 
where $b_{1}$ and $b_{1}^{\prime}$ are the residues at $y=\pm 1$
respectively and $P_n(y)$ is an $n^{th}$ degree polynomial. $C$
represents the analytic part of $\chi$ which is a constant due to
Louville's theorem, since $\chi$ is bounded at infinity as seen from
\eqref{3e59}. As in \S 2.3, we first define the classical momentum
function, which will help us to fix the right form of $\chi$.

\subsection{Definition of the Classical momentum function}

  The classical momentum function, which is given a branch cut in
between the two turning points $x_1$ and $x_2$ in the complex $x$
plane, is given by 
\begin{equation}
p_c = \sqrt{2m(E-V(x))}    \label{3e60a1}
\end{equation}
and is defined to be the branch, which is positive just below the branch cut.
Since all the analysis is performed in the complex $y$ plane
with $\chi$  as the QMF, which has fixed poles at $y = \pm
1$, we use the condition 
\begin{equation}
\lim_{\hbar \rightarrow 0} \chi \rightarrow \chi_c     \label{3e60b}
\end{equation}
instead of \eqref{2c35}. The expression of $\chi_c$ is obtained as,
\begin{equation}   
\chi_c = \frac{i\tilde{p}_c}{\alpha \sqrt{1-y^2}} - \frac{y\hbar}{2(1-y^2)},
\label{3e62b} 
\end{equation}
by putting $q = ip$ in  the 
transformations \eqref{3e58b} and replacing  $p$ with
$\tilde{p_c}$, which  represents the classical momentum function
\eqref{3e60a} in terms 
of $y$ as
\begin{equation}
\tilde{p}_c = \pm i \frac{\sqrt{E+A^2}}{\sqrt{1-y^2}}\left(y^2
+\frac{B^2 -A\alpha -E
  -B(2A-A\alpha\hbar)y}{E+A^2}\right)^{1/2},   \label{3e62b2}
\end{equation} 
where $1 - y^2$ is positive as $y$ lies between $\pm 1$.
Let $y_1$ and $y_2$ be the images of the two turning points $x_1$ and
$x_2$ in the $y$ plane. The mapping from $y = \sin \alpha x $ in
\eqref{3e58}, is such that it maps 
the point just below the branch cut in the $x$ plane to a point just below the
branch cut in the $y$ plane. Hence, we select the branch of
$\tilde{p}_c$, which is positive just below the branch cut. Similar to
the hydrogen atom case, we introduce the variables $\theta_1, r_1$ and
$\theta_2, r_2$, where 
\begin{equation}
y-y_1 = r_1 \exp i\theta_1  \,\,\,\, y-y_2 = r_2 \exp i\theta_2  \label{3e62c}
\end{equation}
as represented in the fig.3.1.\\
\begin{center}
\includegraphics*{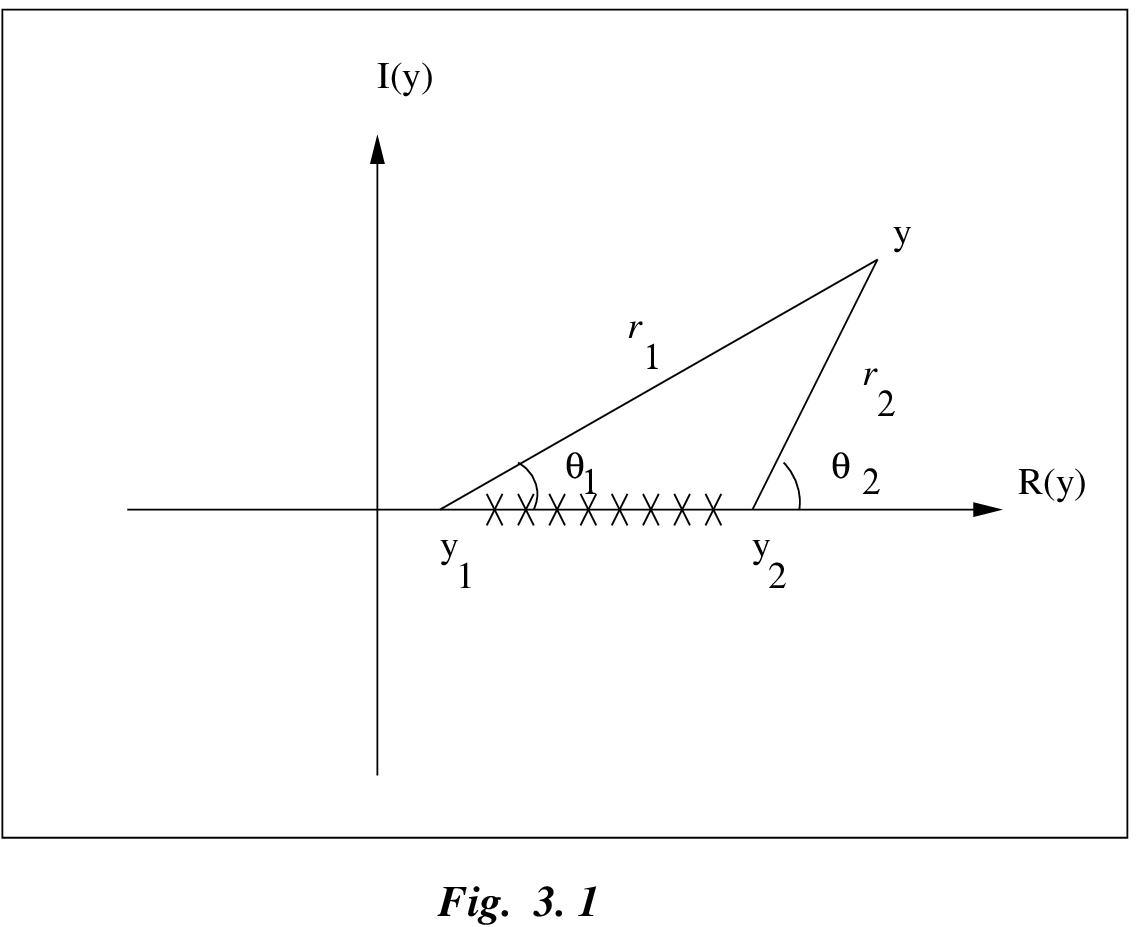}
\end{center}

 The values of $\theta_1$ and $\theta_2$ are
chosen to lie in the interval $(0, 2\pi)$ and $\tilde{p}_c$ can be written
in terms of the above variables as
\begin{equation}
\tilde{p}_c = i\lambda\frac{\sqrt{E+A^2}}{\sqrt{1-y^2}}(r_1
r_2)^{1/2}\exp\left(i\frac{(\theta_1 +\theta_2)}{2}\right),    \label{3e62d}
\end{equation}
where $\lambda = \pm 1$. Proceeding in the same way as done for the
Coulomb problem, we see that for a point just below the branch cut,
$\theta_1 \sim 2\pi$ and $\theta_2 \sim \pi$. Substituting these
values in \eqref{3e62d}, one gets
\begin{equation}
\tilde{p}_c \sim \frac{\lambda}{\sqrt{1-y^2}} (E+A^2)^{1/2} (r_1
r_2)^{1/2} .\label{3e62e}
\end{equation}
For $\tilde{p}_c$ to be positive just below the branch cut, we need to
fix $\lambda =1 $. Thus in the entire complex $y$ plane,
\begin{equation}
\tilde{p}_c = i \frac{\sqrt{E+A^2}}{\sqrt{1-y^2}}(r_1
r_2)^{1/2}\exp\left(i\frac{(\theta_1 +\theta_2)}{2}\right),   \label{3e62f}
\end{equation}
which in turn gives
\begin{equation}
\chi_c = - \frac{\sqrt{E+A^2}}{1-y^2}(r_1
r_2)^{1/2}\exp\left(i\frac{(\theta_1 +\theta_2)}{2}\right) -
\frac{y\hbar}{2(1-y^2)}.    \label{3e62g}
\end{equation}

\subsection{Residues at the fixed poles $ y = \pm 1 $ }
{\bf At \protect \($y = -1$ \protect \)  :}
We calculate the residue of $\chi_c$ at the fixed pole $y
= -1$. In the complex plane the point $y= -1$ corresponds to $\theta_1
\sim \theta_2 \sim \pi$, which gives
\begin{equation}
\chi _c = \frac{\sqrt{E+A^2}}{1-y^2}(r_1r_2)^{1/2}
-\frac{y\hbar}{2(1-y^2)}.   \label{3e62h}
\end{equation}
The residue of $\chi_c$ at $ y = -1$ is
\begin{equation}
\sim \frac{\sqrt{E+A^2}}{2}(r_1r_2)^{1/2} +\frac{\hbar}{4}, \label{3e62i}
\end{equation}  
which is a positive quantity. Now, we evaluate the residue of $\chi$
at $y =-1$,by doing  Laurent expansion of $\chi$ around $y=-1$, as
\begin{equation}
\chi = \frac{b^\prime_1}{y+1} + a_0 +a_1 (y+1)+ ........    \label{3e60a}
\end{equation}
Substituting this in \eqref{3e59} and comparing the coefficients of
$\frac{1}{(y+1)^2}$, one obtains the quadratic equation
\begin{equation}
(b^\prime_1)^2 - \hbar b^\prime_1 + \frac{3 \alpha^2 \hbar^2-(2A +
  2B)^2+4(A + B)\alpha
  \hbar}{16\alpha ^2} =0,   \label{3e60b1}
\end{equation}
which gives the two values of $b^\prime_1$ as 
\begin{equation}
b^\prime_{1} =\frac{(A+B)}{2\alpha}+\frac{\hbar}{4}\,\,, \qquad
-\frac{(A+B)}{2\alpha}+\frac{3\hbar}{4}.    \label{3e61} 
\end{equation}
    Since the residue of $\chi_c$ at $y = -1$ is positive,
\eqref{3e60b}, implies that we select the value of
$b^{\prime}_1$, which is positive in the limit $\hbar \rightarrow
0$. Thus, we pick
\begin{equation}
b^{\prime}_1 =  \left\{ \begin{array}{cc}
                 \frac{A+B}{2\alpha} + \frac{\hbar}{4} {\qquad
                  \mbox{if}\qquad} (A+B) > 0
                  \\ \qquad \mbox{}\qquad \\
		   -\frac{A+B}{2\alpha} +\frac{3\hbar}{4} \qquad
                   \mbox{if}\qquad (A+B)<0.
                 \end{array}  
                \right.   \label{3ea}
\end{equation}
From the above equation, we see that both the values of $b^{\prime}_1$ are
acceptable, but in two different ranges. Comparison of these ranges with
the ranges given in \eqref{3e54} and \eqref{3e55}, show that they
correspond to the different phases of SUSY.\\
\noindent
{\bf At \protect \($y = 1$ \protect \) :}\\
Repeating the above process at the other fixed pole, one 
obtains the residue of $\chi_c$ at $y=1$ as
\begin{equation}
\sim \frac{\sqrt{E+A^2}}{2}(r_1r_2)^{1/2} +\frac{\hbar}{4}, \label{3e61a}
\end{equation}
which is  positive. The two values of the residue of $\chi$ are
\begin{equation}   
b_{1} =\frac{(A - B)}{2\alpha}+\frac{\hbar}{4}\,\,, \qquad-\frac{(A -
  B)}{2\alpha}+\frac{3\hbar}{4}.
\label{3e62}  
\end{equation}     
Comparing equations \eqref{3e61a} and \eqref{3e62}, one sees that
each value of $b_1$ is acceptable in different ranges, {\it i.e}
\begin{equation}
b_1 =  \left\{ \begin{array}{cc}
                 \,\,\,\,\,\,\frac{A-B}{2\alpha} + \frac{\hbar}{4} {\qquad
                  \mbox{if}\qquad} (A-B) > 0
                  \\ \qquad \mbox{}\qquad \\
		   -\frac{A-B}{2\alpha} +\frac{3\hbar}{4} \qquad
                   \mbox{if}\qquad (A-B)<0
                 \end{array}  
                \right.   \label{3e620}
\end{equation}
 But, the present case of interest is $(A-B)>0$, for which the choice
 of residue is 
\begin{equation}
 b_{1} =\frac{(A-B)}{2\alpha}+\frac{\hbar}{4}.    \label{3eb}
\end{equation}
The  range $(A-B)>0$ corresponds to both phases of SUSY, as seen from
\eqref{3e54} and \eqref{3e55}. 
Thus, from the above discussion, one can see that by applying 
the boundary condition \eqref{3e60b} properly, one can obtain both the
ranges of the parameters and simultaneously select the values of the
residues which give physically acceptable solutions. We now proceed to
obtain the eigenvalues and eigenfunctions for both the phases.

\subsection{Case 1 : $(A-B) >0,\quad (A+B) >0 $ } 
For the above   range of parameters, $\chi$ in \eqref{3e60}{} becomes 
\begin{equation}
\chi(y)=\left(\frac{A-B}{2\alpha}+ \frac{\hbar}{4}\right)\frac{1}{y-1} +
\left(\frac{A+B}{2\alpha}+
\frac{\hbar}{4}\right)\frac{1}{y+1}+\hbar \frac{P^{\prime}_n(y)}{P_n(y)} +C.
\label{3e64} 
\end{equation}
where, the suitable  values of $b_1^\prime$ and $b_1$ are taken from
\eqref{3ea} and \eqref{3eb} respectively. Substituting $\chi$ from
the above equation in \eqref{3e59}{} and putting $\hbar =1$, we
compare the constant terms   
and the coefficients of $1/y^2$  terms individually, in the limit of
$y$ going to infinity and obtain $C=0$ and the energy eigenvalue as
\begin{equation}
E_{n}=(A+n\alpha)^{2} -A^2.    \label{3e65}
\end{equation}
 The equation for $P_n(y)$ assumes the form 
\begin{equation}
(1-y^2)P^{\prime\prime} +(\mu -\nu-(\mu +\nu+1)y) P^{\prime}
+n (n+\mu +\nu)P_n(y) = 0     \label{3e66}
\end{equation}
with
\begin{equation}
\mu=\frac{A-B}{\alpha} \,\,,\quad \nu = \frac{A+B}{\alpha},    \label{3e68}
\end{equation}
is seen to be the Jacobi differential equation and hence $P_n(y)$
coincides with $P_{n}^{\mu-1/2,\nu-1/2}(y) $ and  the bound state wave
function for the phase in which SUSY is exact, is
\begin{equation}
\psi_{n}(y)=N(1-y)^{\mu/2}(1+y)^{\nu/2}P_{n}^{\mu-\frac{1}{2},
  \nu-\frac{1}{2}}(y).  \label{3e67} 
\end{equation}
The energy eigenvalues in \eqref{3e65} and the  wave functions agree
with the solutions given in [\ref{3susb}].

\subsection{Case 2 : $(A-B) >0,\quad (A+B) <0$}
The right choice of the residues for this range, as
shown in \S 3. 2. 3 is,
\begin{equation}
b_{1} = \frac{(A-B)}{2\alpha}+\frac{\hbar}{4}
\,\,, \qquad b_{1}^{\prime}=-\frac{(A+B)}{2\alpha}+\frac{3\hbar}{4}.    \label{3e69}
\end{equation}
Substituting $\chi$ from \eqref{3e60}, with the above values of
 the residues in \eqref{3e59}{} with $\hbar=1$, one obtains  the energy
 eigenvalue as    
\begin{equation} 
E=\left(B-(n+\frac{1}{2})\alpha\right)^{2} - A^2    \label{3e70}
\end{equation}
and the differential equation for $P_n(y)$ turns out to be
\begin{equation}
(1-y^2)P^{\prime\prime}_n(y)+((\nu -\mu +1)-y(\nu + \mu
  +2))P^{\prime}_n(y)+n(n+\mu +\nu +1 )P_n(y) = 0     \label{3e71}
\end{equation}
and the bound state wave function is found to be
\begin{equation}
\psi_{n}(y)=N(1-y)^{\frac{\mu}{2}}(1+y)^{\frac{\nu}{2}+
  \frac{1}{2}}P_{n}^{\nu + \frac{1}{2}, \mu-\frac{1}{2}}(y)    \label{3e72}
\end{equation}
where
\begin{equation}
\mu=\frac{A-B}{\alpha} \,\,, \quad \nu=\frac{1}{2}-\frac{A+B}{\alpha}
\label{3e73}
\end{equation}
The Scarf - I potential can be related to the P\"oschl - Teller - I
potential by redefinition of the potential parameters as 
\begin{equation}
x \rightarrow x+\frac{\pi}{2\alpha} ,\,\, A=\delta + \beta,\,\,B=\delta -
\beta.   \label{3e74}
\end{equation}
With this redefinition of the parameters, we see that the bound state wave
functions in \eqref{3e72}{} match with the bound state wave functions of the
broken SUSY phase of the P\"oschl - Teller - I potential given in
[\ref{3dut}]. 

    Thus, we see that the QHJ formalism  in one dimension gives the
accurate expressions for the bound state wave functions, when there are
different phases of SUSY. It may be remarked here that, in the range
$A-B<0,\,\, A+B<0$, SUSY is exact but the roles of $H_{-}$ and $H_{+}$ are
interchanged. In the range $A-B>0,\,\, A+B<0$, SUSY is again broken.

\section{Phases of a periodic potential}
     In this section, we study the Scarf potential 
\begin{equation}
V(x) = - \frac{V_0}{\sin^2(\frac{\pi x}{a})}    \label{3es0}
\end{equation}
with $a$ as the potential period.
The Schr\"{o}dinger equation for this potential, with $\hbar =1$, is
\begin{equation}
\frac{d^2\psi(x)}{dx^2} + \frac{\pi^2}{a^2}\left( \lambda^2 +
\frac{(\frac{1}{4}-s^2)}{\sin^2(\frac{\pi x}{a})} \right)\psi(x) = 0
\label{3es0b}
\end{equation}
with  $\lambda$ and $s$, defined by 
\begin{equation}
 \lambda^2 = \frac{2mEa^2} {\pi^2} ;  \,\,\,\, 
 (\frac{1}{4}-s^2 ) = \frac{2mV_0 a^2}{\pi^2}.   \label{3es4a}
\end{equation}
This potential is known to exhibit two different types of spectra for
different ranges of $s$, unlike the Scarf - I potential, which exhibits
different bound state spectra for different ranges. In the range
$s>1/2$,   
one can see from the  plot in fig. 3.2
\begin{center}
\includegraphics*{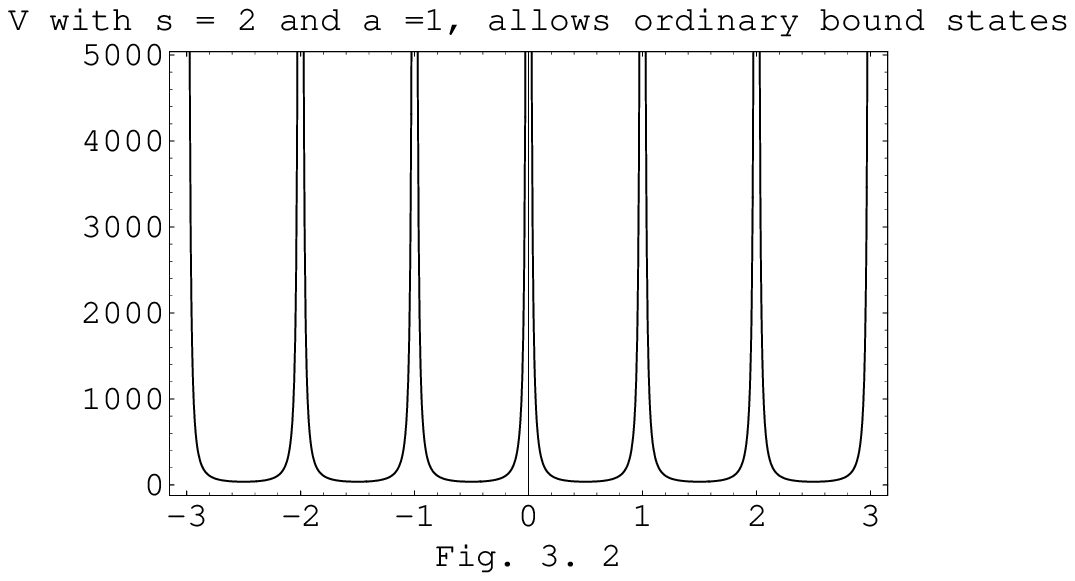}
\end{center}
that the potential is an array of infinite potential wells. Hence,
the particle is confined to one well and the wave function should
vanish at $x = \pm a$. Thus, in the above range, the potential has
bound state spectrum. From the above plot, one can also see that the
minima of the potential is at zero and hence the energy eigenvalues are
always greater than zero.  For $s$, in  the range
$0 < s< 1/2$, 
one can see from the plot in fig 3.3,
\begin{center}
\includegraphics*{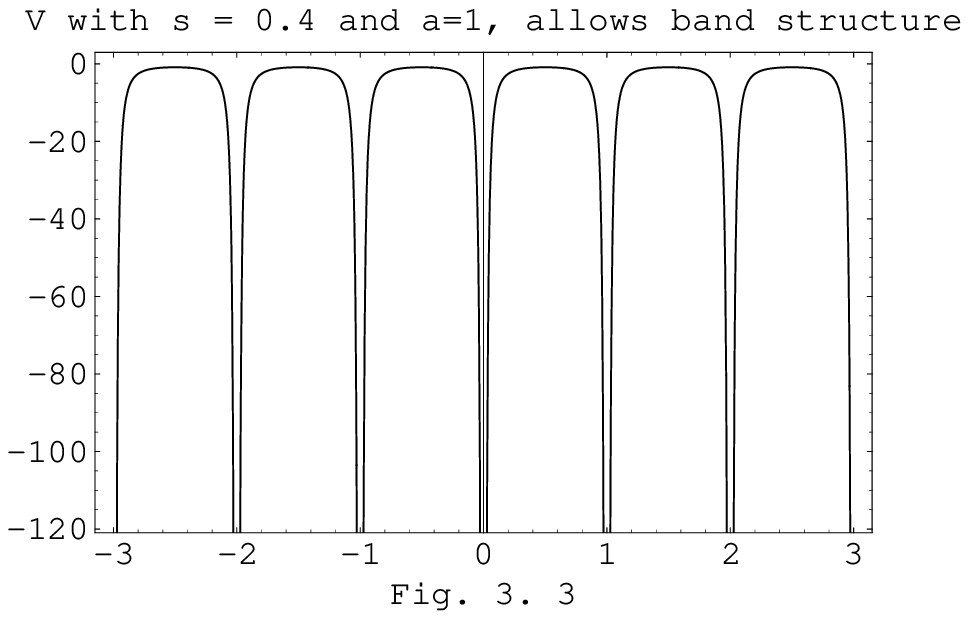}
\end{center}
 that the potential is similar
to that of the potential in a crystal lattice and hence there are bands of
allowed ranges. The particle can escape to infinity and the wave
function need not vanish at $x = \pm a$, but it should not become
infinity anywhere. Unlike the previous case one cannot make any
statement on the energy eigenvalues just by looking at the plot of the
potential.  But in the following sections, we will see that the energy
is greater than zero.
   
     For this potential, we will
show that one can obtain both the ranges and their corresponding
solutions, by doing a QHJ analysis.
The corresponding  QHJ equation for the Scarf potential with $p=-iq$ is
\begin{equation}
q^2 + q^{\prime}+\frac{\pi^2 }{a^2} \left(\lambda^2
+\frac{(\frac{1}{4}-s^2)}{\sin ^2(\frac{\pi     x}{a})}\right) =
0. \label{3es1} 
\end{equation}
Performing the change of variable
\begin{equation}
 y =  \cot \left(\frac{\pi x}{a}\right),  \label{3es1a}
\end{equation}
we make use of the results in \S 2.4. The expression for $df/dx$, 
expressed in $y$, becomes $F(y) = -\pi(1+y^2)/a$, which when
substituted in \eqref{2c43} gives the transformation equations,
\begin{equation}
q = -\frac{\pi }{a}(1+y^2) \phi  \,\,; \,\,\, \phi = \chi
-\frac{y}{1+y^2}. \label{3es3}
\end{equation}
Thus, the QHJ equation \eqref{2c44} for $\chi$, becomes
\begin{equation}
\chi^2 +\frac{d \chi}{dy} + \frac{\lambda^2 -1} {(y^2 +1)^2} +
\frac{(\frac{1}{4} -s^2)}{y^2+1} = 0.    \label{3es4}
\end{equation}

\subsection{Form  of the QMF \protect \( \chi  \protect \) }
      The QMF has $n$ moving poles with residue one on the real
line, corresponding to the nodes of the $n^{th}$ excited state
wave function. From \eqref{3es4}, one can see  that $\chi$ has fixed
poles at $y 
\pm i$. Hence, one can write $\chi$ in the rational form as
\begin{equation}
\chi = \frac{b_1}{y-i} +\frac{b^{\prime}_1}{y+i}
+\frac{P^{\prime}_n(y)}{P_n(y)} + C, \label{3es5}
\end{equation}
where $b_1$ and $b^{\prime}_1$ are the residues at $y =  i$ and $y =-i$
respectively and $P_n(y)$ is an $n^{th}$ degree polynomial. $C$
represents the analytic part of $\chi$ and from \eqref{3es4} one can see
that $\chi$ is bounded for large $y$. Thus, from
Louville's theorem, $C$ is a constant and $\chi$ is a rational function
of $y$. The residues at the fixed poles
$y = \pm i$, when calculated in the same way as in the previous
chapter gives, 
\begin{equation}
b_1 = \frac{1 \pm \lambda}{2}  \,\,\, ; \,\,\, b^{\prime}_1 = \frac{1
  \pm \lambda}{2}.  \label{3es6} 
\end{equation}
We assume that $\chi$ has   
finite number of singularities in the complex plane, which is
equivalent to saying that the point at infinity is an isolated
singularity. As mentioned in the end of \S  2. 4,  one can make use of
the fact that for a rational function, the sum of  all the residues is
equal to zero, to find the eigenvalues, which gives
\begin{equation}
b_1 +b^{\prime}_1 + n = d_1,        \label{3es7}
\end{equation}
where $d_1$ is the residue at infinity and is calculated  by taking 
Laurent expansion of $\chi$ around the point at infinity as 
\begin{equation}
\chi = d_0 +\frac{d_1}{y} + \frac{d_2}{y^2} +........\,\,,\label{3es10}
\end{equation}
which when substituted in the QHJ \eqref{3es4}, gives 
\begin{equation}
d^2_1 -d_1 + (\frac{1}{4} -s^2) = 0 \label{3es11}
\end{equation}
from which, the values of $d_1$ turn out to be
\begin{equation}
  d_1 = \frac{1 \pm 2s}{2}.  \label{3es12}
\end{equation}
Substituting the values of the residues in \eqref{3es7}, one obtains
\begin{equation}
n = - \frac{1}{2} \pm s \mp \lambda,  \label{3es12a}
\end{equation}
which gives the degree of the polynomial $P_n(y)$ in
\eqref{3es5}. From \eqref{3es4a}, 
one can see that if $E<0$, $\lambda$ becomes imaginary. Hence,
\eqref{3es12a} is not satisfied. Thus, from the above equation,  we
have the condition  $E >0$, 
which in turn implies $\lambda >0$ and real. Hence, one can say that
for any range of $s$, the energy eigenvalues are greater than zero.
With this  condition on $\lambda$, we now proceed to 
select the values of residues at the
fixed poles and at infinity, which will give us the
physically acceptable results. 

\subsection{Choice of the residues } 
First, we shall fix the value of the
residue at infinity. From the prior discussion of the potential, we
know that the wave functions should not become infinite  anywhere, in
particular, for $x=  \pm a$, {\it i.e.} the wave function $\psi(y)$
should be finite for large $y$. From \eqref{2c45}, the wave function
in terms of $\chi$ becomes
\begin{equation}
\psi(y) = \exp\left(\int \left(\chi - \frac{y}{1+y^2}\right)dy
\right).   \label{3es13} 
\end{equation}
For large $y$, the leading behaviour of $\chi$  is
obtained as $\chi \sim \frac{d_1}{y}$ from \eqref{3es10}, which when
substituted in \eqref{3es13},  gives
\begin{equation}
\psi(y)  \sim \exp \left( \int \left( \frac{d_1}{y} - \frac{y}{1+y^2}
\right)dy\right)   \label{3es14}
\end{equation}
\begin{equation}
  \sim \frac{y^{d_1}} {( y^2 +1)^{1/2}}. \label{3es15}
\end{equation}
Substituting the value of $d_1$ from \eqref{3es12} in the above
equation, one obtains 
\begin{equation}
\psi(y) \sim  \frac{y^{\frac{1}{2} \pm s}} {(y^2+1)^{1/2}}.
  \label{3es16}
\end{equation}
    For $0 < s < \frac{1}{2}$,
from the above equation, one can see that $\psi \rightarrow 0$ for
 both values of $d_1$ and this range corresponds
 to the case, where the potential exhibits band structure.

  For $s>1/2$,
$\psi(y)\rightarrow 0$, only if $d_1$ takes the value $ 1/2 -s $. Thus, we
fix the value of $d_1 = 1/2 - s$. In this way, the two different ranges of the
potential parameter $s$ emerge simultaneously fixing the values of $d_1$. 
To select the values of $b_1$ and $b^\prime_1$, we note that the
bound state and band edge wave functions of one dimensional
potentials are non- degenerate, which implies, the wave functions are
real. Hence, $\chi$ must be real, for which we must have 
\begin{equation}
b_1 = b^\prime_1.      \label{3es17}
\end{equation}
With the above constraint, \eqref{3es7} becomes
\begin{equation}
2b_1 +n = d_1.    \label{3es17a}
\end{equation}
     Thus, we see that the property, the wave function
should be finite for $ x \rightarrow \infty$, gives the values of
$d_1$ in the two ranges as follows
\begin{equation}
 d_1  =  \left\{ \begin{array}{cc}
         \quad \,\,\,  \frac{1\pm 2s}{2}  \quad\mbox{for}\quad   
               0 < s <1/2 
            \\ \quad \mbox{}\quad 
    \\ \frac{1-2s}{2}  \quad \mbox{for} \quad s >1/2.
           \end{array}
           \right.        \label{3es17ab}
\end{equation}
From the parity constraint, one obtains the restriction on the values
of the residues at the fixed poles as $b_1 = b^{\prime}_1$. Using
these results, we proceed to calculate the solutions for the two ranges.
\subsection{Case 1 : Band spectrum}
    In the range $0<s<1/2$, we have seen that $d_1$ can take both the
values of the residues. Hence, taking all the possible combinations of
the residues, keeping $b_1 = 
b^{\prime}_1$ and substituting them in \eqref{3es7}, we evaluate for
$n$,  which gives the degree of the polynomial $P_n(y)$ and takes the
integer values greater than  or equal to zero. Thus, there will be four
combinations 
forming four different sets with the expression for $n$, given in the
fifth column of table 3.1.
\begin{table}[h]
\caption{All possible combinations of residues for the range $0<s<1/2$}
\vskip0.5cm
\begin{center}
\begin{tabular}{|c|c|c|c|c|c|}
\hline
\multicolumn{1}{|c|}{Set}
&\multicolumn{1}{|c|}{$b_1$}
&\multicolumn{1}{|c|}{$b^{\prime}_1$}
&\multicolumn{1}{|c|}{$d_1$}
&\multicolumn{1}{|c|}{$n = d_1 -b_1-b^{\prime}_1$ }
&\multicolumn{1}{|c|}{remark}\\
\hline
& & & & & \\
1 & $\frac{1-\lambda}{2}$ & $\frac{1-\lambda}{2}$ & $d_1 = 1/2 -s$ &
$\lambda -s -\frac{1}{2}$ &  $\lambda > (s +\frac{1}{2})$\\
& & & & &\\
2 & $\frac{1-\lambda}{2}$ & $\frac{1-\lambda}{2}$ & $d_1 = 1/2 +s$ &
$\lambda +s -\frac{1}{2}$ &  $\lambda > -(s - \frac{1}{2})$\\
& & & & & \\
3 & $\frac{1+\lambda}{2}$ & $\frac{1+\lambda}{2}$ & $d_1 = 1/2 -s$
&$-\lambda -s -\frac{1}{2}$ & not valid\\ 
& & & & &\\
4 & $\frac{1+\lambda}{2}$ & $\frac{1+\lambda}{2}$ & $d_1 = 1/2 +s$&
$-\lambda + s -\frac{1}{2}$ & not valid\\ 
& & & & &\\
\hline
\end{tabular}
\end{center}
\end{table}
Since $n$ needs to be positive, from table 3.1, we  pick only those
sets which give a positive integer value for $n$.
As seen in \S 3.3.1, $\lambda$ is a positive real constant. Thus,
only the sets 1 and 2 will give positive values for $n$ and hence, the
other two sets namely 3 and 4 are ruled out. 
So, taking the values of $b_1, b^\prime_1$ and $d_1$ from
the sets 1 and 2 and substituting them in \eqref{3es17a}, and using the
definition of $\lambda$  and $s$ from \eqref{3es4a}, we obtain the
expressions for the energy eigenvalues corresponding to the two band
edges of the $n^{th}$ band, as 
\begin{equation}
E^{\pm}_n = \frac{\pi^2 }{2ma^2}\left( n+\frac{1}{2} \pm s
\right)^2. \label{3es22} 
\end{equation}
Here, $E_n^{\pm}$ correspond to the upper and lower band energies of
the $n^{th}$ band and they match with the solutions given in
[\ref{3scf}, \ref{3li}].   

    In order to obtain the expression for the wave function, we substitute
$\chi$ from \eqref{3es5} in \eqref{3es13} with $b_1 = b_1^{\prime} $,
which gives
\begin{equation}
\psi(y) = (y^2 +1)^{b_1} \frac{P^{\prime}_n(y)}{P_n(y)}.
  \label{3es23}
\end{equation}
To  evaluate the expression for the polynomial, we
 substitute $\chi$ from \eqref{3es5}, with $b_1 = b^{\prime}_1$, in
 the QHJ equation 
 \eqref{3es4}, which gives the second order differential equation
\begin{eqnarray}
&&P^{\prime\prime}_n(y) + 
 \left(\frac{4b_1 y}{y^2+1}\right)P^{\prime}_n(y)\,\,\, + \nonumber \\ & &
   \qquad \left(\frac{1/4 - s^2}{y^2 +1} + \frac{4b_1^2y^2
 +2b_1(1-y^2) +  \lambda^2 -1}{(y^2  +1)^2}\right)P_n(y) = 0.   \label{3es24}
\end{eqnarray}
The sets 1 and 2 have $b_1 =
 (1-\lambda)/2$, which on substitution in the above equation, gives
\begin{equation}
P^{\prime\prime}_n(y) + \left(\frac{2(1-\lambda)y}{y^2
 +1}\right)P^{\prime}_n(y) +\frac{\frac{1}{4}-s^2
 +\lambda^2 - \lambda}{(y^2 +1)}P_n(y) = 0.   \label{3es24a}
\end{equation}
From \eqref{3es4a} and \eqref{3es22}, one can see that $\lambda$ has
two values $\lambda = n \pm s + \frac{1}{2}$. Substituting these in the
above equation, one obtains two differential equations corresponding to the
two energy eigenvalues $E_n^{\pm}$ as
\begin{equation}
(y^2 +1)P_n^{\prime\prime}(y) + (1 - 2n \mp 2s)y
  P_n^{\prime}(y) + n(n\pm 2s)P_n(y) = 0. \label{3es24b}
\end{equation}
Defining $y=it$, the above equation takes the form of the well known Jacobi
differential equation
\begin{equation}
(1-t^2)P^{\prime\prime}_n(t) + (\nu_1 -\nu_2 -t(\nu_1 +\nu_2
  +2))P^{\prime}_n(t) +n(n +\nu_1 +\nu_2 +1)P_n(t) = 0,  \label{3es23a}   
\end{equation} 
with $\nu_1 = \nu_2 = -n \mp s -1/2$, for the corresponding two $\lambda$
values.
Hence, the polynomial $P_n(t)$ will be proportional to the Jacobi
Polynomial and the  expression for the two band edge wave
functions for the $n^{th}$ band will be  
\begin{equation}
\psi(y) = (y^2 +1)^{\frac{\lambda}{2}}P^{\nu_1, \nu_2}_n(-iy),\label{3es20}
\end{equation}
with respective $\nu_1, \nu_2$ values corresponding to $\lambda = n \pm 2
+1/2$.

\subsection{Case 2 : Bound state spectrum}
    We proceed in the same way as done in case 1 {\it i.e.}
take all possible combinations of $b_1, b^{\prime}_1$ and $d_1$,
keeping $b_1 =b^{\prime}_1 $ in 
\eqref{3es17a}. Since,  $d_1$ can take only single value $1/2 -s$, 
only two sets are possible. Out of these, the set corresponding to $b_1 =
 b^{\prime}_1 = (1-\lambda)/2$ alone, will give a positive value for
$n$. Thus,  substituting these values of residues
in \eqref{3es17a},
one obtains the  expression for the  energy eigenvalue as
\begin{equation}
E_n = \frac{\pi^2 }{2ma^2}\left(\frac{1}{2} + n
+\sqrt{\frac{1}{4}-\frac{2mV_0a^2}{\pi^2 \hbar^2}}\right)^2, \label{3e17}
\end{equation}
where $n$ can take values 0, 1, 2... and is left with the Jacobi differential
equation in terms of $t$ as 
\begin{equation}
(1-t^2)P^{\prime\prime}_n(t) -2t(-n-s+\frac{1}{2})P^{\prime}_n(t) -n(n
  +2s)P_n(t).   \label{3es19}
\end{equation}
The expression for the wave function from \eqref{3es13} becomes
\begin{equation}
\psi(y) = (y^2 +1)^{-\frac{\lambda}{2}}P^{s_1, s_2}_n(-iy),\label{3es20a}
\end{equation}
where $s_1 = s_2 = -n-s-1/2$.

    Note that by subtracting off the ground
state energy and redefining $\alpha = -1\, ,\,\, A
= -(s+1/2)  $ and $x \rightarrow \pi x/a$ in
\begin{equation}
V = A(A-\alpha)\cosec^2\alpha x - A^2,       \label{3es21}
\end{equation}
the supersymmetic potential [\ref{3susb}], one can see that the above
expression and its solutions match with those obtained here.
\newpage
{\bf References}

\begin{enumerate}

\item{\label{3susb}} F. Cooper, A. Khare  and U. Sukhatme  
{\it Supersymmetry in Quantum Mechanics} (World Scientific, Singapore,
2001) and references therein.

\item {\label{3supr}} F. Cooper, A. Khare and U. Sukhatme, {\it
  Phys. Rep.} {\bf 251}, 267 (1995).

\item {\label{3bph}} R. S. Bhalla, A. K. Kapoor and P. K. Panigrahi  {\it
    Int. J. Mod. Phys. A} {\bf 12}, No. {\bf 10}, 1875 (1997) and
    references therein.

\item{\label{3scf}} F. L. Scarf, {\it Phys. Rev}, {\bf 112}, 1137
  (1958) and references therein.   

\item{\label{3li}} H. Li and D. Kusnezov, {\it Phys. Rev. Lett.} {\bf
  83}, 1283 (1999).

\item {\label{3sree}} S. Sree Ranjani, K. G. Geojo, A. K Kapoor and
  P. K. Panigrahi, to be published in {\it Mod. Phys. Lett. A.} {\bf
  19}, No. {\bf 19}, 1457 (2004);
  preprint quant - ph/0211168. 

\item {\label{3dut}} R. Dutt, A. Khare and U. Sukhatme, {\it
  Phys. Lett. A.} {\bf 174}, 363 (1993).

\end{enumerate}

\chapter{Exactly Solvable Periodic Potentials}
\section{Introduction}
       This chapter is devoted to the study of potentials, which are ES
and periodic in nature. It is well known that periodic potentials play a
significant role in condensed matter physics.
The energy spectrum of periodic potentials is unique due to the
existence of energy bands and the solutions of the
Schr\"odinger equation have the Bloch form:
\begin{equation}
\psi(x) = u(x) \exp(ik.x)           \label{e1}
\end{equation}
Here, $u(x)$ has the periodicity of the potential function. These
features of the periodic potentials are usually illustrated using
the Kronig - Penney model [\ref{4kit}], in which the crystal
lattice is approximated by a periodic array of square wells.
Imposing the Bloch condition on the solutions of the Kronig-Penney
model, one is led to a transcendental equation, which has to be
solved, in order to obtain the band edge solutions.

Recently, there has been considerable interest in the study
[\ref{4kuz} - \ref{4osc}] and construction 
[\ref{4asis}, \ref{4tka}, \ref{4khare}, \ref{4d2}] of   
new periodic potentials, especially those belonging to the elliptic
class. The manifestation of periodic lattices in BEC has also
renewed interest in this area [\ref{4bec1},\ref{4opl}]. This gave us
enough motivation to check, if we could apply the QHJ formalism to these
potentials. In the previous chapter we have discussed the Scarf
potential [\ref{4sc}, \ref{4sc4}], $ V(x)  = -\frac{a^2(\frac{1}{4}-s^2)}{2m
\sin^{2}(\frac{\pi  x}{2})}$, which exhibits band structure in a
finite domain $ 0 < s < 1/2 $ and obtained the
band edge solutions of this potential.  

       Here, we study the ES case of Lam{\'e} and the associated
Lam{\'e} potentials, which belong to the class of elliptic potentials
and obtain the band edge solutions [\ref{4akk}]. These potentials are
usually expressed in terms of the Jacobi elliptic functions $ \sn
(x,m),\cn (x,m)$ and $\dn (x,m)$ [\ref{4han}, \ref{4grad}], where the 
parameter $m$ is known as the elliptic modulus, whose value lies
between 0 and 1.  

   The Lam{\'e} potential
\begin{equation}
V(x) = j(j+1)m \, \sn ^{2}(x,m),        \label{4e2}
\end{equation}
with $j$ being an integer, has $2j+1$ bands followed by the continuum.
This potential has been extensively studied in [\ref{4ars} -
\ref{4whit}]. It is interesting to note that the Schr\"{o}dinger
equation with the Lam{\'e} potential, 
\begin{equation}
\frac{d^{2}\psi(x)}{dx^{2}} +\frac{2m}{\hbar^{2}}( E - j(j+1)m \, \sn
^{2}(x,m))\psi(x) = 0   \label{4e4}
\end{equation}
is a result of separating the Laplace equation in the ellipsoidal
coordinates and is referred to as the Lam{\'e} equation
[\ref{4ars}, \ref{4whit}]. This has unique analytical properties
[\ref{4ars}], which render it 
useful in the study of topics ranging from astrophysics [\ref{4fin}
- \ref{4bac}], condensed matter physics [\ref{4gli}] to chaos
[\ref{4brac1},\ref{4brac2}]. 

    In comparison to the Lam{\'e} potential, the associated Lam{\'e} potential
\begin{equation}
V(x) = pm \, \sn^{2}(x,m)+qm \frac{\cn^{2}(x,m)}{\dn^{2}(x,m)}, \label{4e3}
\end{equation}
with $p =a(a+1)$ and $q = b(b+1)$, did not have a systematic study 
till recently and has a few scattered references in the mathematical
literature. This potential is ES, when $a = b =j$ with $j$ being an
integer, and QES when $ a \neq b$ with $a$, $b$ being real. A
systematic study of the band structure of this potential has been done
for various cases of $a$ and $b$ by Khare and Sukhatme [\ref{4uday}] and 
the algebraic aspects of the QES associated Lam{\'e}
potential have been investigated in [\ref{4tur1} - \ref{4alh}].

   The band edge eigenfunctions and the eigenvalues of both these potentials
have been found using the conventional techniques for solving the ordinary
differential equations [\ref{4ars}, \ref{4whit}]. One can obtain the
solutions, using the same method used to solve the band structure case
of the Scarf potential in the previous chapter.  

   In \S 4.2, we show, how one can obtain the form of
the wave functions for the Lam{\'e} potential for 
$j$ being an integer, followed by \S 4.3, where we take up a special case of $
j=2$ and obtain explicit band edge wave functions and energies. In
\S 4.4, the form of the wave functions for the  general
associated Lam{\'e} potential (ES case) is derived, followed by the
specific example, where of $a = b = 1$, in \S 4.5. 
\section{General Lam{\'e} potential}
  The QHJ equation in terms of $q$ for the Lam{\'e} potential is
\begin{equation}
q^2 + \frac{dq}{dx} + E -j(j+1)m\, \sn ^{2}(x,m) = 0.     \label{4e10}
\end{equation}
As discussed in \S 2.4, we perform a change of variable
\begin{equation}
t = \sn(x,m),      \label{4e11}
\end{equation}
which gives $F(t) = \sqrt{(1-t^2)(1-mt^2)}$. The elliptic function is
periodic with $\sn(x,m) = \sn (x+4 K(m),m)$, where $K(m)$ is
$\int_0 ^1 dt/\sqrt{(1-t^2)(1-mt^2)}$, the
complete elliptic integral of the first kind [\ref{4han},
\ref{4grad}, \ref{4chu}]. We would like to point out here, that
the Lam{\'e} equation can be written in five forms, namely, two algebraic, one
trigonometric, one Weierstrassian and one Jacobi [\ref{4ars}] 
depending on the change of variable. Of all these change of 
variables, we found \eqref{4e11}, which gives the Jacobi form of this 
Lam{\'e} equation, to be the best suited for our method. It enables 
one to write the QHJ equation in a form, which can be 
easily analyzed using the QHJ formalism. Another added advantage of 
this particular choice is, that it maps half of the period
parallelogram $ 0 \leq x <2 K(m) $ 
of the Jacobi elliptic $\sn(x,m)$ function to the  
upper half complex plane [\ref{4chu}]. Using the transformation 
equations \eqref{2c43}, one gets the equation for $\chi$ as
\begin{equation}
\chi^2 +\frac{d\chi}{dt} +\frac{(mt)^{2}+2m}{4(1-mt^{2})^2} +
\frac{t^{2}+2}{4(1-t^2)^2}
+\frac{2E-2j(j+1)mt^{2}-mt^{2}}{2(1-t^2)(1-mt^2)} = 0  \label{4e14}
\end{equation}
with
\begin{equation}
q = \sqrt{(1-t^2)(1-mt^2)}\phi \, ,\qquad  \phi = \chi
+\frac{1}{2}\left(\frac{mt}{1-mt^2} + \frac{t}{1-t^2} \right).   \label{4e15}
\end{equation}
The following properties of the Jacobi elliptic functions [\ref{4han}]
\begin{equation}
\frac{d}{dx}\sn(x,m) = cn(x,m)dn(x,m)    \label{4e12}
\end{equation}
and
\begin{equation}
\sn^{2}(x,m)+cn^{2}(x,m)=1,\,\,  \sn^{2}(x,m) + mdn^{2}(x,m) = 1
\label{4e13}
\end{equation}
were used. For all our further calculations we will regard $\chi$ as the QMF
instead of $p$. Using \eqref{4e14}, in place of the original QHJ
equation \eqref{4e10}, one can obtain the expressions for the
wave functions by analyzing the singularity structure of $\chi$.

\subsection{Form of QMF $\chi$ } 

   Equation \eqref{4e14}{} shows that $\chi$ has fixed poles at $t = \pm
1$ and $t=\pm 1/ \sqrt{m}$. We assume, there are a finite number of
moving poles in the complex plane. Hence, we make an assumption that
the point at $\infty$ is an isolated singular point and there are
no singularities of the QMF except for those mentioned
above. Therefore, proceeding in the same way as done in previous chapters,
we can write $\chi$, separating it into its singular and analytical
parts as
\begin{equation}
\chi =
\frac{b_{1}}{t-1}+\frac{b^{\prime}_{1}}{t+1}+\frac{d_{1}}{t-1/\sqrt{m}}+\frac{d^{\prime}_{1}}{t+1/\sqrt{m}}  
+\frac{P^{\prime}_n(t)}{P_n(t)} + Q(t). \label{4e18}
\end{equation}
Here, $Q(t)$ is the analytic part of $\chi$ and the rational terms
represent the singular parts. The coefficients $b_1,\,\,
b^{\prime}_{1}$ and
 $d_1,\,\, d^{\prime}_{1}$ are the residues at $ t=\pm 1$ and $t
= \pm 1/\sqrt{m}$ respectively. The term $P^{\prime}_n(t)/P_n(t)$
represents the singular part coming from the $n$ moving poles with
residue one. $Q(t)$ represents the analytic part of $\chi$ and from
\eqref{4e14}, one can see that $\chi$ is bounded at infinity. Hence,
from Louville's theorem, $Q(t)$ is a constant. 

    The residues $b_1$, $b_{1}^{\prime}$, $d_{1}$ and $d_{1}^{\prime}$
at the fixed poles can be determined by substituting the
Laurent expansion of $\chi $ around these poles in \eqref{4e14}.  
The two values of the residues  at all the fixed poles 
turn out to be
\begin{equation}
 \frac{3}{4} \,, \quad \frac{1}{4}.        \label{4e18c}
\end{equation}
Note that all the residues are independent of the
potential parameter $j$. Hence, irrespective of the value of $j$
in the potential, the residues at the fixed poles will take only
the above values for any Lam{\'e} potential.

   We need to select the  values of the residues, which give rise to acceptable
wave functions. Similar to the case of Scarf potential, there is no
way of selecting the residues, hence,
we accept  both the values i.e. $3/4$ and $1/4$ for each of
the residues in the present case . The only
requirement that restricts possible combinations of values is that
of parity, which arises due the fact that the band edge wave functions
are non -  degenerate. $\psi(t)$ having definite parity implies $\psi(-t)
= \pm \psi(t)$, which in turn gives
$\chi(-t) = -\chi(t)$. This condition rules out the possibilities  $ b_1 \neq
b^{\prime}_1$ and $ d_1 \neq d^{\prime}_1$. So, the possible
combinations of  $b_1$, $b_{1}^{\prime}$, $d_{1}$ and
$d_{1}^{\prime}$, which are allowed, are those which satisfy
$ b_1 = b^{\prime}_1$ and $ d_1 = d^{\prime}_1$.

   We have assumed that $\chi$ has finite number of singularities in
the complex plane and  hence, the point at infinity is an isolated 
singularity of $\chi$. The form of \eqref{4e14}{} suggests that $\chi$
is bounded at $\infty$. Therefore, $\chi$ can be expressed as
\begin{equation}
\chi(t) = \lambda_0 +\frac{\lambda_1}{t} +\frac{\lambda_2}{t^2} +....
\label{4e19}
\end{equation}
and the coefficients $\lambda_k$'s are fixed using the QHJ equation
\eqref{4e14}{} and one gets
$\lambda_0 = 0$ ,
and two values for $\lambda_1$ as
\begin{equation}
\lambda_1 = j+1 ,\,\,  -j.    \label{4e21}
\end{equation}
Making use of the fact, that for a rational function, the sum of the
residues is equal to zero, we get
\begin{equation}
2b_{1} + 2d_{1}+n = \lambda_1,      \label{4e22}
\end{equation}
where, $b_1 = b^\prime_1$ and $d_1 = d^\prime_1$ have been used.
As the left hand side is positive, it is clear that the only
choice $\lambda_1 = j+1 $ is consistent with the above equation, thus
fixing the value of residue at infinity.
    
   Thus from \eqref{4e22}{}, one gets the expression for $n$ as
\begin{equation}
n = j+1 -2b_1 -2d_1 ,   \label{4e23}
\end{equation}
which gives the degree of the polynomial $P_n(t)$. So making  use of all the
residue values at the fixed poles keeping the constraint $b_1 =
b^\prime_1$ and $d_1 =  d^\prime_1$, one obtains four different
sets and corresponding four different values of $n$ from \eqref{4e23},
as listed out in table 4.1 
\begin{landscape}
\begin{table}[h]
\caption{Number of zeros of the wave function corresponding to the
  number of moving poles $n$ of $\chi$, form of $\psi(x)$,  number of
  linearly independent solutions and the total  number of zeros of
  $\psi(x)$. } 
\vskip1.5cm
\begin{center}
\begin{tabular}{| c| c| c| c| c| c| c| c| }
\hline
\multicolumn{1}{|c|}{set}
&\multicolumn{1}{|c|}{$b_1$}
&\multicolumn{1}{|c|}{$d_1$}
&\multicolumn{1}{|c|}{n}
&\multicolumn{1}{|c|}{$\psi$ in terms of x}
&\multicolumn{2}{|c|}{Number of LI  solutions}& Total zeros  \\
&&&&&\multicolumn{1}{|c|}{$j=2N+1$}
&\multicolumn{1}{|c|}{$j=2N$ }
&\multicolumn{1}{|c|}{of $\psi(x)$\footnotemark}  
 \\ \hline
    &       &       &     &                &       &       &    \\            
$1$ & $1/4$ & $1/4$ & $j$ & $P_{j}(\sn x)$ & $N+1$ & $N+1$ & $j$    \\
    &       &       &     &                &       &       &     \\  
$2$ & $3/4$ & $1/4$ & $j-1$ & $\cn x P_{j-1}(\sn x)$ & $N+1$ & $N$ & $j$  \\
    &       &       &     &                &       &       &     \\  
$3$ & $1/4$ & $3/4$ & $j-1$ & $\dn x P_{j-1}(\sn x)$ & $N+1$ & $N$ & $j-1$  \\
    &       &       &     &                &       &       &     \\  
$4$ & $3/4$ & $3/4$ & $j-2$ & $\cn x \dn xP_{j-2}(\sn x)$ & $N$ & $N$
 & $j-1$  \\
    &       &       &     &                &       &       &     \\  
\hline
\end{tabular}
\end{center}
\end{table}
\footnotetext{These constitute of both real and complex zeros.}
\end{landscape}
\subsection{Explicit forms of the wave functions}
 One can obtain the form of the wave functions using \eqref{2c45}. 
Substituting the relations given in \eqref{4e15}{} in \eqref{2c45}{},
one gets the expression for the wave function in terms of $\chi$
as
\begin{equation}
\psi(t) = \exp \left(\int \left(\chi(t) +
\frac{1}{2}\left(\frac{mt}{1-mt^2}+\frac{t}{1-t^2}\right)\right)dt \right).
\label{4e24}
\end{equation}
Substituting $\chi$ from \eqref{4e18} gives
\begin{equation}
          \psi(t) = \exp \left(\int
        \left(\frac{(1-4b_1)t}{2(1-t^2)}+\frac{(1-4d_1)mt}{2(1-mt^2)}
        +\frac{P^{\prime}_n(t)}{P_n(t)}\right)dt\right),    \label{4e25}
\end{equation}
which when simplified and written in terms of the original variable
$x$, gives
\begin{equation}
\psi(x) = (\cn x)^{\alpha} (\dn x)^{\beta} P_{n}(\sn x),    \label{4e26}
\end{equation}
where, $\alpha = \frac{4b_1 - 1}{2}$ and $\beta = \frac{4d_1
  -1}{2}$. For the sake of simplicity, the elliptic modulus $m$ of
the Jacobi elliptic functions will be suppressed here onwards. The
four different combinations of the residues (given as sets 1 to 4
) give rise to four different forms of the band edge wave functions as
listed in the table 4.1.

    The parity constraint $\chi(-t) = -\chi(t)$ restricts the polynomial
$P_n (t)$ to have either only odd or only even powers of $t \equiv \sn
    x$. In the sixth and the seventh columns of table 4.1, the number
    of linearly 
independent solutions, for the two cases, $j$ being odd and $j$ being even, are
given in terms of a positive integer $N$. For odd $j$, $N= (j-1)/2$
and for even $j$, $N = j/2$. In both the cases, the total number of
solutions for a particular $j$ is equal to $2j +1$. It is easy to see that
the four forms and the number of solutions of a particular form
obtained here are in agreement with those already known
[\ref{4ars}]. For a given set of $b_1$ and $d_1$, $n$ is fixed  
using \eqref{4e23}{} and the differential equation for the unknown
polynomial $P_n (t)$ can be obtained by substituting $\chi(t)$
from \eqref{4e18}{} in \eqref{4e14}{}, which gives $Q = 0$ and
\begin{equation}  
P^{\prime\prime}_{n}(t) +4t\left(\frac{b_1}{t^{2} -1} +
\frac{md_1}{mt^{2}-1}\right)P^{\prime}_{n}(t) + G(t)P_n(t) = 0,    \label{4e27}
\end{equation}
where
\begin{eqnarray*}
G(t)  &=&\frac{t^{2}(4b_{1}^{2} - 2b_{1} + 1/4)+1/2 -
                         2b_{1}}{(t^{2}-1)^2}
     \\&& +\ \frac{(mt)^{2}(4d_{1}^{2} - 2d_{1} + 1/4)+m/2 -
                       2md_{1}}{(mt^{2}-1)^2}  
      \\&& +\  \frac{2E -2j(j+1)mt^{2}
                     +(16mb_{1}d_{1} - 1)mt^{2}}{2(t^{2}-1)(mt^{2}-1)}.
\end{eqnarray*}
 For each set of residues given in table 4.1, this differential equation is
equivalent to a system of $n$ linear equations for the
coefficients of different powers of $t$ in $P_{n}(t)$. The energy
eigenvalues are obtained by setting the corresponding determinant
equal to zero. We illustrate this process by obtaining the
eigenvalues and eigenfunctions explicitly for the case $j = 2$.

\section{Lam{\'e} potential with $j=2$ }

   Using the procedure described in the previous section, we obtain
the eigenvalues and  the eigenfunctions for the supersymmetric
potential [\ref{4khare}]
\begin{equation}
V_{-}(x) = 6m \, \sn^{2}x - 2m -2 +2\delta,       \label{4e28}
\end{equation}
where $\delta =  \sqrt{1-m+m^2}$. This potential is same as the Lam{\'e}
potential in \eqref{4e2}{} with $j = 2$, except for an additive constant
$2\delta -2m -2$, which has been added to make the lowest energy equal
to zero.
The equation for $\chi(t)$ is
\begin{equation}
\chi^{2} +\frac{d\chi}{dt} + \frac{(mt)^{2}+2m}{4(1-mt^{2})^{2}}+\frac{t^2
  +2}{4(1-t^2)^{2}} +\frac{2E+4(m+1-\delta)-13mt^2}{2(1-t^2)(1-mt^2)} = 0.
  \label{4e29}
\end{equation}
From \eqref{4e23}{}, the number of moving poles $n$, for $j =2$, are
\begin{equation}
n = 3 - 2b_1 -2d_1,     \label{4e30}
\end{equation}
where the values of $b_1$ and $d_1$ are obtained from 
and \eqref{4e18c}{}. For each set of $b_1$, $d_1$ and
$n$, one can write the form and the number of linearly independent
solutions by substituting $j=2$, which gives $N=1$ in table 4.1.
Hence for the case $j=2$, the form  and the number of solutions are as given in
table 4.2.

 To get the unknown polynomial part $P_{n}(t)$ of the wave function
and the band edge energies, one needs to substitute the different
sets of $b_1$, $d_1$ and $n$ from table 4.2 in the differential
equation
\begin{equation}
P^{\prime\prime}_{n}(t) +4t\left(\frac{b_1}{t^{2} -1} +
\frac{md_1}{mt^{2}-1}\right)P^{\prime}_{n}(t) + G(t)P_n(t) = 0,
\label{4e31}
\end{equation}
where
\begin{eqnarray*}
G(t) &=&  \frac{t^{2}(4b_{1}^{2} - 2b_{1} + 1/4)+1/2 -
  2b_{1}}{(t^{2}-1)^2} 
   \\ && +\ 
 \frac{(mt)^{2}(4d_{1}^{2} - 2d_{1} + 1/4)+m/2 -
  2md_{1}}{(mt^{2}-1)^2}   
  \\&& +\  \frac{2E +4(m+1-\delta)
  +(16b_{1}d_{1}-13)mt^{2}}{2(t^{2}-1)(mt^{2}-1)}.  
\end{eqnarray*}
\begin{table}[h]
\centering
\caption{ The form of the wave functions and the number of
linearly independent  solutions for the Lam{\'e} potential with $j=2$.}
\vskip1cm
\begin{tabular}{|c|c|c|c|c|c|}
\hline 
\multicolumn{1}{|c|}{$set$} &\multicolumn{1}{|c|}{$b_1$}
&\multicolumn{1}{|c|}{$d_1$} &\multicolumn{1}{|c|}{n}
&\multicolumn{1}{|c|}{$\psi$ in terms of x}
&\multicolumn{1}{|c|}{Number of LI solutions} \\
\hline
$1$ & $1/4$ & $1/4$ & $2$ & $P_{2}(\sn x)$ &  $2$    \\
    &       &       &     &                &       \\
$2$ & $3/4$ & $1/4$ & $1$ & $\cn x P_{1}(\sn x)$  & $1$  \\
    &       &       &     &                &     \\
$3$ & $1/4$ & $3/4$ & $1$ & $\dn x P_{1}(\sn x)$  & $1$  \\
    &       &       &     &                       &    \\
$4$ & $3/4$ & $3/4$ & $0$ & $\cn x \dn x P_{0}(\sn x)$ & $1$  \\
    &       &       &     &                  &     \\
\hline
\end{tabular}
\end{table}
Thus, for various sets of residues, one has the band edge energies
and the wave functions as:

\noindent
{\bf Set 1 : $b_1 = 1/4$ , $d_1 = 1/4 $, $ n = 2$} \\
Taking $P_2 =  At^{2}+Bt+C$, the parity constraint implies
\begin{equation}
B =0.    \label{4e33}
\end{equation}
Substituting $b_1 , d_1$ and $ P_2$ in \eqref{4e31}{}, one gets a
$2 \times 2$ matrix equation for $A$ and $C$ as follows
\begin{equation}
\left( \begin{array}{cc}
         E-2m-2-2\delta    &  -6m   \\
         2       &  E+2m+2-2\delta
\end{array}   \right)  \left( \begin{array}{c}
     A \\ C
\end{array}  \right)   = 0.    \label{4e34}
\end{equation}
Equating the determinant of the matrix in \eqref{4e34}{} to zero,
one gets the two values for the energy as
 \begin{equation}
E_I = 4\delta \,,\qquad E_{II} = 0,      \label{4e35}
\end{equation}
which in turn give the polynomial as
\begin{equation}
P_I(t) = m + 1 - \delta - 3mt^2  \,,\qquad P_{II}(t) = m + 1 +\delta -3mt^2.   \label{4e36}
\end{equation}
From \eqref{4e26}{}, we see that the band edge wave functions in
terms of $x$ will be
\begin{equation}
\psi_{I}(x) = m + 1 - \delta - 3m \, \sn^{2}x \,,\qquad \psi_{II}(x) = m +
1 +\delta -3m \, \sn^{2}(x).  \label{4e37}
\end{equation}
\noindent
{\bf Set 2 : $b_1 = 3/4$ , $d_1 = 1/4 $, $ n = 1$} \\
  Taking $P_1 = t-t_0 $ and substituting in \eqref{4e31}{}, one gets
  $t_0 =0$ and the band edge eigenvalue as
\begin{equation}
E = 2\delta - m +2.    \label{4e38}
\end{equation}
The band edge wave function becomes
\begin{equation}
\psi(x) = \cn x \,\sn x.     \label{4e40}
\end{equation}
\noindent
{\bf Set 3 : $b_1 = 1/4$ , $d_1 = 3/4 $, $ n = 1$}\\
Proceeding in the same way as in set 2, one gets
\begin{equation}
E = 2\delta +2m-1 \,,\qquad  \psi(x) = \dn x \, \sn x.  \label{4e41}
\end{equation}
\noindent
{\bf Set 4 : $b_1 = 3/4$ , $d_1 = 3/4 $, $ n = 0$} \\
   In this case, $P_0(t)$ is a constant and one gets
the band edge energy and the wave function to be
\begin{equation}
E = 2\delta -m-1   \,,\qquad   \psi(x) = \cn x \, \dn x.  \label{43}
\end{equation}
Thus, one obtains five band edge wave functions and their
corresponding energies, which agree with those given in
[\ref{4khare}].

\section{Associated Lam{\'e} potential}
The associated Lam{\'e} potential is ES when  $a = b =j$.
In this case, the potential expression becomes
\begin{equation}
V(x) = j(j+1) m \left(\sn^{2}x +\frac{\cn^{2}x}{\dn^{2}x}\right).
\label{4e43a}
\end{equation}
The QHJ equation with the associated Lam{\'e} potential is
\begin{equation}
q^2 + q^{\prime} + E -m j(j+1)\left(
\sn^{2}x+\frac{\cn^{2}x}{\dn^{2}x}\right) = 0.      \label{4e45}
\end{equation}
Doing the change of variable $t \equiv \sn x$ and proceeding in the same way
as in the case of general Lam{\'e} potential described in \S 4.2,
one gets the equation for $\chi(t)$ as,
\begin{eqnarray}
\chi^{2} +\frac{d\chi}{dt} &+& \frac{(mt)^{2}+2m(1-2j(j+1))}{4(1-mt^{2})^2}
  + \frac{2+t^2}{4(1-t^2)^2} \nonumber \\
&&+\frac{2E-mt^2(1+2j(j+1))}{2(1-t^2)(1-mt^2)}  = 0.   \label{4e46}
\end{eqnarray}

\subsection{Form  of QMF ($\chi$)} Similar to the Lam{\'e}
potential, $\chi(t)$ has fixed poles at $t = \pm 1$ and $t
= \pm 1/ \sqrt{m}$ along with $n$ moving poles in the entire complex $t$
plane. Further, we assume that the point at infinity is an isolated
singularity. $\chi$ has no other singular points, except  at $t = \pm
1$ and $t= \pm 1/ \sqrt{m}$. Following the same procedure used in
previous chapters, one obtains the residues at $t =\pm 1$ as
\begin{equation}
b_1 = \frac{3}{4},\,\,\frac{1}{4} \,,\qquad b_{1}^{\prime}
=\frac{3}{4},\,\,\frac{1}{4},    \label{4e47}
\end{equation}
which is independent of $j$. The residues at $ t = \pm 1/
\sqrt{m}$ turn out to be
\begin{equation}
d_1 = \frac{3+2j}{4},\,\,\frac{1-2j}{4} \,,\qquad d_1^{\prime} =
\frac{3+2j}{4},\,\,\frac{1-2j}{4},    \label{4e48}
\end{equation}
 which are $j$ dependent.
Knowing the singularity structure of $\chi$, one can write it as
\begin{equation}
\chi =
\frac{b_{1}}{t-1}+\frac{b^{\prime}_{1}}{t+1}+\frac{d_{1}}{t-1/\sqrt{m}}+\frac{d^{\prime}_{1}}{t+1/\sqrt{m}}
+\frac{P^{\prime}_{n}(t)}{P_n(t)} + C,    \label{4e49}
\end{equation}
valid for all $t$ similar to the Lam{\'e} potential.
Assuming that the point at infinity is an isolated singularity, we
calculate the residue at infinity by expanding $\chi$ in  Laurent
expansion for large $t$ as 
\begin{equation}
\chi = \lambda_0 +\frac{\lambda_1}{t} +\frac{\lambda_2}{t^2} +....
\label{4e50}
\end{equation}
 and substituting it in the the QHJ equation \eqref{4e46}{}, one
 obtains the two values for $\lambda_1$ as
\begin{equation}
\lambda_1 = j+1 ,\,\, -j.    \label{4e51}
\end{equation}
For a rational function, the sum of all its residues is equal to
zero. Hence, for $\chi$
\begin{equation}
b_1 +b^{\prime}_1 +d_1 +d^{\prime}_1 + n = \lambda_1.    \label{4e52a} 
\end{equation}
As in the Lam\'e case, the parity requirement implies $ b_1
=b_{1}^{\prime}$and  $ d_1 =d_{1}^{\prime}$ and we get
\begin{equation}
2b_1 + 2d_1 +n = \lambda_1.      \label{4e53}
\end{equation}
In this case, both the values of $\lambda_1$ are allowed because $d_{1}$
and $d_{1}^{\prime}$ can take negative values unlike the case of Lam{\'e}
potential, where $\lambda_1 = -j$ was ruled out. Thus, one has two cases
$\lambda_1 = j +1$ and $\lambda_1 = -j$. Note that the potential is
invariant under the transformation $ j \rightarrow -j-1$ and one has
two different values of $j$, one positive ($j = j^\prime >0$) and
another negative ($j = -j^\prime-1 <0$), leading to the same expression
for the potential described by \eqref{4e3} and hence, the same answers for
the eigenfunctions and eigenvalues. Therefore, it is sufficient to
restrict $j$ to positive values alone. We consider only the case
$\lambda_1 = j+1$ and neglect the other value $\lambda_1 = -j$.
For $\lambda_1 = j+1$, the values of $n$, which describe the number of
zeros, for different combinations of $b_1$ and $d_1$ values, are given
in table 4.3. 

The sets 3 and 4 give $ n = -1 $ and $n=-2$ respectively and will
not be considered,  as $n$ should be greater than or equal to 0. The
sets 1 and 2 will give positive values of $n$, only if $j$ is positive.
Since the singularity structure of the associated Lam\'e potential is same as
the Lam\'e potential, the form of the wave functions will be same as
that given in \eqref{4e26}. As in the Lam\'e case, the number of
linearly independent solutions are given in the table 4.3 in terms of
$N$, for $j$ being odd or even integer.
\section{Associated Lam{\'e} potential with $j = 1$}
We perform the calculation with the supersymmetric potential
\begin{equation}
  V_{-}(x) =  2m \,\sn^{2}x +2m \frac{\cn^{2}x}{\dn^{2}x} - 2- m + 2
  \sqrt{1-m}, 
  \label{4e54}
\end{equation}
corresponding to $j =1$.
\begin{landscape}
\begin{table}[h]
\caption{Values of $n$ and the form of the wave functions, for
  different combinations of $b_1$ and $d_1$, for $\lambda_1 = j+1$. }
\vskip1.5cm
\begin{center}
\begin{tabular}{|c|c|c|c|c|c|c|c|}
\hline 
\multicolumn{1}{|c|}{set}
&\multicolumn{1}{|c|}{$b_1$}
&\multicolumn{1}{|c|}{$d_1$}
&\multicolumn{1}{|c|}{$n$}
&\multicolumn{1}{|c|}{$\psi(x)$ in terms of $x$}
&\multicolumn{2}{|c|}{Number of LI  solutions}& Total zeros  \\
&&&&&\multicolumn{1}{|c|}{$j=2N+1$}
&\multicolumn{1}{|c|}{$j=2N$ }
&\multicolumn{1}{|c|}{ of $\psi(x)$\footnotemark} 
\\ \hline
    &       &             &       &           &    &     &     \\
$1$ & $1/4$ & $(1-2j)/4$  & $2j$  & $(\dn
x)^{-j} P_{2j}(\sn x)$      &  $ 2N+2$    & $2N+1$ &   $2j$  \\
    &       &             &       &           &    &     &        \\
$2$ & $3/4$ & $(1-2j)/4$  & $2j-1$& $\cn x(\dn x)^{-j} P_{2j-1}(\sn x)$      &  $2N+1$     & $2N$  &    $2j$ \\
    &       &             &       &        &     &     &            \\      
$3$ &$ 1/4$ & $(3+2j)/4$  & -1    &   -    & -   & -   & -           \\
    &       &             &       &        &     &     &            \\
$4$ & $3/4$ & $(3+2j)/4$  & -2    &    -   &  -  & -   & -           \\
    &       &             &       &        &     &     &             \\
 \hline
\end{tabular}
\end{center}
\end{table}
\footnotetext{These constitute of real and complex zeros.}
\end{landscape}
Proceeding in the same way  as
in the previous sections, the equation for $\chi(t)$ is found to be
\begin{eqnarray}
\chi^{2} +\frac{d\chi}{dt} &+&\frac{(mt)^{2}-6m}{4(mt^{2}-1)^2}
+\frac{t^{2}+2}{4(t^{2}-1)^2} \nonumber \\
&+& \frac{2E+4+2m-4\sqrt{1-m}-5mt^{2}}{2(1-mt^{2})(1-t^{2})} = 0
\label{4e55}
\end{eqnarray}
and the form of $\chi$ is
\begin{equation}
\chi =
\frac{b_{1}}{t-1}+\frac{b^{\prime}_{1}}{t+1}+\frac{d_{1}}{t-1/\sqrt{m}}+\frac{d^{\prime}_{1}}{t+1/\sqrt{m}}
+\frac{P^{\prime}_{n}(t)}{P_n(t)} +C.    \label{4e56}
\end{equation}
Substituting \eqref{4e56} in \eqref{4e55}{}, one obtains $C = 0$  and
is left with the differential equation
\begin{equation}
P^{\prime\prime}_{n}(t) +4t\left(\frac{b_1}{t^{2} -1} +
\frac{md_1}{mt^{2}-1}\right)P^{\prime}_{n}(t) + G(t)P_n(t) = 0,
\end{equation}   \label{4e57}
where
\begin{eqnarray*}
G(t) =&& \frac{t^{2}(4b_{1}^{2} - 2b_{1} + 1/4)+1/4 -
  2b_{1}}{(t^{2}-1)^2}\\
 && +\ \frac{(mt)^{2}(4d_{1}^{2} - 2d_{1} + 1/4)+-3m/2 -
  2md_{1}}{(mt^{2}-1)^2} \\ && +\   
\frac{2E+4+2m-4\sqrt{1-m}+(16b_{1}d_{1}-5)mt^{2}}{2(t^{2}-1)(mt^{2}-1)}.
\end{eqnarray*}

Substituting $j =1$ in table 4.3,  one gets the band edge
eigenfunctions and eigenvalues from set 1 and set 2  as follows.\\
\noindent
{\bf Set 1 : $b_1 = 3/4$ , $d_1 = -1/4 $, $ n = 1$} \\
 This combination gives only one solution with
\begin{equation}
E = 2-m+2\sqrt{1-m}  \,,\qquad   \psi(x) = \frac{\cn x \, \sn x}{\dn x}.
\label{4e59}
\end{equation}
\noindent
{\bf Set 2 : $b_1 = 1/4$ , $d_1 = -1/4 $, $ n = 2$} \\
This combination gives two solutions. The band edge energies are
\begin{equation}
E_I = 0  \,,\qquad E_{II} = 4\sqrt{1-m}     \label{4e61}
\end{equation}
and the corresponding wave functions are
\begin{equation}
\psi_{I}(x) = \frac{1}{m}\left(\dn x +\frac{\sqrt{1-m}}{\dn x}\right)
  \,,\,
  \psi_{II}(x) = \frac{1}{m}\left(\dn x -\frac{\sqrt{1-m}}{\dn x}\right),   \label{4e62}
\end{equation}
which match with the wave functions in [\ref{4khare}]. Thus, in this
chapter, we have  shown that the QHJ formalism successfully yields the band
edge eigenvalues and eigenfunctions for ES periodic potentials.

For all the models studied, the parameter $n$, which appears in the
exact quantization condition \eqref{2c38} described in chapter II,
gives the number 
of moving poles of the QMF, which in turn correspond to the zeros of the wave
function. For ES models [\ref{4sree}], these moving 
poles were found to be confined to the classical region alone. Thus, the
$n^{th}$ excited state has $n$ real zeros. The
parameter $n$ characterizes the energy level and increasing $n$
increases the energy.  

  For the QES models, the same $n$ appears  as a parameter in the
potential and of the possible states, only a few  states, which are
determined by $n$,  can
be obtained analytically. The QMF for all these states will have
the same number of moving poles, which are both real and complex.
Correspondingly the  wavefunctions for each state will have the
same number  $(\equiv n)$ of zeroes (both real and complex), of which
the number of real
zeros is in accordance with the oscillation  theorem. The number
of real zeros increases as one goes from the lowest state to the
highest state possible.

   On the other hand, the exactly solvable {\em periodic} potential
models show quite a different kind of distribution of the moving
poles in the complex plane. Like QES models, the moving poles of
the QMF consist of both real and complex poles. But unlike the QES
models, where the number of poles of QMF for  all the known states
is same, for ES periodic potential models  one finds groups of
solutions with number of moving poles remaining same within a
group, but varying from one group to another group of solutions.
This point becomes clear from the form of the solutions listed in
table 4.1 for the Lam{\'e} potential. Different groups of solutions
are precisely the different sets of the solutions listed in table
4.1. In each set, the degree of the polynomial $P_{n}(\sn x)$  is
same and the total number of zeros of the band edge wave functions
in the interval $0 \leq x < 2K(m)$ are same. For example, for the
$N+1$ linearly independent solutions belonging to set 1, the total
number of zeros of 
the wavefunctions will be $j$. However, for the solutions belonging
to set 2, there will be $j-1$ zeros of $P_{n}(\sn x)$ and  one zero
corresponding to $ \cn x$, thus a total of $j$ zeros. For all the sets,
the total
number of zeros are given in the last column of table 4.1. However,
the number of real zeros in a group increases with the increasing
energy and correspondingly the number of complex zeros will
decrease. For all values of $j$, we will have utmost four such
groups. A similar pattern is observed for ES associated Lam{\'e}
potential.
\newpage
{\bf References}

\begin{enumerate}

\item {\label{4kit}} C. Kittel, {\it Introduction to Solid state
  Physics}, 191 (Seventh edition, Wiley Eastern Limited, New Delhi, 1995).

\item {\label{4kuz}} H. Li and D. Kusnezov, {\it Phys. Rev. Lett.}
  {\bf 83}, 1283 (1999).

\item {\label{4d1}} G. V. Dunne and M. Shifman, {\it Annals of Phys.}
{\bf 299}, 143 (2002).  

\item {\label{4asis}} A. Ganguly, {\it J. Math. Phys.} {\bf 43}, 1980
  (2002); preprint math - ph/0207028.

\item {\label{4tka}} V. M. Tkachuk and O. Voznyak, {\it Phys. Lett. A.}
{\bf301}, 177 (2002). 

\item {\label{4pav}} P. Ivanov, {\it J. Phys A : Math. Gen.} {\bf 34}, 8145
  (2001); preprint quant - ph/0008008. 

\item {\label{4osc}} O. Rosas - 0rtiz, appeared in Proceedings of
  the IV workshop on {\it Gravitational and Mathematical Physics},
  N. Bret\'{o}n {\it et. al} (Eds),Chapala Jalisco, Mexico (2001);
  preprint math - ph/ 0302189; {\it Rev. mex. Fis.} {\bf 49}, 145 (2003). 
 
\item {\label{4khare}} A.Khare and U. Sukhatme, {\it J. Math. Phys.} {\bf
  40}, 5473 (1999); preprint quant - ph/9906044.

\item {\label{4d2}} G. V. Dunne and J. Feinberg, {\it Phys. Rev.} {\bf
 D57} 1271 (1998). 

\item {\label{4bec1}} J. C. Bronski, L. D. Carr, B. Deconinck and
J. N. Kutz, {\it Phys. Rev. Lett.} {\bf 86}, 1402 (2001).

\item {\label{4opl}} C. Fort, F. S. Cataliotti, L. Fallani, F.
Ferlaino, P. Maddaloni and M. Inguscio, {\it Phys. Rev. Lett.} {\bf 90},
140405 (2003). 

\item{\label{4sc}} F. L. Scarf, {\it Phys. Rev.} {\bf D57}, 1271 (1998).

\item {\label{4sc4}} F. L. Scarf, {\it Phys. Rev.} {\bf 112}, 1137 (1958) .

\item {\label{4akk}} S. Sree Ranjani, A. K. Kapoor and P. K. Panigrahi,
  to be published in {\it Mod. Phys. Lett. A.} {\bf 19}, No. {\bf 27},
  2047 (2004); preprint quant  - ph/0312041 

\item {\label{4han}} H. Hancock, {\it Theory of Elliptic Functions}, (Dover
  Publications, Inc, New York, 1958).

\item{\label{4grad}} I. S. Gradshteyn and I. M. Ryshik, {\it Table of
  Integrals, Series and Products}, (Academic Press, 1965).

\item {\label{4ars}} F. M. Arscot, {\it Periodic Differential
  Equations} (Pergamon, Oxford, 1964).

\item {\label{4mag}} W. Magnus and S. Winkler, {\it Hills
Equation} (Interscience Publishers, New York, 1966).

\item {\label{4whit}} E. Whittaker and G. N. Watson, {\it A Course of Modern
  Analysis} (Cambridge Univ. Press, Cambridge, 1963).

\item {\label{4fin}} F. Finkel, A. Gonz\'{a}lez - L\'{o}pez, A. L. Maroto and
  M. A. Rodriguez,{\it Phys. Rev.} {\bf D62} (2000) 103515; preprint
  hep - ph/0006117. 

\item {\label{4kan}} R. Kantowski and R. C. Thomas; preprint astro -
  ph/0011176. 

\item {\label{4mok}} O. I. Mokhov, preprint math. DG/0201224.

\item {\label{4mar}} M. Bouhmadi - L\'{o}pez, L. J. Garay and
P. F. Gonz\'{a}lez  - Diaz, {\it Phys. Rev.} {\bf D66}, 083504 (2002);
preprint gr - qc/0204072.  

\item {\label{4ale}} A. V. Razumov and M. V. Saveliev, Published in
  Proceedings of the International Conference {\it Selected Topics of
  Theoretical and Modern Mathematical Physics (SIMI - 96)}, Tbilisi,
  Georgia (1996); preprint solv - int/9612004.

\item {\label{4dav}} D. J. Fernandez C, B. Mielnik, O. Rosas - Ortiz and
 B. F. Samsonov, {\it J. Phys. A.} {\bf 35}, 4279 (2002); preprint
 quant - ph/0303051.  

\item {\label{4bac}} I. Bacus, A. Brandhuber and K. Sfetsos, Contribution
  to the proceedings of the TMR meeting {\it Quantum Aspects of Guage Theories,
  Supersymmetry and Unification}, Paris (1999);  preprint hep - th/0002092.

\item {\label{4gli}} M. D. Glinchuk, E. A. Eliseev and
  V. A. Stephanovich; preprint cond - mat/0103083. 

\item {\label{4brac1}} M. Brack, M. Mehta and K. Tanaka, {\it J. Phys. A.}
  {\bf34}, 8199 (2001); preprint nlin. CD/0105048.

\item {\label{4brac2}} M. Brack, S. N. Fedotkin, A. G. Magner and
  M. Mehta, {\it J. Phys. A.} {\bf 36}, 1095 (2003); preprint
  nlin. CD/0207043.  

\item {\label{4uday}}  A. Khare and U. Sukhatme, {\it J. Math. Phys.} {\bf 42}
  5652 (2001); preprint quant - ph/0105044.

 \item {\label{4tur1}} A. V. Turbiner, Commun. {\it Math. Phys.}
 {\bf118}, 467 (1988). 

\item {\label{4shif}} M. A. Shifman, {\it Contemp. Math.} {\bf 160},
  237 (1998). 

\item {\label{4tur2}} A. V. Turbiner, {\it J. Phys. A.} {\bf 22} LI (1989).

\item {\label{4Gonz}} F. Finkel, A. Gonz\'{a}lez - L\'{o}pez and
  M. A. Rodriguez, {\it J. Math. Phys.} {\bf 37}, 3954 (1996). 

\item {\label{4alh}} Y. Alhasad , F. G{\"u}rsey and F. Iachello,
  {\it Phys. Rev. Lett.} {\bf 50}, 873 (1983).

\item {\label{4chu}} R. V. Churchill and J. W. Brown, {\it Complex
  Variables and Applications} (McGraw - Hill Publishing Company, New
  York, 1990).
\item {\label{4sree}} S. Sree Ranjani, K. G. Geojo, A. K Kapoor and
  P. K. Panigrahi {\it Mod. Phys. Lett. A.} {\bf 19}, No. {\bf 19},
  1457 (2004); preprint quant - ph/0211168.
\end{enumerate}

\chapter{ Periodic Quasi Exactly Solvable Models }

\section{Introduction}
      In chapter I, we have mentioned that the QHJ analysis
 of ordinary QES models {\it i.e.} potentials with ordinary bound
 state spectrum, successfully yielded the QES conditions along with the
 eigenvalues  and the eigenfunctions [\ref{5ge},\ref{5geth}]. The
 present chapter is a study of QES periodic potentials. Here, we
 discuss the QES case of the associated Lam\'e potential [\ref{5qes}]. The QHJ
 analysis works on certain
assumptions on the singularity structure of the QMF, hence, it is essential to
see the validity of these assumptions by analyzing as many
different models as possible. Therfore, we study the QES associated
 Lam\'e potential, even though the method used is very much similar to
 that used in the previous chapter.
 
  The expression for the associated Lam\'e potential is
\begin{equation}
V(x) = a(a+1)m \sn^2(x,m) +b(b+1)m \frac{\cn^2(x,m)}{\dn^2(x,m)}.
\label{5e1} 
\end{equation}
Physically, this potential is described as
a periodic lattice of period $K(m)$ with the basis composed of
two different atoms, which are alternately placed. For a systematic study of
the associated Lam\'e potential, one is referred to [\ref{5su}, \ref{5khar}]. 
This potential is ES, when $a=b$ and has been analyzed in
the  previous chapter, QES
properties when $a \ne b$. There are three different QES cases
depending on whether $a$ and $b$ are integers or non - integers in 
[\ref{5su}] and are given below.\\
 {\bf Case (i) :} For $a$, $b$ being unequal integers with $a>b>0$,
there will be $a$ bound bands followed by a continuum band. if $a-b$
is odd(even) integer, then there will be $b$ doubly degenerate band edges of
period $2K(4K)$ with the corresponding band gaps zero, which cannot
be obtained analytically.\\
{\bf Case (ii) :} For $a$, $b$ being half - integers, but $a+b$ and
$a-b$ being integers, there will be infinite number of bands with
band edge wave functions of period $2K$ or $4K$ depending on whether $a-b$ is
an odd or even. Of these infinite bands, there will be $ a-b$ bands whose
band edges are non-degenerate with period $2K(4K)$ and $b+\frac{1}{2}$
doubly degenerate states of period $2K(4K)$ and can be obtained analytically.\\
{\bf Case (iii) :} For $a$ being an integer and $b$ being a half
integer or vice versa, one can  obtain  some exact analytical results
for mid - band states. 
    
    For all these above mentioned cases, the information regarding the
    existence 
of the missing states, mid band states etc. can be acquired by appealing to
the oscillation theorem [\ref{5ma}, \ref{5ars}]. 
 
     We proceed with the analysis of the associated Lam\'e potential
with the same change of variable and the transformation
equations used in chapter IV. With the 
same assumptions on the singularity structure of $p$, we proceed to
obtain the QES  condition and the forms of the band edge wave
functions for the general  associated Lam\'e potential in the next
section. In section \S 5.3, we  
analyze case (i), where both $a$, $b$ are integers taking the values 2
and 1 respectively. Case (ii), where $a$ and $b$ are both half
integers with the values 7/2 and 1/2 respectively, is analyzed in section
\S 5.4.

\section{QES condition and the forms of the wave functions}
The QHJ equation for the associated Lam\'e potential, putting $\hbar =
2m = 1$, is 
\begin{equation}
p^2 - ip^{\prime} = \left( E - a(a + 1)m \, \sn^2(x) - b(b + 1)m \,
  \frac{\cn^2(x)}{\dn^2(x)}\right).  \label{5e5}
\end{equation}
Note that with the transformation
$b \rightarrow - b - 1$ or $a \rightarrow - a - 1$, the potential
does not change. Hence, for all our studies, without
losing generality, we take $a,b$ positive and $a > b$.
Proceeding in the same way as in chapter IV, defining $p \equiv -iq$
and using the change of variable  $ t \equiv \sn (x) $, one obtains the
equation for $\chi$ as 
\begin{equation}
\chi^2 + \frac{d\chi}{dt} + \frac{m^2t^2 + 2m(1-2b(b + 1))}{4(1 - mt^2)^2} +
\frac{2 + t^2}{4(1 - t^2)^2} + \frac{2E - mt^2(1 - 2a(a + 1))}{2(1 - t^2)(1 -
  mt^2)} = 0.      \label{5e8}
\end{equation}
with the help of the transformation equations
\begin{equation}
q = \sqrt{(1 - t^2)(1 - m \, t^2)}\phi \,\, ,\,\, \phi = \chi +
  \frac{1}{2}\left(\frac{m \,t}{1 - m \,t^2}+\frac{t}{1 - t^2}\right).
  \label{5e7} 
\end{equation}
Like all the earlier investigations, we
shall treat $\chi$ as the QMF and \eqref{5e8} as the QHJ equation.

\subsection{Form of QMF ($\chi$)  }
From \eqref{5e8}, we see that $\chi$ has fixed poles at $ t = \pm
1$ and $t = \pm \frac{1}{\sqrt{m}}$. In addition to the fixed poles, 
the QMF is assumed to have finite number of moving poles and no other
singular points 
in  complex plane. Hence, one writes $\chi$ as a sum of the
singular and analytical parts as follows 
\begin{equation}
\chi = \frac{b_1}{t - 1} + \frac{b_1^\prime}{t + 1} + \frac{d_1}{t -
  \frac{1}{\sqrt m}} + \frac{d^{\prime}_1}{t + \frac{1}{\sqrt m}} +
  \frac{P_n^\prime(t)}{P_n(t)} + Q(t),    \label{5e9}
\end{equation}
where $b_1, b_1^\prime$ and $d_1, d_1^\prime$ are the residues at $t = \pm 1$
and $t = \pm \frac{1}{\sqrt m}$ respectively, which need to be
calculated. $P_n(t)$ is an $n^{th}$ degree polynomial with
 $ \frac{P_n^\prime}{P_n} = \sum_{k=1}^n{\frac{1}{t-t_k}}$,
being the summation  of terms coming from the $n$ moving poles with
residue one. The function $Q(t)$ is analytic and since $\chi$ is
bounded at infinity, from Louville's theorem, 
it is a constant, say $C$. The residues at the fixed poles can be calculated
by taking the Laurent expression around each individual pole and substituting
them in \eqref{5e8} as in previous chapters. 
The two values of the residues $b_1$ ,$ b^{\prime}_1 $ at $t = \pm 1$ are
\begin{equation}
b_1 = \frac{3}{4}\,\,,\,\,\frac{1}{4}  \,\,\,\,\,\,\,\,\,\,\, \,\,\,
b^{\prime}_1 = \frac{3}{4}\,\,,\,\,\frac{1}{4}  \label{5e11} 
\end{equation}
and at $t = \pm \frac{1}{\sqrt m}$, one has the residue values
\begin{equation}
d_1 = \frac{3}{4} + \frac{b}{2}\,\,,\,\,\frac{1}{4} - \frac{b}{2}
\,\,\,\,\,\,\,\,\,\,\,d^{\prime}_1 = \frac{3}{4} + \frac{b}{2}\,\,,
\,\,\frac{1}{4} - \frac{b}{2}.  \label{5e12} 
\end{equation}
  As in chapter IV,  we consider both the values of the residues. We
demand $b_1 = b^{\prime}_1$ and $ d_1 = d_1^\prime $, a condition
coming from the parity constraint $ \chi (t) = - \chi (t)$.
We assume that the point at infinity is an isolated singularity and hence
using the fact,  that the sum of
all the residues of a rational function is zero and the restriction $b_1 =
 b_1^\prime$ and $d_1 = d_1^\prime$, we obtain
\begin{equation}
2b_1 + 2d_1 + n = \lambda_1,    \label{5e16}
\end{equation}
 where $\lambda_1$ is residue  at infinity. 
Since $\chi$ has an isolated singular point at  infinity, one
 can expand  $\chi$ in Laurent series around the point at infinity as
\begin{equation}
\chi(t) = \lambda_0 +\frac{\lambda_1}{t} +\frac{\lambda_2}{t^2} +....
\label{5e14}
\end{equation}
Substituting \eqref{5e14}
in \eqref{5e8} and comparing the various powers of $t$,
one obtains
\begin{equation}
\lambda_1 = a+1    ,\,\,\,  -a.    \label{5e15}
\end{equation}
Taking various combinations of $b_1$ and $d_1$ from \eqref{5e11}
and \eqref{5e12}, substituting them in \eqref{5e16}, one obtains the
QES condition from each combination as given in table 5.1 for $\lambda =
a+1$. 
\begin{table}[h]
\caption{The quasi - exact solvability condition from the four
  permitted combinations of $b_1$ and $d_1$}
\vskip1cm
\begin{center}
\begin{tabular}{|c|c|c|c|c|}
\hline
\multicolumn{1}{|c|}{set} &\multicolumn{1}{|c|}{$b_1$}
&\multicolumn{1}{|c|}{$d_1$} &\multicolumn{1}{|c|}{$2b_{1} +2d_{1}
+n = \lambda_{1}$} &\multicolumn{1}{|c|}{QES condition}\\ 
\hline
    &       &                           &             &       \\
$1$ & $3/4$ & $\frac{3}{4}+\frac{b}{2}$ & $2+b+n = a$ & $b-a = -n -2$  \\
    &       &                           &             &             \\
$2$ & $3/4$ & $\frac{1}{4}-\frac{b}{2}$ & $1-b+n = a$ & $a+b+1 = n+2 $ \\
    &       &                           &             &            \\
$3$ & $1/4$ & $\frac{3}{4}+\frac{b}{2}$ & $1+b+n = a$ & $b-a = -n-1$   \\
    &       &                           &             &              \\
$4$ & $1/4$ & $\frac{1}{4}-\frac{b}{2}$ & $-b+n = a$  & $a+b = n $ \\
&&&&\\
\hline
\end{tabular}
\end{center}
\end{table}
Thus, one sees that all the allowed combinations of residues give
one of the 
forms of QES condition [\ref{5khar}], where $n = 0, 1, 2...$. Note that the
other value of residue at infinity {\it i.e.} 
$\lambda_1 = -a$, when substituted instead of $a + 1$ in
\eqref{5e16}, gives the QES condition for negative values of $a, b$ {\it i.e.}
for $a \rightarrow - a - 1,\, b \rightarrow -b - 1$ in $b_1,\,d_1$.

\subsection{Forms of wave function }
Since the change of variable and the transformation equations are same
as in chapter IV, we get the same expression for the wave function as
in \eqref{4e25} and \eqref{4e26} {\it i.e.}
\begin{equation} 
\psi(x) = \exp \left (\int{\left(\chi + \frac{1}{2}\left(\frac{mt}{1-mt^2} +
  \frac{t}{1-t^2}\right) \right)dt} \right),  \label{5e18}
\end{equation}
which on substitution of $\chi$ from \eqref{5e9} 
and written  in terms of the original variable $x$, becomes
\begin{equation}
\psi(x) = (\cn x)^\alpha (\dn x)^\beta P_n (\sn x),  \label{5e20}
\end{equation}
where $\alpha = \frac{4b_1 - 1}{2}, \,\,\,  \beta = \frac{4d_1 -
1}{2} $. Hence, for each set of $b_1,\,d_1$, one gets a wave
function given by \eqref{5e20}. The degree $n$ of this polynomial,
which is obtained from \eqref{5e16}, as
\begin{equation}
n = a + 1 - 2b_1 - 2d_1,    \label{5e22}
\end{equation}
which is in terms of either $a + b$ or $a - b$,  as evident from table 5.1.
The forms of the wave function can be found and are as given in table 5.2 and table 5.3, for the two different cases, when $a+b$ and $ a-b$
are odd and even separately.\\
\begin{landscape}
\vskip0.5cm
\noindent
{\bf Case I} {\it Both $a+b$, $a-b$ are even} : We introduce $N =
\frac{a+b}{2}$ and $ M =\frac{a-b}{2}$, where $M$ and $N$ are
integers, and obtain the forms of the wave functions in table 5.2, in
terms of $M$ and $N$ for the four sets of combinations of $b_1$ and
$d_1$ in table 5.1.
\vskip0.5cm
\begin{table}[h]
\caption{The form of the wave functions for the four sets of
residue combinations when $a+b$ and $a-b$ are even and equal to
$2N$ and $2M$ respectively.}
\vskip1cm
\begin{center}
\begin{tabular}{|c|c|c|c|c|c|c|}
\hline
\multicolumn{1}{|c|}{set}
 &\multicolumn{1}{|c|}{$b_1$}
&\multicolumn{1}{|c|}{$d_1$}
 &\multicolumn{1}{|c|}{$n =\lambda_{1}- 2b_{1} -2d_{1}$}
&\multicolumn{1}{|c|}{$n(M,N)$} 
&\multicolumn{1}{|c|}{wave function $\psi(x)$}
&\multicolumn{1}{|c|}{ LI solutions}
\\ \hline
    &       &                           &         &        &    &     \\
$1$ & $3/4$ & $\frac{3}{4}+\frac{b}{2}$ & $a-b-2$ & $2M-2$ & $\cn x (\dn x)^{1+b}P_{2M-2}(\sn x)$    & $M$ \\
&&&&&&\\
$2$ & $3/4$ & $\frac{1}{4}-\frac{b}{2}$ & $a+b-1$ & $2N-1$ & $\frac{\cn x}{(\dn x)^b}P_{2N-1}(\sn x)$& $N$ \\
&&&&&&\\
$3$ & $1/4$ & $\frac{3}{4}+\frac{b}{2}$ & $a-b-1$ & $2M-1$ & $(\dn x)^{b+1}P_{2M-1}(\sn x)$          & $M$ \\
&&&&&&\\
$4$ & $1/4$ & $\frac{1}{4}-\frac{b}{2}$ & $a+b$   & $2N$   & $ \frac{P_{2N}(\sn x)}{(\dn x)^b}$      & $N+1$\\
&&&&&&\\
\hline
\end{tabular}
\end{center}
\end{table}
\end{landscape}
\begin{landscape}
\noindent
{\bf Case II} {\it Both $a+b$, $a-b$ odd} : Introducing,
$N^{\prime}=\frac{a+b}{2}$ and $M^{\prime} = \frac{a-b}{2}$, where
 $M^{\prime}$ and $N^{\prime}$ are integers, we obtain the wave
 functions in table 5.3, in terms of $M^{\prime}$ and $N^{\prime}$ 
for the four sets of combinations of $b_1$ and $d_1$ in table 5.1.
\vskip0.5cm
\begin{table}[h]
\caption{The form of the wave functions for the four sets of
residue combinations when $a+b$ and $a-b$ are odd and equal to
$2N^{\prime} +1$ and $2M^{\prime}+1$ respectively.}
\vskip1cm
\begin{center}
\begin{tabular}{|c|c|c|c|c|c|c|}
\hline
\multicolumn{1}{|c|}{set}
 &\multicolumn{1}{|c|}{$b_1$}
&\multicolumn{1}{|c|}{$d_1$}
 &\multicolumn{1}{|c|}{$n= \lambda_1 - 2b_1 - 2d_1$}
&\multicolumn{1}{|c|}{$n(M,N)$}
&\multicolumn{1}{|c|}{wave function $\psi(x)$ }
&\multicolumn{1}{|c|}{ LI solutions }\\ 
\hline
&&&&&&\\
$1$ & $3/4$ & $\frac{3}{4}+\frac{b}{2}$ & $a-b-2$ & $2M^{\prime}-1$ & $\cn x (\dn x)^{1+b}P_{2M^{\prime}-1}(\sn x)$    & $M^{\prime}$ \\
&&&&&&\\
$2$ & $3/4$ & $\frac{1}{4}-\frac{b}{2}$ & $a+b-1$ & $2N^{\prime}  $ & $\frac{\cn x}{(\dn x)^b}P_{2N^{\prime}}(\sn x)$  & $N^{\prime}+1$ \\
&&&&&&\\
$3$ & $1/4$ & $\frac{3}{4}+\frac{b}{2}$ & $a-b-1$ & $2M^{\prime}  $ & $(\dn x)^{b+1}P_{2M^{\prime}}(\sn x)$            & $M^{\prime}+1$ \\
&&&&&&\\
$4$ & $1/4$ & $\frac{1}{4}-\frac{b}{2}$ & $a+b$   & $2N^{\prime}+1$ & $ \frac{P_{2N^{\prime}+1}(\sn x)}{(\dn x)^b}$    & $N^{\prime}+1$\\
&&&&&&\\
\hline
\end{tabular}
\end{center}
\end{table}
\end{landscape}
   From the forms of the wave functions in tables 5.2 and 5.3, one observes 
that the number of linearly independent solutions is different for
the two cases. The unknown polynomial in the wave function can be
obtained by substituting $\chi$ from \eqref{5e9} in the QHJ
equation \eqref{5e8}, which gives
\begin{equation}
P_{n}^{\prime\prime}(t) + 4P_{n}(t)\left(\frac{b_1 t}{t^2 -1} +
\frac{md_1 t}{mt^2 -1}\right)+G(t)P_{n}(t) =0    \label{5e23}
\end{equation}
where 
\begin{eqnarray}
G(t) &=&\frac{t^2 (4b_{1}^{2}-2b_1 +\frac{1}{4})-2b_1 
+\frac{1}{2}}{(t^2 -1)^2}\nonumber \\ &+& \frac{m^2 t^2 (4d_1 ^2 -2d_1
+\frac{1}{4})- 2md_1 + m(\frac{1-2b(b+1)}{2})}{(mt^2 -1)^2} 
\nonumber \\
&+&\frac{2E +(16b_1d_1 -1 -2a(a+1))mt^2}{2(1-t^2)(1-mt^2) }.\nonumber
\end{eqnarray}
 The above differential equation is equivalent to a system of $n$
 linear equations for the coefficients of different powers of $t$ in
 $P_n(t)$. The energy eigenvalues are obtained by setting the
 corresponding determinant equal to zero. In the next section, we
 obtain the band edge wave functions for the associated Lam{\'e}
 potential with $a=2$, $b = 1$

\section{Associated Lam\'e potential with \protect\( $a, b$ \protect
 \) as integers}
     For this case, we consider the associated 
Lam{\'e} potential with $a = 2,\,b = 1$. Similar to the previous
chapter, we work with SUSY potential
\begin{equation}
V_-(x) = 6m \,\sn^2 x + 2m \frac{{\cn^2 x}}{{\dn^2 x}}
  - 4m.   \label{5e24}
\end{equation}
This potential is same as \eqref{5e1} with $a=2$ and $b=1$,
except that a constant has been added to make the
lowest energy equal to  zero. The QHJ equation in terms
of $\chi$ is
\begin{equation}
\chi^2 + \chi^{\prime} + \frac{m^2 t^2 - 6m}{4(1 - mt^2)^2} +
\frac{2 + t^2}{4(1 - t^2)^2} + \frac{2E +8m -13mt^2}{2(1 - t^2)(1-
mt^2)} = 0.      \label{5e25}
\end{equation}
Apart from $n$ moving poles, $\chi$ has poles at $t=\pm 1$ and
$t=\pm 1/ \sqrt{m}$. As in the previous section, one can write
$\chi$ with the parity constraint as
\begin{equation}
\chi = \frac{2b_1t}{t^2 -1} +\frac{2md_1t}{t^2 - \frac{1}{m}}+
\frac{P_{n}^{\prime}(t)}{P_n (t)},   \label{5e26}
\end{equation}
which gives the form of $\chi$ in the entire complex plane, where
$P_n(t)$ is yet to be determined.
Note that for this potential the combination $a+b$ and $a-b$ are both
odd {\it i.e.} 3 and 1 respectively. Hence, we use table 5.3 to
obtain all the information regarding the residues at the fixed
poles, number of moving poles of $\chi$, number of linearly
independent solutions and their form for each set etc. by
taking the values of $a=2$, $b=1$, which give $M^{\prime}=0$ and
$N^{\prime}=1$. 
The unknown polynomial in the wave function can be obtained from
\eqref{5e23}, where $G(t)$ for this potential turns out to be
\begin{eqnarray}
G(t) &=& \frac{t^2 (4b_{1}^{2}-2b_1 +\frac{1}{4})-2b_1
+\frac{1}{2}}{(t^2 -1)^2} \nonumber \\ &+& \frac{m^2 t^2 (4d_1 ^2 -2d_1
+\frac{1}{4})- 2md_1 - \frac{3m}{2}}{(mt^2 -1)^2}   \nonumber \\
&+&\frac{2E +8m +(16b_1 d_1 -13)mt^2}{2(1-t^2)(1-mt^2)}.
\label{5e27}
\end{eqnarray}
Using \eqref{5e23} with $G(t)$ from \eqref{5e27} and proceeding in the same way
as is in the previous section, one gets the explicit expressions for the
eigenfunctions and the eigenvalues as given in table 5.4.
\begin{landscape}
\begin{table}[h]
\caption{The residues, the value of $n$, number of linearly
  independent solutions, the band edge eigenfunctions and eigenvalues
  are as follows. Here, $a+b =3$ and $a-b =1$, which give $N^{\prime}
  =1$ and $M^{\prime}= 0$.} 
\vskip2cm
\begin{center}
\begin{tabular}{|c|c|c|c|c|c|c|}
\hline
\multicolumn{1}{|c|}{set}
&\multicolumn{1}{|c|}{$b_1$}
&\multicolumn{1}{|c|}{$d_1$}
&\multicolumn{1}{c}{$n$}
&\multicolumn{1}{|c|}{LI Solution}
&\multicolumn{1}{|c|}{Eigenfunction
 $\psi(x)$ }
&\multicolumn{1}{|c|}{Eigenvalues}
\\ \hline
&&&&&&\\
$1$ & $3/4$ & $5/4$  & -1 & - &                                                -      &        -                 \\
&&&&&&\\
$2$ & $3/4$ & $-1/4$ & 2 &  2 & $\frac{\cn x}{\dn x}(3m\sn ^2 x -2 \pm \sqrt{4-3m})$  &$ 5-3m \pm 2\sqrt{4-3m}$  \\
&&&&&&\\
$3$ & $1/4$ & $5/4$  & 0  & 1 & $\dn ^2 x                                $            & 0                        \\
&&&&&&\\
$4$ & $1/4$ & $-1/4$ & 2 &  2 & $ \frac{\sn x}{\dn x}(3m\sn ^2 x-2-m\pm \sqrt{4-5m+m^2}$ & $5-2m \pm 2\sqrt{m^2 -5m +4}$\\
&&&&&&\\
\hline
\end{tabular}
\end{center}
\end{table}
\end{landscape}
 From the table 5. 4, we see that the first set of residues gives $n
 = -1$, which   
will not be considered as $n$ cannot be negative. Thus, this
particular case of associated Lam\'e potential has 5 band edge solutions,
which can be obtained analytically out of an infinite number of
possible states.

\section{Associated Lam\'e potential with \protect\($a and b$
 \protect\) as half integers }  
The potential studied here is the
supersymmetric associated Lam{\'e} potential with $a = 7/2 \,, b=
1/2$, whose expression is
\begin{equation}
V_{-} = \frac{63}{4}m\, \sn ^2 x +\frac{3}{4}m\frac{\cn ^2 x}{\dn
^2 x} -2 - \frac{29}{4}m + \delta_9     \label{5e28}
\end{equation}
where $\delta_9 = \sqrt{4-4m +25m^2}$.
\begin{equation}
\chi ^2 + \chi ^{\prime} +\frac{2+t^2}{4(t^2-1)^2}+\frac{m^2
t^2-m}{4(mt^2-1)^2}+\frac{4E+8+29m-4\delta_9-65mt^2}{4(t^2
-1)(mt^2-1)} =0  \label{5e29}
\end{equation}
Note that for this case, $a+b = 4$  and
$a-b=3$ are even and odd respectively. Hence, for such cases, one
needs to use  sets 2 and 4 from table 5.2 and sets 1 and 3 from table 5.3,
in order to get the four groups of the eigenfunctions. The solutions for this
potential are given in table 5.5. 

We see that there is a degeneracy in
the band edge 
energy eigenvalue $14-7m+\delta_9$ and all the solutions agree with
the known solutions [\ref{5khar}]. In this study, we have
demonstrated the applicability of QHJ formalism to 
QES periodic potentials. We have been successful in obtaining the
quasi - exact solvability condition and band edge solutions for cases 1
and 2. Case 3, requires some additional work and is not discussed here. 
\begin{landscape}
\begin{table}[h]
\caption{The residues, the value of $n$, number of linearly independent
solutions, the band edge eigenfunctions and eigenvalues are as
follows, with $\delta = \sqrt{4 - 4 m + 25 m^2}$. Here, $a+b =4$ and
  $a-b =3$, which give $N =2$ and $M^{\prime}= 1$.}
\vskip1.5cm
\begin{center}
\begin{tabular}{|c|c|c| c c c|c|c|c|}
\hline
\multicolumn{1}{|c|}{set}
 &\multicolumn{1}{|c|}{$b_1$}
&\multicolumn{1}{|c|}{$d_1$}
&\multicolumn{1}{|c}{     }
 &\multicolumn{1}{c}{$n$}
&\multicolumn{1}{c}{     }
&\multicolumn{1}{|c|}{ LI solutions }
&\multicolumn{1}{|c|}{Eigenfunction $\psi(x)$ }
&\multicolumn{1}{|c|}{Eigenvalues}
\\ \hline
 1  & $3/4$ & 1 & & 1 & & 1 &  $\cn x (\dn x)^{3/2}\sn x$                & $\delta_9 -m +2$  \\
    &       &   & &   & &   &                                 &   \\
 2  & $3/4$ & 0 & & 3 & & 2 &  $\cn x (\dn x)^{3/2}\sn x $               & $\delta_9 -m +2 $ \\
    &       &   & &   & &   &  $\cn x (\dn x)^{-1/2}\sn x (1-2\sn ^2 x)$ &$14 -7m+\delta_9$  \\
    &       &   & &   & &   &               &   \\
 3  & $1/4$ & 1 & & 3 & & 2 &$(\dn x)^{3/2}(12m\,\sn ^2x-5m-2-\delta_9)$ & 0               \\
    &       &   &   & & &   &$(\dn x)^{3/2}(12m\,\sn ^2x-5m-2+\delta_9)$ &$ 2\delta_9$      \\
    &       &   &   & & & &  &  \\
 4  & $1/4$ & 0 & & 4 & & 3 &$(\dn x)^{3/2}(12m\,\sn ^2x-5m-2-\delta_9)$ & 0               \\
    &       &   & &   & &   &$(\dn x)^{3/2}(12m\,\sn ^2x-5m-2+\delta_9)$ & $2\delta_9$  \\
    &       &   & &   & &   & $1-8\sn ^2 x\cn^2 x $                      & $14-7m +\delta_9$    \\
\hline
\end{tabular}
\end{center}
\end{table}
\end{landscape}

     Comparing the two different models of periodic potentials, namely
the ES and QES, we find that the singularity structure of the QMF in
both the cases is similar. For a potential with fixed values of $a$ and
$b$, these four conditions give four groups of solutions as seen as in
table 5.2 and 5.3.  A comparative analysis of periodic QES models with
their non - periodic counterparts, reveals a few differences. For ordinary
QES models, one obtains only one QES condition [\ref{5ge}] in terms of
$n$, which appears in the quantization condition
\eqref{2c43}. Specifying the value of $n$, allows one to pick a
potential from a family of 
potentials. Whereas, in the case of the periodic potentials, in
general, one 
obtains four conditions as given in table 5.1. 
These four conditions can also be viewed as a
constraint between $a$ and $b$. If we select the value of $n$ and the
parameter $a$, one obtains four different values of $b$. For example,
we choose the value $a=7/2$ and $n=4$ and substituting these values in
the 4 conditions in table 5.1, one obtains the values of $b$ as $-5/2,\,
3/2,\, -3/2,\, 1/2$. Among these, the values $-5/2,\, 3/2$ correspond to $q
= 15/4$ and $-3/2,\, 1/2$, correspond $q = 3/4$. Thus, we have two
potentials corresponding to distinct $(p,q)$ values, namely $(63/4, 15/4)$
and $ (63/4, 3/4)$. The form of the solutions for these potentials can
be obtained from tables 5.2 and 5.3. Bot these potentials will have a
group of levels which have a QMF with four moving poles.

\newpage
{\bf References}

\begin{enumerate}

\item {\label{5ge}} K. G. Geojo, S. Sree Ranjani and A. K. Kapoor,
  {\it J. Phys. A : Math. Gen.} {\bf 36}, 4591 (2003); preprint quant
  - ph/0207036. 

\item {\label{5geth}} K. G. Geojo, {\it Quantum - Hamilton Jacobi
  Study of Wave functions and energy spectrum of solvable and quasi -
  exactly solvable models}, {\it Ph. D. thesis} submitted to University of
  Hyderabad (2004).

\item{\label{5qes}}  S. Sree Ranjani, A. K. Kapoor and
  P. K. Panigrahi; preprint quant - ph/0403196. 

\item {\label{5su}}  A. Khare and U. sukhatme, {\it J. Math. Phys.} {\bf 42},
 5652 (2001); preprint quant - ph/0105044.

\item {\label{5khar}} A.Khare and U. Sukhatme, {\it J. Math. Phys.} {\bf
  40}, 5473 (1999); preprint quant - ph/9906044.

\item {\label{5ma}} W. Magnus and S. Winkler, {\it Hills
Equation} (Interscience Publishers, New York, 1966).

\item{\label{5ars}} F. M. Arscot, {\it Periodic Differential equations}
  (Pergamon, Oxford, 1964).




\item {\label{5ha}} H. Hancock, {\it Theory of Elliptic Functions}, (Dover
  Publications, Inc, New York, 1958).


\item {\label{5akk}} S. Sree Ranjani, A. K. Kapoor and
  P. K. Panigrahi, {\it Mod. Phys. Lett. A.} {\bf 19}, No. {\bf 27},
  2047 (2004); preprint quant - ph/0312041.

\end{enumerate}

\chapter{ PT symmetric Hamiltonians }
\section{Introduction}
       Complex Hamiltonians possessing real eigenvalues have
attracted considerable attention in the current literature [\ref{6zno}
  - \ref{6can}]. These
quantal systems are not well understood because of their recent
origin and apart from being counter intuitive. These Hamiltonians
are characterized by a PT symmetry, which is discrete parity $(x
\rightarrow - x)$ followed by
time reversal $(i \rightarrow - i)$ symmetry. In the case, when
the wave functions are also PT symmetric, the eigenvalues are real
and the violation of PT symmetry by the wave function leads to
eigenvalues which are complex conjugate pairs. Apart from
identifying new Hamiltonians belonging to this class, the role of
various discrete symmetries is also under thorough investigation.

      The presence of complex potentials in these systems, makes
them ideal candidates to be probed using the QHJ formalism, since this
approach has been formulated in the complex domain[\ref{6lea6}, \ref{6pad6}]. 
It is extremely interesting to
investigate the properties of the QMF of the PT symmetric
Hamiltonians and to find out their differences and
similarities with ES real potentials and also
the QES ones.
     
    In this chapter, we investigate the structure of the
QMF of a class of PT symmetric Hamiltonians consisting of ES
and QES models, for which the  eigenfunctions and the eigenvalues are
simultaneously obtained. The differences and similarities of these
novel systems, with their ordinary ES and QES counter parts are
clearly brought out.
    
     Application of the QHJ formalism to the PT symmetric potentials
requires  an approach  different from that used earlier. In the
absence of a generalization of the  
oscillation theorem, it is not clear whether the quantization rule
\eqref{2c38} is valid. If so what contour  should be used. In this
case, the quantization condition still holds for a contour which
encloses all the moving poles. We shall assume this to be the case for
the PT symmetric models to be taken up in this chapter. Here we will
analyze two PT symmetric potentials namely,   
the Khare - Mandal potential in \S 6.2 and the complex Scarf potential
 $ V(x) = -A \sech ^{2}x - iB \tanh x\,\, \sech x $, where $A > 0$ in
\S 6.3.
\section{Khare - Mandal model}
    The potential expression for the Khare- Mandal model is $V(x) =
- (\zeta \cosh2x - iM)^2$. This potential has either complex or real
eigenvalues depending on whether $M$ is odd or even [\ref{6kha},
 \ref{6bag}]. Using the QHJ 
formalism, we obtain the QES condition, when $M$ is odd and $M$ is  even. We
also obtain the explicit expressions for the eigenvalues and
eigenfunctions for the cases $M=3$ and $M=2$. 
The QHJ equation in terms of $q \equiv ip$,
keeping $\hbar = 2m =1$ and 
 after the change of variable $
t \equiv \cosh 2x$ becomes
\begin{equation}
q^2 + 2 \sqrt{t^2 - 1} \frac{dq}{dt} + E + (\zeta t - i M)^2  =0.
\label{6e5} 
\end{equation}  
The transformation equations in \eqref{2c43} for this change of variable become
\begin{equation}
q = 2(\sqrt{t^2 - 1})\phi   ,\,\,\, \phi=\chi - \frac{t}{2(t^2 -1)},
\label{6e6}  \end{equation}  
which transforms \eqref{6e5} into
\begin{equation}
\chi^2 +\frac{d\chi}{dt} +\frac{t^2 +2}{4(t^2 -1)^2} + \frac{E + (\zeta 
  t - iM)^2}{4(t^2 -1)}  =0,   \label{6e7}
\end{equation}  
which will be treated as the the QHJ equation and $\chi$ as the QMF..  

\subsection{Form of QMF ($\chi$)}
 We shall assume that $\chi$ has a finite number of moving poles 
in the complex $t$ plane and that the point at infinity is an isolated
 singularity. Besides the moving poles, $\chi$ has fixed poles at $t = \pm 1$. 
It is seen from \eqref{6e7} that the function $\chi$ is bounded at 
$t=\infty$. Assuming that $\chi$ has only these above mentioned singularities,
we separate the singular part of $\chi$ and write it in the  following
form.
\begin{equation}
\chi = \frac{b_1}{t-1}+ \frac{b^{\prime}_1}{t+1}+
\frac{P^{\prime}_n(t)}{P_n(t)} + C,    \label{6e8}
\end{equation}  
where $b_1$ and $b^{\prime}_1$ are the residues at fixed poles $t=\pm
1$ and  $P_n(t)$ is a polynomial of degree $n$. $C$ gives the analytic part of
$\chi$ and is a constant due to the Louville's theorem. From
\eqref{6e7}, one can sec that $\chi$ goes as $\pm i\xi/2$, for large
$t$, which are the values of $C$.

   To find the residues at the fixed poles $t= \pm 1$, we
expand $\chi$ in Laurent series around each pole 
and proceeding in the same way as
in the earlier chapters, one gets the following two values for $b_1$ and
$b_1^\prime$ as 
\begin{equation}
b_1 = \frac{3}{4},\,\,\, \frac{1}{4},\,\,\,\,     
\,\,\,b^{\prime}_1 = \frac{3}{4} ,\,\,\ \frac{1}{4}.      \label{6e10}
\end{equation}  
It has been assumed that the point at infinity is an isolated
singularity. 
In order to find the residue of $\chi$ at infinity, one expands $\chi$
in terms of the Laurent series for large $t$ as follows,
\begin{equation}
\chi = a_0 +\frac{\lambda_1}{t} + \frac{\lambda_2}{t^2} +........
\label{6e12}
\end{equation}  
 and substitution of which in \eqref{6e7}, gives the residue at
 $\infty$ as 
\begin{equation}
\lambda_1 = \frac{im \zeta }{4a_0}\label{6e13} 
\end{equation}  
along with $a_0 = \pm \frac{i \zeta }{2}$, due to which the leading
behaviour of $\chi$ at infinity takes two values, which turns out to be
\begin{equation}
\lambda_1 = \frac{M}{2}, \qquad \frac{-M}{2}.     \label{6e14}
\end{equation}  
Hence, using the fact, that for a rational function, the sum of all the
residues is equal to zero, one obtains
\begin{equation}
b_1 + b^{\prime} + n = \lambda_1      \label{6e15}
\end{equation}  
From \eqref{6e10}, we see that the right hand side of
\eqref{6e15} is positive. Hence, for \eqref{6e15} to be true, we
choose only the 
positive value of $\lambda_1 = \frac{M}{2}$, which fixes the
value of residue at infinity.  This means, we choose
$C = a_0 =  +i\xi/2$, since $a_0$ gives the leading behaviour of $\chi$
at infinity. It should be noted that there is no way of choosing a particular
value of residue at the fixed 
poles, since one does not have information regarding the square
integrability of the solutions. Hence, one needs to consider both the
values of $b_1$ and 
$b^{\prime}_1$. Thus taking all possible combinations of $b_1$ and
$b^{\prime}_{1}$ in \eqref{6e15}, one 
obtains the QES condition for each combination along with a constraint
on $M$ as given in table 6.1.
\begin{table}[h]
\caption{The QES condition and the number of moving
  poles of $\chi$ for each combination of $b_{1}$ and $b^{\prime}_1$.  } 
\vskip0.5cm
\begin{center}
\begin{tabular}{|c| c| c| c| c| c|}
\hline
\multicolumn{1}{|c|}{set}
&\multicolumn{1}{|c|}{$b_1$}
&\multicolumn{1}{|c|}{$d_1$}
&\multicolumn{1}{|c|}{$n = \lambda -b_1 - b^{\prime}_1$}
&\multicolumn{1}{|c|}{Condition on M}
&\multicolumn{1}{|c|}{QES condition}
\\ \hline
  &              &              &                             &  & \\ 
1 & $\frac{1}{4}$ & $\frac{1}{4}$ & $\frac{M}{2}-\frac{1}{2}$ & $ M =
odd,\, M \ge 1 $    & $M =2n+1$  \\
&&&&&\\
2 & $\frac{3}{4}$ & $\frac{3}{4}$ & $\frac{M}{2}-\frac{3}{2}$ & $ M =
odd,\, M  \ge 3 $   & $ M= 2n +3$ \\
&&&&&\\
3 & $\frac{3}{4}$ & $\frac{1}{4}$ & $\frac{M}{2}-1$              &$ M =
even,\, M \ge 2$    & $M = 2n +2$ \\
&&&&&\\
4 & $\frac{1}{4}$ & $\frac{3}{4}$ & $\frac{M}{2}-1$           &$ M =
even,\, M \ge 2$    & $M = 2n +2$ \\ 
 &  &             &               &                           &   \\
\hline
\end{tabular}
\end{center}
\end{table}
From the table 6.1, we see that sets 1 and 2 are
valid, only when $M$ is odd and sets 3 and 4 are valid, only when $M$ is even.

\subsection{Forms of the wave function}
    Doing the change of variable and writing $p$ in terms of $\chi$ in
\eqref{2c45}, one gets the equation for $\psi(t)$ as
\begin{equation}
\psi(t) = \exp \int \left( \frac{b_1}{t-1} + \frac{b^{\prime}_1}{t+1}
+ \frac{P^{\prime}_n(t)}{P_n(t)} +\frac{i \zeta }{2} -
\frac{t}{2(t^2-1)} \right) dt.   \label{6e17}
\end{equation}  
Hence, one can substitute the values of $b_1$ and $b_1^{\prime}$ from
sets 1 and 2 in 
\eqref{6e17}, if $M$ is odd and sets 3 and 4 if $M$ is even, to obtain
the form of the wavefunction. The expression for the wave function is
in terms of the unknown  polynomial $P_n(t)$, where $n$ gives the
number of zeros of $P_n(t)$. In order to obtain the polynomial, we
substitute $\chi$ from \eqref{6e8} in \eqref{6e7} to get
\begin{eqnarray}
&&\frac{P^{\prime\prime}_n(t)}{P_n(t)} + \frac{2P^{\prime}_n(t) }{P_n(t)}\left(
\frac{b_1}{t-1} + \frac{b^{\prime}_1}{t+1}+ \frac{i \zeta }{2}\right)
+\frac{b^{2}_1 - b_1}{(t-1)^2} +\frac{(b^{\prime})^{2}_1 -
  b^{\prime}_1}{(t+1)^2} +\frac{t^2 +2}{4(t^2-1)^2} \nonumber \\  
&&\,+\frac{E+(\zeta t-iM)^2 +8b_1 b^{\prime}_1 - 4\zeta  ^2 (t^2
    -1)}{4(t^2 -1)} +i \zeta  \left( 
\frac{b_1}{t-1} +\frac{b^{\prime}_1}{t+1}\right) =0.   \label{6e18}   
\end{eqnarray}
 This turns out to be equivalent to $n$ linear homogeneous equations,
 for the coefficients of 
 different powers of $t$ in $P_n(t)$. The energy eigenvalues are obtained
 by setting the corresponding determinant equal to zero. The explicit
 eigenvalues and eigenfunctions are obtained for $M=3$ and $M=2$
 cases. 
\subsection{Case 1 : \( \protect $M = 3$ \protect \)}
    Here, $M$ is odd, so we can use sets 1 and 2 from table 6.1
 and get the required results as below.\\
{\it Set 1 :} $ b_1 =\frac{1}{4}$, $b^{\prime}_1 = \frac{1}{4}$ and
 $n=1$.\\
  This implies $P_n$ is a first degree polynomial, say, $Bt
 +C$. Substituting these values in \eqref{6e18} and comparing various
 powers of $t$, one obtains a $2 \times 2$ matrix for $B$ and $C$ as
 follows

\begin{equation}
\left( \begin{array}{cc}
         1+ \frac{E-9+ \zeta  ^2}{4}    &  -i \zeta    \\
        -i \zeta         &  \frac{E-9+\zeta  ^2}{4}
\end{array}   \right)  \left( \begin{array}{c}
     B \\ C
\end{array}  \right)   = 0.    \label{6e19}
\end{equation}
Equating the determinant of this matrix to zero, one obtains the two
values for energy and the polynomials as
\begin{equation}
E= 7- \zeta  ^2 \pm 2 \sqrt{1 - 4 \zeta  ^2}, \qquad P_1 = \frac{B}{2}(2t - \frac{i}{\zeta }(1 \pm \sqrt{1-4\zeta ^2})).  \label{6e20}
\end{equation}  
Substituting the values of $b_1$, $b^{\prime}_1$ and $P_1$ in
\eqref{6e17}, gives the two eigenfunctions corresponding to the two
eigenvalues as
\begin{equation}
\psi(x) = e^{{\frac{i \zeta }{2}\cosh2x}}  \left(2\cosh2x
    -\frac{i}{\zeta }(1 \pm  \sqrt{1-4\zeta ^2})\right).    \label{6e22} 
\end{equation}  
{\it Set 2:} $ b_1 =\frac{3}{4}$, $b^{\prime}_1 = \frac{3}{4}$ and
 $n=0$.\\
Here, we see that $n=0$ implies $P_n(t)$ is a constant. Substituting these
 values in \eqref{6e18} and proceeding in the same as before, one obtains
\begin{equation}
E= 5- \zeta  ^2 ,\qquad  \psi(x) =e^{{\frac{i \zeta }{2} \cosh 2x}}
\sinh 2x,
\label{6e23}   
\end{equation}  
which match with the known results  [\ref{6kha}, \ref{6bag}].

\subsection{Case 2 : \( \protect $ M=2$ \protect \)}
In this case, one makes use of sets 3 and 4 in table 6.1 and proceed in
the same way as was done for case 1.\\
{\it Set 3 :} $ b_1 =\frac{1}{4}$, $b^{\prime}_1 = \frac{3}{4}$ and
 $n=0$.\\
The eigenvalues and eigenfunctions obtained are
\begin{equation}
E = 3- \zeta ^2  +2i \zeta  \, , \qquad   \psi(x) = e^{{\frac{i
      \zeta }{2}\cosh 2x}} (\cosh 2x +1)^{1/2}.    \label{6e24}
\end{equation}  
Similarly, for\\
{\it Set 4 :} $ b_1 =\frac{3}{4}$, $b^{\prime}_1 = \frac{1}{4}$ and
 $n=0$,\\ 
one obtains
\begin{equation}
E = 3- \zeta ^2  - 2i \zeta   \, ,\qquad   \psi(x) = e^{{\frac{i
      \zeta }{2}\cosh 2x}} (\cosh 2x -1)^{1/2}.    \label{6e25}
\end{equation}  
These match with the solutions given in [\ref{6bag}]. Thus, for any given
positive value of $M$, odd or even, one can obtain the eigenvalues and
eigenfunctions for the Khare - Mandal potential. In the next section,
we go over to the study of complex Scarf -II potential.  

\section{Complex Scarf - II potential}
  The expression for the potential is 
\begin{equation}
V_{-} = - A \sech^{2}x - iB \sech x \tanh x.   \label{6e26} 
\end{equation}  
The QHJ equation in terms of $q$, where $q=\frac{d \psi}{dx}$ is
\begin{equation}
q^2 +\frac{dq}{dx} + E + A \sech^{2}x + iB \sech x \tanh x =0.    \label{6e27}
\end{equation}  
Doing the change of variable $y= i \sinh x$ and proceeding in the same
as was done for other the potentials, one obtains the  QHJ equation for
$\chi$, with $F(y) = i \sqrt{1 - y^2}$, as follows
\begin{equation}
\chi^2 +\frac{d\chi}{dy} + \frac{2+y^2}{4(1-y^2)^2} - \frac{E}{1-y^2}
- \frac{A + B y}{(1-y^2)^2}  =0,    \label{6e28}
\end{equation}  
where the transformation equations \eqref{2c43} are
\begin{equation}
 q = i
(\sqrt{1-y^2})\phi \, , \qquad \chi = \left( \phi - \frac{y}{2(1-y^2)} \right).    \label{6e29}
\end{equation}  
Along with the $n$ moving poles with residue one, $\chi$ has poles at
$y= \pm 1$. We assume that except for these poles, there are no other
singularities and the point at infinity is an isolated singularity.
Proceeding in the same way as in previous section, we obtain the
residues at the poles and at infinity as follows.
The residue at $y=-1$
\begin{equation}
b^{\prime}_1 = \frac{1}{2} \pm \frac{1}{2}\sqrt{\frac{1}{4} + A -B}.   \label{6e30}
\end{equation}  
The residue at $y = 1$
\begin{equation}
b_1 = \frac{1}{2} \pm \frac{1}{2}\sqrt{\frac{1}{4} + A +B}   \label{6e31}
\end{equation}  
and the residue at infinity is
\begin{equation}
\lambda = \frac{1}{2} \pm i\sqrt{E}.     \label{6e32}
\end{equation}  
As seen in the earlier section, one can write $\chi$ in terms of its
analytic and singular parts as
\begin{equation}
\chi = \frac{b_1}{y-1}+ \frac{b^{\prime}_1}{y+1}+
\frac{P^{\prime}_n}{P_n} + C    \label{6e33}
\end{equation}  
where, $C$ is an analytic part of $\chi$ and is a constant due to
Louville's theorem and it turns out to be zero. 
Using the fact, that the sum of all the residues is zero for a rational
function, we obtain the expression for the energy eigenvalue  as
\begin{equation}
-E = (b_1 + b^{\prime}_1 +n - \frac{1}{2})^2.     \label{6e35}
\end{equation}  
The wave function  in terms of $\chi$ can be written using
\eqref{2c45} as
\begin{equation}
\psi(y) = \exp \left( \int \left( \frac{b_1}{y-1}+\frac{b^{\prime}_1}{y+1}
+\frac{P^{\prime}_n}{P_n} + \frac{y}{2(1-y^2)} \right)\right) dy,
  \label{6e36}  
\end{equation} 
which is equal to 
\begin{equation}
\psi(y) = (y-1)^{b_1 - \frac{1}{4}} (y+1)^{b^{\prime}_1 - \frac{1}{4}}
P_n(y).     \label{6e37}
\end{equation} 
To obtain the polynomial $P_n(y)$, one needs to solve,
\eqref{6e28} by substituting $\chi$ from \eqref{6e33}, which gives the following differential equation
\begin{equation}
P^{\prime\prime}_n + 2P^{\prime}_n \left(\frac{b_1}{y-1} +
\frac{b^{\prime}_n}{y+1}\right) + G(y)P_n = 0,    \label{6e38} 
\end{equation}  
where
\begin{eqnarray}
G(y) &=& \frac{(4(b^2 _1 -b_1 +b^{\prime 2}-b^{\prime}_1)) 
+1 + 4E + 8b_1 b_1^{\prime} )y^2}{(y^2-1)^2} \nonumber \\
& &+ \frac{ 2y(4(b^2 _1 -b_1
  -b^{\prime 2}_1+b^{\prime}_1) - 2B))}{(y^2-1)^2} \nonumber \\
& &+\frac{(4(b^2 _1 - b_1 + b^{\prime 2} -b^{\prime}_1) +2 - 4A
  -4E - 8b_1 b_1^{\prime} )}{(y^2-1)^2}. \nonumber
\end{eqnarray}
 Substituting the expression for $E$ from \eqref{6e35} in\eqref{6e38}
  gives,
\begin{equation}
(1-y^2)P_n^{\prime\prime} +  (2(b^{\prime}_1 - b_1) - 2 (b_1 +
  b^{\prime}_1)y)P_n^{\prime} +n(n+2(b_1 +b^{\prime}_1 -1) +1 )P_n =0    \label{6e39}
\end{equation}  
which is in the form of the Jacobi differential equation and hence the
polynomial $P_n (y)$ is proportional to the Jacobi
polynomial $ P^{2b^{\prime}_1 -1, 2b_1 -1} _ n (y)$. Thus, the complete expression for the wave function in
terms of the residues and the Jacobi polynomial in the $x$ variable
becomes
\begin{equation}
\psi(x) = (i \sinh x -1)^{b_1 - \frac{1}{4}} (i \sinh x +1)^{b^{\prime}_1
    -\frac{1}{4}} P ^{\,\, 2b^{\prime}_1 -1, 2b_1 -1} _n (i \sinh x).    \label{6e40}  
\end{equation}  
   Note that in this whole
process, we had written the expression of the eigenvalues and the
eigenfunctions in terms of $b_1$ and $b^{\prime}_1$, which have two
values. No particular value has been chosen. Hence, we need to choose
one value of each residue to remove 
this ambiguity. For this case, unlike the Khare-Mandal potential, the
solutions are known to be square integrable. Therefore, we  make use
of square integrability to choose one of 
the values of residues. We consider two different ranges of
the potential parameters $A$ and $B$ and for each range, obtain the
solutions.

\subsection{ Case 1 : $ |B| > A + \frac{1}{4}$}

With this restriction on $A$ and $B$, the residues at $y = \pm 1$
becomes $b^{\prime}_1 = \frac{1}{2} \pm \frac{i}{2}\sqrt{B-A-\frac{1}{4}}$
and $b_1 = \frac{1}{2} \pm
\frac{1}{2}\sqrt{A+B+\frac{1}{4}}$ and thus, the wave function in \eqref{6e40}
becomes 
\begin{equation}
\psi(x) = (i \sinh x -1)^{\frac{1}{4}\pm
  \frac{s}{2}} (i \sinh x +1)^{\frac{1}{4} \pm
  \frac{ir}{2}} P ^{\,\,\pm ir, \pm s } _n (i \sinh x), \label{6e42} 
\end{equation}
where $r = \sqrt{B-A-\frac{1}{4}}$ and $s =
\sqrt{A+B+\frac{1}{4}}$. 
From the above equation, one can see that for $x \rightarrow \infty$,
$\psi(x) \rightarrow 0$, for 
\begin{equation}
b^{\prime}_1 = \frac{1}{2} \pm \frac{i}{2}\sqrt{B-A -\frac{1}{4}} \,\,\,\,
\,\,\,\,\, b_1 = \frac{1}{2} - \frac{1}{2}\sqrt{A+B+\frac{1}{4}},
\label{6e444} 
\end{equation}
with  $n$ limited to lie between $ 0 \leq
 n < \frac{1}{2}\sqrt{A+B+\frac{1}{4}} - \frac{1}{2}$.
Thus, for this parameter range, expression for  $\psi(x)$ will be
\begin{equation}
\psi(x) = (i \sinh x -1)^{\frac{1} {4} - \frac{s}{2}} (i \sinh x
  +1)^{\frac{1}{4} \pm \frac{ir}{2}} P^{\pm ir, -s}_n (i \sinh x)
  \label{6e44a} 
\end{equation}
 and the corresponding expression for energy from \eqref{6e35} will be
\begin{equation}
E = -\left(n + \frac{1}{2}-\frac{1}{2}\left(\sqrt{A+B+\frac{1}{4}}
+\pm i \sqrt{B-A-\frac{1}{4}}\right)\right)^2.    \label{6e45}
\end{equation}

\subsection{ Case 2 : $|B| \le A +\frac{1}{4}$}
In this case, the wave function in \eqref{6e40} is square integrable, if
\begin{equation}
b^{\prime}_1 = \frac{1}{2}- \frac{1}{2}\sqrt{\frac{1}{4} +A-B} \, ,\qquad
b_1 = \frac{1}{2}- \frac{1}{2}\sqrt{\frac{1}{4} +A+B}.
\label{6e46}
\end{equation}
The expression for the wave function is given by,
\begin{equation}
\psi(x) = (i \sinh x -1)^{\frac{1}{4} -\frac{s}{2}} (i \sinh x
+1)^{\frac{1}{4} - \frac{t}{2}} P^{-t, -s}_n (i \sinh x),
\label{6e47}
\end{equation}
where $t = \sqrt{\frac{1}{4} + A -B}$ and $ s = \sqrt{\frac{1}{4} +A
  +B}$ and the  energy eigenvalue expression in \eqref{6e35} becomes
\begin{equation}
E = -\left( n+\frac{1}{2} -\frac{1}{2}\left
(\sqrt{\frac{1}{4}+A-B} - \sqrt{\frac{1}{4}+A+B} \right ) \right )
^2,   \label{6e45a}
\end{equation}
with  $n$ restricted by 
\begin{equation}
0 \leq n < \frac{1}{2}\sqrt{\frac{1}{4} +A -B } +
\frac{1}{2}\sqrt{A+B+\frac{1}{4}} - \frac{1}{2}.    \label{6e46a}
\end{equation}

     In conclusion, PT symmetric potentials belonging to the QES and ES
class have been investigated through the QHJ formalism. The QES
solvable Khare-Mandal potential has complex or real eigenvalues,
depending on whether the
 potential parameter $M$ is odd or even.
The singularity structure of the QMF for these two cases is
different.  For the case, when $M$ is odd, one observes from table
6.1, that the solutions fall into two groups, which consist of
solutions coming from sets 1 and 2. For a solution belonging to a
particular group, the number of singularities of the QMF are fixed
and consists of both real and complex locations. This kind of
singularity structure of the QMF has been observed in the study of
periodic potentials [\ref{6akk}, \ref{6qpp}]. Though the solutions, for $M$
even, fall into two groups coming from sets 3 and 4,  they  all
have the same number of singularities, which again can consist of
complex and real poles. This singularity structure is same as that
observed in the ordinary QES models [\ref{6geo}].

    Coming to the case of exactly solvable PT symmetric potential, the
location of the moving poles can be either real or complex. In the
specific example of complex Scarf potential, it turns out that all
the moving poles are off the real line. In contrast for the
ordinary ES models [\ref{6sree}], the moving poles are always
real. For both the 
cases, the number of moving poles of the QMF characterize the
energy eigenvalues.\\

\noindent
{\bf References}

\begin{enumerate}

\item{\label{6zno}} M. Znojil, {\it J. Phys. A : Math. Gen.} {\bf 33},
  4561 (2000); preprint quant - ph/9912027.

\item{\label{6kha}} A. Khare and B. P. Mandal; preprint quanth - ph/0004019.

\item{\label{6zaf}} Z. Ahmed, {\it Phys.Lett. A.} {\bf 282}, 343 (2001).

\item{\label{6bag}} B. Bagchi, S. Mallik, C. Quesne and
  R. Roychoudhury; preprint quant - ph/0107095.  

\item{\label{6lev}} G. L{\'e}vai and M. Znojil, {\it
  Mod. Phys. Lett. A.} {\bf 16}, 1973 (2001); preprint quant - ph/0110064.

\item{\label{6gl}}  G. L{\'e}vai and M. Znojil, {\it J. Phys. A.} {\bf
    35}, 8793 (2002); preprint quant - ph/0206013.

\item{\label{6oz}} O. Yesiltas, M. Simsek, R. Sever and C. Tezcan,
  {\it Phys. Scripta}, {\bf 67}, 472 (2003); preprint hep - ph/0303014. 

\item{\label{6ba}} B. Bagchi, C. Quesne and M. Znojil, {\it Mod. Phys.
  Lett. A.} {\bf 16}, 2047 (2001); preprint quant - ph/0108096.

\item{\label{6can}} B. Bagchi, F. Cannata and C. Quesne, {\it
  Phys. Lett. A.} {\bf 269}, 79 (2000); preprint quant - ph/0003085.

\item {\label{6lea6}} R. A. Leacock and M. J. Padgett, {\it
  Phys. Rev. Lett.} {\bf 50}, 3 (1983). 

\item {\label{6pad6}} R. A. Leacock and M. J. Padgett, {\it
  Phys. Rev.} {\bf D28}, 2491 (1983).

\item{\label{6akk}} S. Sree Ranjani, A. K. Kapoor and P. K. Panigrahi,
 to be published in {\it Mod. Phys. Lett. A.} {\bf 19}, No. {\bf 27},
 2047 (2004); preprint quant - ph/0312041 

\item{\label{6qpp}}  S. Sree Ranjani, A. K. Kapoor and
  P. K. Panigrahi; quant - ph/0403054.

\item {\label{6geo}} K. G. Geogo, S. Sree Ranjani and A. K. Kapoor,
  {\it J. Phys. A : Math. Gen.} {\bf 36}, 4591 (2003); preprint quant
  - ph/0207036. 

\item {\label{6sree}} S. Sree Ranjani, K. G. Geojo, A. K Kapoor and
  P. K. Panigrahi, to be published in {\it Mod. Phys. Lett. A.} {\bf
  19}, No. {\bf 19}, 1457 (2004);  preprint quant - ph/0211168. 
\end{enumerate}

\chapter{Conclusions}

           In this  thesis, we have studied various potential models
with special spectral properties using QHJ formalism. The various
models, being potentials which exhibit different spectra
for different ranges of parameters, potentials with band structure and
potentials 
with both real and complex eigenvalues. 
For all the models studied, we obtained the eigenvalues and the
eigenfunctions successfully. Thus, the QHJ formalism provides an
alternate method to solve various types of ES and QES models. The
added advantage of this method is, that it is very simple and makes use
of well known complex variable theorems. 
  
  The most important steps in obtaining the solutions, for both ES and QES
models, have been, the choice of the change of
variable and use of the singularity structure of
the QMF in the complex plane. We did not attempt to derive the
singularity structure of the QMF within the QHJ formalism. However, we
assumed that the singularity structure is very simple {\it viz.} {\it
the QMF has a finite number of moving poles}. With this
assumption, we could obtain the bound state and band edge solutions in a
most straight forward fashion, even if the equations to be solved  were
not of a simple form.  The fact that one arrived at correct solutions
in all the cases, supports a conjecture that the assumption may 
be true for all ES and QES models. It should be remarked that this
assumption is equivalent to the property that `the QMF', becomes a
rational function after the suitable transformations. Stated
differently, this assumption is equivalent to, the point at infinity
being an isolated singular point. In this work  our study has been
limited to one dimensional models. It will be interesting to
investigate if any of these forms of the assumption remain valid for higher
dimensional models.

In all the models, the integer $n$
appearing in the exact quantization condition is the number of moving
poles of the QMF. For ES models [\ref{7sree}] and for Scarf potential, these
poles are located 
on the real axis only, with no moving poles off the real axis. Given a
value of $n$, there is a unique energy level and
the corresponding  eigenfunction has only real, $n$, zeros.
However, for the QES models [\ref{7geo}], $n$ appears  as a parameter in the
potential and of the infinite possible states, only a finite number,
determined by $n$, of states can be obtained
analytically. In the case of QES models, the moving poles appear at
complex locations also.  Specifying a value of $n$, selects a
potential within a 
family and does not pick up a unique level for the potential. In fact,  
all the analytically solvable  eigenfunctions have the same number  $(\equiv
n)$ of zeros, of which the number of real zeros are in accordance with
the oscillation  theorem [\ref{7in}]. 

 The periodic potentials in chapters IV [\ref{7akk}], however, show a richer 
distribution of the moving poles in the complex plane. 
 Like ordinary QES models, the moving poles of the QMF are located both on
and off the  real axis.
Here, in this case, selecting a value of the integer $n$, picks up a
subset of solutions all having $n$ zeros just like QES
models. However, unlike the QES models, varying the 
integer $n$ does not give a new potential, instead it gives a different
subset of levels. 

    The case of QES periodic potentials [\ref{7ppqes}] of chapter V
is very similar to the ES periodic potentials, as far as the
singularity structure of the QMF is concerned.

The above conclusions have been arrived in the context of specific
models studied. However, we expect them to remain generally valid for other one
dimensional, periodic and aperiodic, ES and QES potentials. The number
of PT symmetric models [\ref{7pt}] studied here is too small to allow
any similar generalizations.

   Our study exhausts analytically solvable bound state and band edge
solutions for  both ES and QES cases of one dimensional potentials.   
For problems which are not ES,  making a proper guess about the singularity
 structure of the QMF, does not appear 
to be possible.  A study
of the distribution of the moving poles in the complex plane may have
a relation to the classical properties of the system. Such a study is
interesting in its own right 
and can be best done numerically. We expect that, here also, the
established results in the complex variable theory will provide a
useful scheme to obtain numerical, possibly  approximate analytical
solutions for the wave functions and energies. Finally, we hope  that
a useful extension to continuous energy states in one dimension and to
non separable models in higher dimensions can be found.\\
 
\noindent
{\bf References}

\begin{enumerate}

\item{\label{7sree}}  S. Sree Ranjani, K. G. Geojo, A. K Kapoor and
  P. K. Panigrahi, to be published in {\it Mod. Phys. Lett. A.} {\bf
  19}, No. {\bf 19}, 1457 (2004); preprint quant - ph/0211168.  

\item{\label{7geo}}  K. G. Geojo, S. Sree Ranjani and A. K. Kapoor,
  {\it J. Phys. A : Math. Gen.} {\bf 36}, 4591 (2003); quant - ph/0207036.

\item{\label{7in}} E. L. Ince, {\it Ordinary Differential Equations}
  (Dover Publications, Inc, New York, 1956).

\item {\label{7akk}} S. Sree Ranjani, A. K. Kapoor and P. K. Panigrahi,
  to be published in {\it Mod. Phys. Lett. A.} {\bf 19}, No. {\bf 27},
  2047 (2004); preprint quant  - ph/0312041 

\item{\label{7ppqes}}  S. Sree Ranjani, A. K. Kapoor and P. K. Panigrahi;
  preprint quant - ph/0403196.

\item{\label{7pt}}  S. Sree Ranjani, A. K. Kapoor and P. K. Panigrahi;
  preprint quant - ph/0403054. 

\end{enumerate}

\end{document}